\documentclass[aps,pra,twocolumn,superscriptaddress]{revtex4-2}
\usepackage{graphicx}    % needed for figures
\usepackage{dcolumn}    % needed for some tables
\usepackage{amssymb}   % for math
\usepackage{amsmath}    % for multiline equation
\usepackage{amsthm}
\usepackage{commath}   % for absolute value
\usepackage{subfigure}    % for subfigure
\usepackage{braket}
\usepackage{natbib}
\usepackage{float}
\usepackage{color}
\usepackage{siunitx}
\usepackage[colorlinks=true, allcolors=blue]{hyperref}
\usepackage{hyphenat}
\usepackage{notes2bib}
\usepackage{MnSymbol}
\usepackage{tabularx}
\usepackage{array}
\usepackage{dsfont}
\usepackage{physics}
\usepackage{empheq}
\usepackage{makecell}

\newcommand{\bk}{\text{bk}}
\newcommand{\hc}{\text{h.c.}}
\newcommand{\eff}{\text{eff}}
\newcommand{\da}{\dagger}
\newcommand{\bknh}{\text{bk,NH}}
\newcommand{\nh}{\text{NH}}
\newcommand{\jal}{{(j,\alpha)}}
\newcommand{\tcaL}{\tilde{\mathcal{L}}}
\newcommand{\pp}{{\prime\prime}}

\graphicspath{{Figures/}}

\begin{document}

\title{Enhancing Microwave-Optical Bell Pairs Generation for Quantum Transduction Using Kerr Nonlinearity}

\author{Fangxin Li}
\email{fangxinli@uchicago.edu}
\affiliation{Department of Physics, University of Chicago, Chicago, IL 60637, USA}

\author{Ming Yuan}
\affiliation{Pritzker School of Molecular Engineering, University of Chicago, Chicago, IL 60637, USA}

\author{Zhaoyou Wang}
\affiliation{Pritzker School of Molecular Engineering, University of Chicago, Chicago, IL 60637, USA}

\author{Changchun Zhong}
\affiliation{Pritzker School of Molecular Engineering, University of Chicago, Chicago, IL 60637, USA}
\affiliation{Department of Physics, Xi'an Jiaotong University, Xi'an, Shaanxi 710049, China}

\author{Liang Jiang}
\email{liang.jiang@uchicago.edu}
\affiliation{Pritzker School of Molecular Engineering, University of Chicago, Chicago, IL 60637, USA}

\date{\today}

\begin{abstract}

Microwave-optical quantum transduction can be achieved via quantum teleportation using microwave-optical photon Bell pairs. The standard spontaneous parametric down-conversion (SPDC) has to trade off between generation fidelity and probability due to unwanted higher-excitation pairs in the output. In this work, we propose a pulsed SPDC scheme that employs strong Kerr nonlinearity in the microwave mode. This nonlinearity causes significant detuning of higher excitations due to the anharmonicity of energy levels, and the system can be pulse-driven to produce single-photon pairs in the output. Our pulsed nonlinear approach can generate high-fidelity Bell pairs with high probability, alleviating the trade-off between fidelity and probability inherent in traditional SPDC schemes. We optimize both the pulse width and driving strength, demonstrating that our protocol outperforms the SPDC scheme in a realistic setting of finite nonlinearity and intrinsic photon loss.

\end{abstract}

\maketitle

\emph{Introduction}\textemdash
Modern quantum networks are envisioned as consisting of widely distributed computational nodes connected by flying optical photons \cite{Kimble2008,Cirac1997}. At the current stage of development, local computational nodes have been realized across a variety of physical platforms. To integrate platforms operating at disparate frequency ranges, quantum transduction—which coherently transfers quantum states between microwave and optical photons \cite{Lauk2020,Lambert2020, Han2021}—is a crucial component in constructing future quantum networks. 

Microwave-to-optical quantum transduction can be realized through effective beam splitter coupling, which requires intrinsic nonlinearity in the material and a pump to supply the large energy difference \cite{Andrews2014, Fan2018, Rueda2019, Jiang2020, Mirhosseini2020}. Alternatively, we may first generate entangled microwave-optical photon pairs and then complete the state transfer via quantum teleportation, which consumes a microwave-optical photon Bell pair \cite{Zhong2020PRL, Krastanov2021, Wu2021, PhysRevApplied.22.064047}. In comparison, this entanglement-based quantum transduction protocol has demonstrated a larger quantum capacity region for information transmission in a practical parameter space \cite{Zhong2022}. In addition, the non-classical correlation generated between the microwave and optical photons can be integrated into the DLCZ protocol for linking remote superconducting processing nodes \cite{Duan2001}. 

The first step in the entanglement-based quantum transduction is to generate entangled microwave-optical Bell pairs, which has been demonstrated by recent experiments utilizing spontaneous parametric down-conversion (SPDC) interactions in platforms such as electro-optical and piezo-optomechanical systems \cite{Sahu2023,Meesala2024,PhysRevX.14.031055}. However, the microwave-optical photon Bell pairs produced in current experiments are limited in fidelity and generation rate due to undesirable higher-order processes generating multiple excitation pairs. 

\begin{figure}[b]
\centering
\includegraphics[width=1\columnwidth]{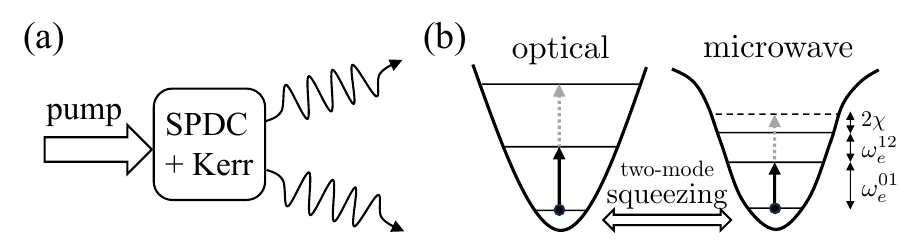}
\caption{(a) SPDC entanglement generation with strong Kerr nonlinearity. (b) Inhibition of higher excitation pair due to energy detuning $2\chi$ from Kerr non-linearity. \label{Fig1}}
\end{figure}

In this paper, we propose deterministically generating single microwave-optical photon Bell pairs by introducing strong Kerr nonlinearity in the microwave mode in addition to the SPDC interaction between the optical and microwave modes. Due to the Kerr-induced anharmonicity in the energy levels, the higher-excitation states become detuned from the laser pump, and the only on-resonant excitation produces single-photon pairs. As illustrated in Fig.~\ref{Fig1}, we can blockade higher excitations. Numerical and analytical calculations of the output state, based on the input-output theory with quantum pulses \cite{Kiilerich2019,Kiilerich2020}, show that our scheme can produce single-photon pairs with close-to-unity fidelity, exceeding the highest possible Bell fidelity of a two-mode squeezed vacuum (TMSV) state from SPDC. We further protect fidelity against photon loss using dual-rail encoding and post-selection, where we blockade unwanted excitations with additional cross-Kerr nonlinearity. In the presence of intrinsic photon loss, we optimize the pulse width and driving strength. With experimentally feasible nonlinearity, our scheme can significantly outperform the traditional SPDC scheme by achieving both higher fidelity and an enhanced generation rate. 

\begin{figure}[t]
\centering
\includegraphics[width=\columnwidth]{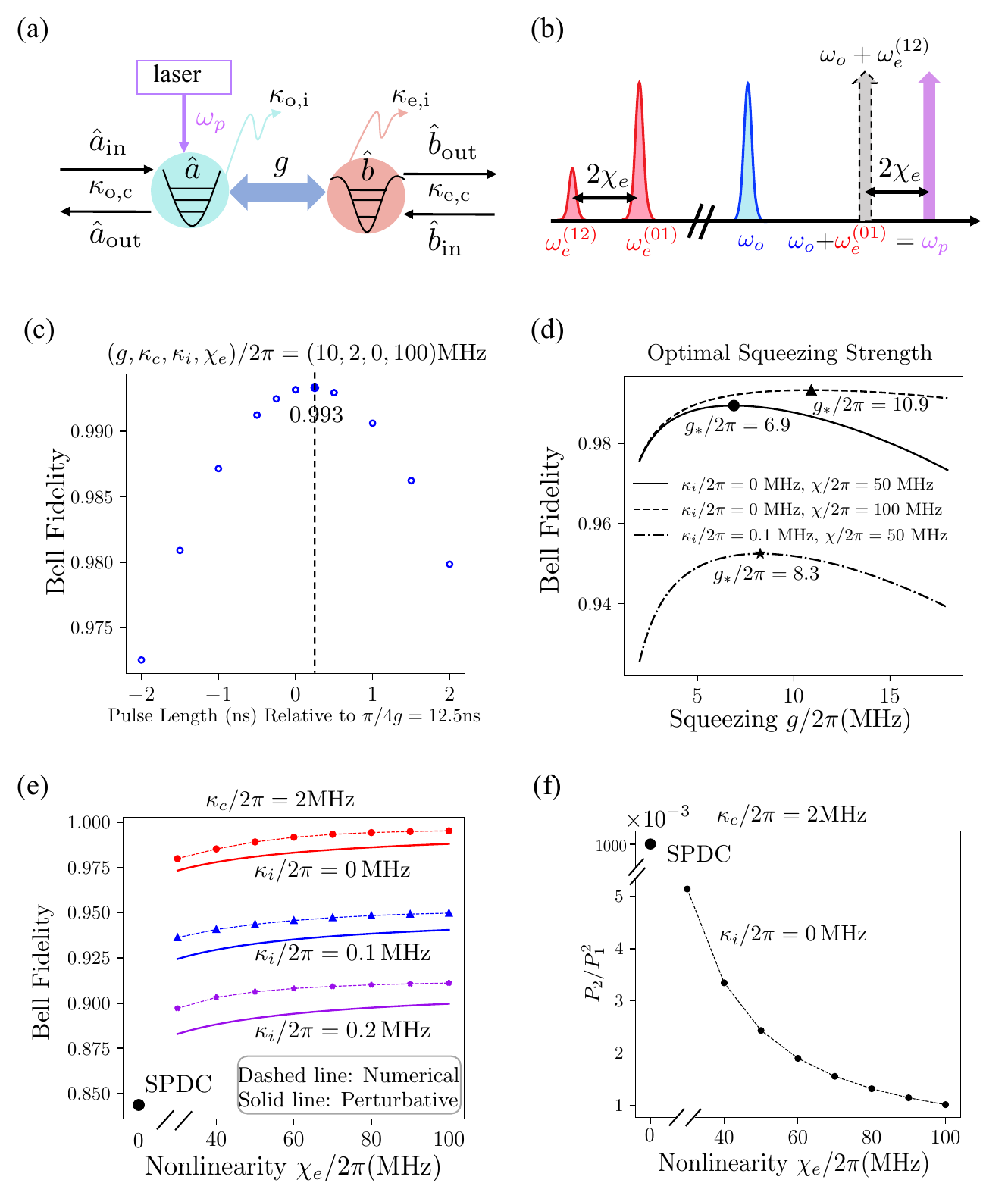}
\caption{(a) Illustration of entanglement generation with self-Kerr nonlinearity. The intra-cavity modes $\hat{a}$ and $\hat{b}$ each have their respective input and output noise operators ($\hat{a}_{\text{in}},\hat{a}_{\text{out}},\hat{b}_{\text{in}},\hat{b}_{\text{out}}$) and coupling rates, $\kappa_{\text{o,c}}$ and $\kappa_{\text{e,c}}$. The optical mode $\hat{a}$ is coupled to the strongly anharmonic microwave mode $\hat{b}$ through a two-mode squeezing interaction induced by a laser drive at frequency $\omega_p$. The system has intrinsic loss rates $\kappa_{\text{o,i}}$ and $\kappa_{\text{e,i}}$. (b) Spectral landscape of our scheme. Due to the Kerr nonlinearity, the second-excited-state transition of the microwave mode is detuned by $2\chi_e$. If we were to pump on resonance $\omega_0+\omega_e^{(12)}$, the virtual pump (shown in gray) would be detuned by $2\chi_e$. (c) Optimal pulse length (solid circle) plotted relative to the ideal Rabi $\pi/2$-pulse at $\tau_{\pi/2} = \pi/4g$. (d) Fidelity versus driving strength $g$, plotted using three different sets of parameters (solid, dashed, and dot-dashed lines), with the optimal $g_*$ values marked by a circle, star, and triangle, respectively. (e) Optimized fidelity for different intrinsic loss $\kappa_i$ and Kerr nonlinearity $\chi_e$ values. The numerical result is shown as dashed dots, and the analytical result is shown as solid lines. The best fidelity of the SPDC is shown by the black dot at the lower left corner $F\approx 0.84$. (f) Suppression of the higher excitation with increasing Kerr nonlinearity. (c) and (f) are plotted using data from numerical simulations. (d) is plotted using Eq.~(\ref{empirical}). (e) includes both numerical data and analytical results from a perturbation calculation. \label{Fig2}}
\end{figure}

\emph{Two-mode squeezing with self-Kerr nonlinearity.}\textemdash\label{sec:EO-model}
We first consider a two-mode squeezed electro-optical system with strong microwave self-Kerr nonlinearity induced by a transmon. A schematic of our system is shown in Fig.~\ref{Fig2}(a). 
We denote $\hat{a}$ and $\hat{b}$ as the optical and microwave mode operators, $\omega_o$ and $\omega_e$ as the corresponding mode frequencies, and $\chi_e$ as the strength of the self-Kerr nonlinearity on the microwave side. With pump laser at frequency $\omega_p = \omega_o + \omega_e$, the Hamiltonian in the co-rotating frame of the two modes can be written as 
\begin{equation}\label{entangler}
\hat{H}_{\text{ent}}(t)= \begin{cases}
 -g(\hat{a}^\dagger\hat{b}^\dagger+\hat{a}\hat{b})- \chi_e \hat{b}^\dagger \hat{b}^\dagger \hat{b}\hat{b}& \text { if } 0<t<\tau \\ - \chi_e \hat{b}^\dagger \hat{b}^\dagger \hat{b}\hat{b} & \text { otherwise }\end{cases},
\end{equation}
where $\tau$ is the pulse length of a parametric square-pulse pump. The driving strength $g$ creates squeezing, and the system is initially in the vacuum state. When $\chi_e = 0$, the Hamiltonian generates the TMSV state, which is used in the current entanglement-based transduction protocol \cite{Zhong2020PRL}. 

Due to the Kerr nonlinearity in the system, higher-excitation states are far-detuned. In particular, for a given microwave excitation number $n$, the energy levels are shifted on top of $\omega_e$ by $\chi_e n(n-1)$. As an example, Fig.~\ref{Fig2}(b) shows that for pump frequency at $\omega_p = \omega_o +  \omega_e^{(01)}$ resonant with the ground-to-first-excited state transition of the microwave mode $\omega_e^{(01)} = \omega_e$, the microwave double-excitation transition $\omega_e^{(12)}$ is detuned by $2\chi_e$. In the infinite nonlinearity limit, the photon number blockade is perfect, and the system will behave like a two-level system. The two-mode squeezed Hamiltonian in this case drives the Rabi oscillation between the $\ket{0}_o\ket{0}_e$ and $\ket{1}_o\ket{1}_e$ states.

To create a Bell state $\frac{1}{\sqrt{2}}(\ket{0}_o\ket{0}_e+\ket{1}_o\ket{1}_e)$, the Rabi-flopping intuition suggests that the optimal driving time should correspond to an effective $\pi/2$-pulse on the vacuum. In our case, assuming no loss and perfect blockade, the effective $\pi/2$-pulse time is achieved at $\tau_{\pi/2} = \pi/4g$. 

To generate high-fidelity Bell pairs, our protocol operates in the regime with $\kappa \ll g \ll \chi_e$, where $\kappa$ characterizes the loss in the system. To be specific, the total loss rate of the optical (microwave) mode is the sum of the coupling loss rate $\kappa_{o,c}$ ($\kappa_{e,c}$) and the intrinsic loss rate $\kappa_{o,i}$ ($\kappa_{e,i}$). For simplicity, we assume that the optical and microwave modes have the same loss rate $\kappa_{o,c} = \kappa_{e,c} = \kappa_c$ and $\kappa_{o,i} = \kappa_{e,i} = \kappa_i $, so that they also share the same total loss rate $\kappa = \kappa_c + \kappa_i$. Having the same loss rate on the optical and microwave sides is not required for our scheme to perform as long as $\kappa_o+\kappa_e\ll\chi_e$. 

If we violate the hierarchy $\kappa \ll g \ll \chi_e$ by having $\kappa>\chi_e$, then the blockade cannot be established because the large linewidth $\kappa$ can still facilitate the creation of off-resonant photons. This intuition is backed up by calculating the photon population through adiabatically eliminating the lossy mode (see Appendix~\ref{AE}). We also require $\kappa < g$ to preserve maximal coherence in the state. 

Given the Hamiltonian in Eq.~(\ref{entangler}), loss rates $\kappa_i$ and $\kappa_c$, we compute the output state of our pulse-driven system. Due to the Kerr nonlinearity, photons from the microwave and optical cavities occupy multiple temporal modes. We apply the input-output theory developed in \cite{Kiilerich2019,Kiilerich2020}, which allows us to select a temporal mode from the emission and construct the state by cascading it into an empty downstream cavity with time-varying coupling. This process can be solved numerically by setting up cascaded Lindblad equations using QuTiP \cite{lambert2024qutip5quantumtoolbox}. In the limit $\kappa \ll g \ll \chi_e$, the output state can also be obtained analytically using perturbation theory (see Appendix~\ref{PT}). 

The numerical and analytical results complement our intuition in the presence of loss and finite nonlinearity. Fig.~\ref{Fig2}(c) shows that we can numerically find an optimal pulse length near $\tau=\pi/4g$ that achieves a higher fidelity.

To estimate the impact of loss and imperfect blockade on the fidelity of the output state, we consider their lowest-order contributions. At the lowest order, we are allowed to treat the two mechanisms independently and sum their respective contributions. Imperfect blockade is characterized by leakage into higher excitation states when the system is driven by the two-mode squeezed pump. The leading-order contribution of this process to the infidelity scales as \( g^2 / \chi_e^2 \). Loss contributes to infidelity in three different ways. First, a photon may come out of the cavity early during the drive, contributing an error on the order of \( \kappa_c / g \). Second, a photon may be lost internally during the drive, introducing a correction on the order of \( \kappa_i / g \). Third, after the pump is turned off, a photon may be lost while coming out of the cavity, altering the output state by a factor of \( \kappa_i / \kappa \). These contributions are formalized in the perturbative calculation in Appendix~\ref{PT}, which shows that the leading-order infidelity can be written in the following functional form:
\begin{align}\label{empirical}
1-F \approx  C_1\frac{\kappa_c}{g} + C_2
 \frac{\kappa_i}{g}  + C_3 \frac{\kappa_i}{\kappa} + C_4\frac{g^2}{\chi_e^2}. 
\end{align}
To obtain an empirical estimation of the infidelity, the coefficients can be fixed by numerically scanning across the parameter space $\{g, \kappa_c ,\kappa_i, \chi_e\}$ while optimizing the pulse duration. These coefficients are extracted by fitting the numerical data (see Appendix~\ref{NS}), and the parameter $g$ can be optimized for different loss rates and nonlinearities, as shown in Fig.~\ref{Fig2}(d).     

In Fig.~\ref{Fig2}(e), we show the best-achievable fidelity dependence on $\{\kappa_c, \kappa_i, \chi_e\}$, where the values of $\tau$ and $g$ are numerically optimized. The Bell fidelity increases steadily with Kerr nonlinearity but decreases rapidly with rising intrinsic loss (shown by the comparison between red, blue, and purple lines in the plot). In the figure, we also include the analytical result, shown by solid lines, obtained through perturbation theory. In general, the perturbation theory provides a lower bound on the numerical results. They differ by $1\%$ for large $\chi_e$, which is on the sub-leading order of $(\kappa/g)^2$. The best-achievable Bell fidelity ($\approx 0.84$) by SPDC is marked by a black circle at the lower-left corner.

In Fig.~\ref{Fig2}(f), we showcase the photon blockade effect with increasing Kerr nonlinearity. The higher excitation blockade is measured by the ratio between the double-pair population $P_2$ and the single-pair population squared $(P_1)^2$. For the TMSV state, $P_2/P_1^2 =1$. With increasing Kerr nonlinearity, this ratio is suppressed by a factor of $1000$, which implies that the cavity emission consists of almost purely single-photon pairs. 

Together, Figs.~\ref{Fig2}(e) and \ref{Fig2}(f) illustrate a well-established strong Kerr blockade, but a high susceptibility to intrinsic photon loss. In the case of perfect blockade, we estimate using Eq.~(\ref{empirical}) and find that when $\kappa_i/\kappa_c > 0.3$, the best-achievable fidelity of the system in Eq.~(\ref{entangler}) becomes worse than that of SPDC.

\emph{Dual-rail encoding.}\textemdash
To address intrinsic excitation loss, we may use dual-rail encoding, which allows for heralding of excitation loss errors.

\begin{figure*}[t]
\centering
\includegraphics[width=1\textwidth]{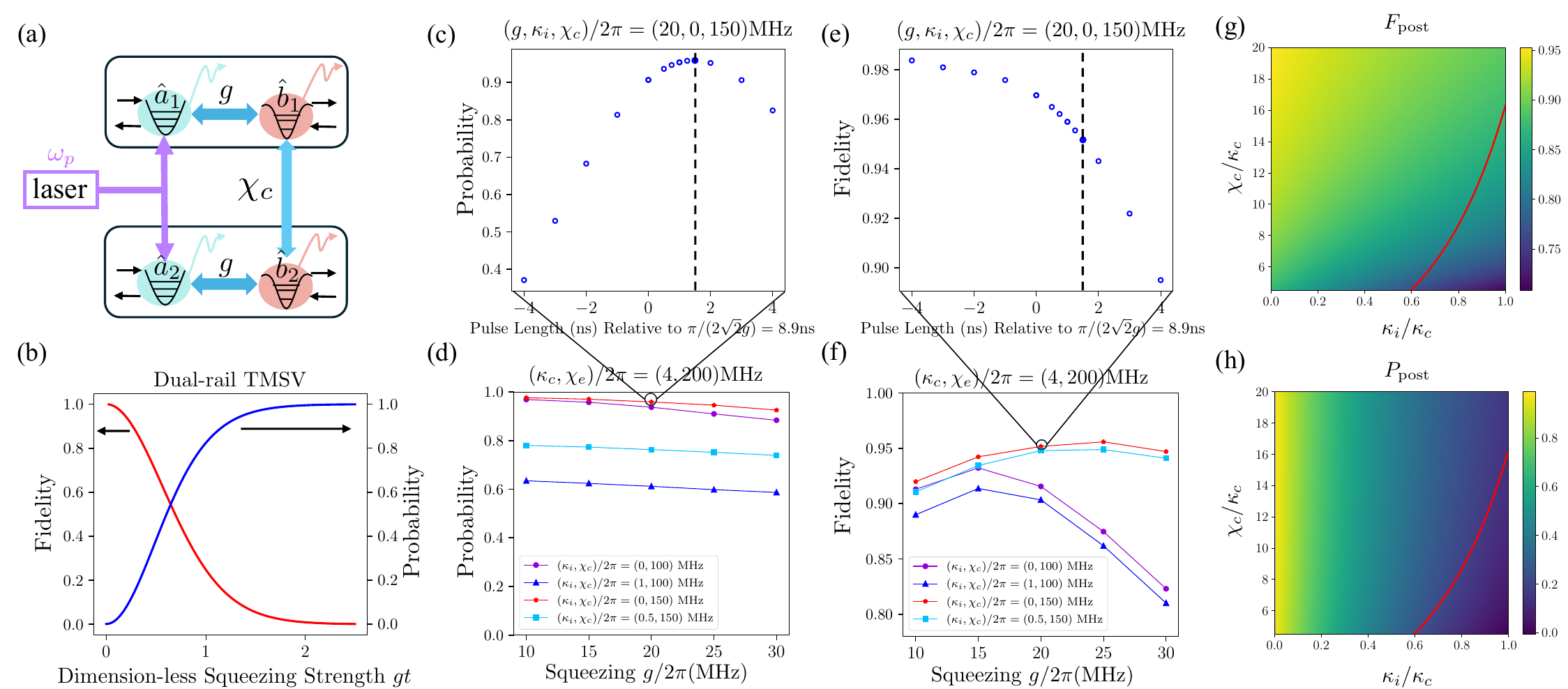}
\caption{(a) Schematic of dual-rail entanglement generation with strong cross-Kerr nonlinearity between the two microwave modes. (b) The dual-rail SPDC scheme exhibits an intrinsic trade-off between fidelity and probability. (c) The optimal pulse length (marked by a solid circle) for generation probability, plotted along the axis relative to the Rabi $\pi$-pulse in our system, $\tau_{\pi} = \pi/(2\sqrt{2}g).$ (d) Probability as a function of driving strength across different parameter sets. Each point is numerically optimized with respect to pulse duration. (e) No optimal pulse length exists for fidelity in the vicinity of a $\pi$-pulse. The solid circle corresponds to the pulse length that maximizes probability in (c). (f) Fidelity as a function of driving strength exhibits local maxima for different parameter sets. Each point corresponds to the optimal pulse length for maximizing probability. (g) Fidelity plotted using Eq.~(\ref{dualF}), at the optimal driving strength. (h) Probability plotted using Eq.~(\ref{dualP}), at the optimal driving strength. The red line in (g) and (h) represents the boundary, above which our scheme outperforms the SPDC scheme in both fidelity and success probability. \label{Fig3}}
\end{figure*}

The Hamiltonian of our dual-rail scheme consists of two identical copies of the Hamiltonian in Eq.~(\ref{entangler}) and a cross Kerr nonlinearity $\chi_c$ between the two microwave modes. In practice, the cross-Kerr can be realized by hybridizing the two microwave cavity modes through a common nonlinear Josephson device \cite{Gao2019,Blais2021,Ye2021, Chakram2022}. 
\begin{equation}\label{dual}
\hat{H}_{\text{dual-rail}}=  \hat{H}_{\text{ent}}^{(1)} + \hat{H}_{\text{ent}}^{(2)} + \chi_c \hat{b}_1^\dagger \hat{b}_1 \hat{b}_2^\dagger \hat{b}_2.
\end{equation}
For clarity in the subsequent discussion, we assume that the two copies have identical parameters. While this is not a requirement, it is crucial for both copies to be pumped synchronously. We also impose the hierarchy of terms $\kappa \ll g \ll \chi_e \sim \chi_c $ in analogy to the single-rail model.  The scheme is illustrated in Fig.~\ref{Fig3}(a).

In addition to the photon blockade induced by the self-Kerr interaction, the cross-Kerr interaction imposes an energy penalty when both microwave modes simultaneously contain photons. This effect is crucial for generating Bell pairs in the dual-rail basis $\ket{\Psi^+}=1/\sqrt{2}\left(\ket{00}_{o1,e1}\ket{11}_{o2,e2} + \ket{11}_{o1,e1}\ket{00}_{o2,e2} \right)$. Given $\hat{H}_{\text{dual-rail}}$, the set of lowest energy states connected by the two-mode squeezed drive includes the vacuum, the symmetric and anti-symmetric states $1/\sqrt{2}\left(\ket{00}_{o1,e1}\ket{11}_{o2,e2} \pm \ket{11}_{o1,e1}\ket{00}_{o2,e2} \right)$ as well as $\ket{11}_{o1,e1}\ket{11}_{o2,e2} $. With the four-photon state detuned by the cross-Kerr nonlinearity and the anti-symmetric state decoupled from the Hamiltonian, the system undergoes Rabi oscillations between the vacuum and the symmetric state, which is the desired Bell state $\ket{\Psi^+}$. Hence, a $\pi$-pulse applied to the vacuum state exactly creates $\ket{\Psi^+}$. 

In the open system setting, we denote the output state from the dual-rail Hamiltonian in Eq.~(\ref{dual}) as $\rho_{\text{out}}$. This state is protected against photon loss by post-selecting out vacuum events during photon detection. The post-selection process can be modeled by the projector: 
\begin{equation}
    \hat{\Pi} = \left(\mathds{1}_{oe} - \ket{00}_o\bra{00}\right) \left(\mathds{1}_{oe}-\ket{00}_e\bra{00}\right),
\end{equation}
and the post-selected state $
\rho_{\text{post}} = \frac{\hat{\Pi} \rho_{\text{out}} \hat{\Pi}} {\text{tr}(\rho_{\text{out}} \hat{\Pi})}$ is generated with probability 
\begin{equation}
\begin{split}
    P_{\text{post}} = \text{tr}(\rho_{\text{out}} \hat{\Pi}).
\end{split}
\end{equation}
While post-selection makes the state generation probabilistic, it improves the fidelity $F_{\text{post}}$ between the post-selected state and the dual-rail Bell state by removing the vacuum component. 

In the low loss limit ($\kappa \rightarrow 0$), photon blockade enables the deterministic generation of the dual-rail Bell state. In contrast, the standard SPDC scheme cannot achieve deterministic dual-rail Bell state generation, due to the presence of both vacuum components and high photon number states \cite{Zhong2020PRL, Krastanov2021}. Increasing the generation probability via stronger pumping leads to a decrease in fidelity due to the presence of higher excitation states. This trade-off is illustrated in Fig.~\ref{Fig3}(b), where both the fidelity and post-selection probability of the dual-rail TMSV state are plotted as functions of the squeezing strength.

When the loss in the system is non-negligible and the nonlinearity is finite, we numerically explore the performance of our dual-rail scheme.

Following the intuition from Rabi oscillations in a closed system, there exists an optimal pulse duration $\tau_*$ at which the system is driven farthest from the vacuum. Numerical results reveal that an optimal pulse duration exists for maximizing the generation probability but not the generation fidelity. Fig.~\ref{Fig3}(c) shows the optimal pulse duration for the generation probability in the vicinity of a closed-system prediction of $\tau_\pi= \pi/(2\sqrt{2}g)$. In the same neighborhood, Fig.~\ref{Fig3}(e) reveals no optimal pulse duration for the fidelity. This is because the post-selection removes vacuum components, so the Rabi oscillation between the vacuum and the first excited state does not affect the post-selected fidelity. The fidelity depends on pulse duration due to oscillations in unsuppressed higher excitation states, as well as the mixedness of the state due to intrinsic photon loss, both of which are subleading effects. In the following discussion, we take $\tau = \tau_*$ to be the optimal time for generation probability. 

To estimate the generation probability and fidelity, we identify the dominant error mechanisms as state leakage and photon loss during Rabi oscillations, analogous to the single-rail case. Accordingly, we introduce dimensionless small parameters, as defined in Eq.~(\ref{empirical}), to capture the leading-order corrections. The maximum achievable generation probability is limited by intrinsic photon loss—quantified by \( \kappa_i / g \) and \( \kappa_i / \kappa \)—and by leakage into higher excitation states, which induces imperfect Rabi oscillations between the ground state and the first-excited state. These contributions are characterized by \( g^2 / \chi_e^2 \) and \( g^2 / \chi_c^2 \). In contrast to fidelity, the generation probability is not reduced by \( \kappa_c / g \), as this represents the photon emission from the cavity during pumping, not dissipative loss. Numerical simulations, such as the one shown in Fig.~\ref{Fig3}(d), reveal a strong dependence on intrinsic loss. By scanning over the parameters \( \{g, \kappa_c, \kappa_i, \chi_c\} \) while taking \( \chi_e \rightarrow \infty \), we obtain an empirical estimate of the generation probability based on numerical fitting:

\begin{equation}\label{dualP}
    1-P_{\text{post}} \approx 0.23\frac{\kappa_i}{g}+ 1.51\frac{\kappa_i}{\kappa} + 0.85\frac{g^2}{\chi_c^2}.
\end{equation}

The fidelity conditioned on post-selection is protected against photon loss through projection onto the non-vacuum subspace. Since intrinsic photon loss during the pump-off period—quantified by \( \kappa_i / \kappa \)—is the dominant source of vacuum population, its effect is significantly suppressed by post-selection. As a result, fidelity is primarily limited by leakage into higher excitation states—characterized by \( g^2 / \chi_c^2 \) and \( g^2 / \chi_e^2 \)—and by photon loss during the drive. Figure~\ref{Fig3}(f) shows a rapid decline in fidelity as the ratio \( g^2 / \chi_c^2 \) increases. Numerical fitting yields a negligible coefficient for \( \kappa_i / \kappa \), and the resulting infidelity is estimated by:

\begin{equation}\label{dualF}
    1-F_{\text{post}} \approx 0.07\frac{\kappa_c}{g} + 0.17\frac{\kappa_i}{g} + 1.29\frac{g^2}{\chi_c^2}.
\end{equation}
Minimizing the overall infidelity with respect to $g$ gives the optimal $g_* \approx \left( \chi_c^2(0.07\kappa_c + 0.17 \cdot \kappa_i)/2.58 \right)^{1/3}$. 

Even in the presence of photon loss and limited nonlinearity, our scheme can still outperform the dual-rail SPDC scheme in an experimentally feasible regime. In the extreme case of weak nonlinearity and large intrinsic loss, the system has low generation fidelity and probability, which can equally be achieved with the SPDC scheme. To identify the parameter regime where our scheme outperforms the SPDC counterpart in both fidelity and success probability, we evaluate Eqs.~(\ref{dualP}) and (\ref{dualF}) using the optimized coupling strength $g=g_*$ \footnote{In principle, there also exists an optimal driving strength for maximizing the generation probability, given by \( g = \left( 0.09 \chi_c^2 \kappa_i \right)^{1/3} \). However, Fig.~\ref{Fig3}(d)(f) indicates that fidelity is more sensitive to variations in \( g \) than the probability is. More rigorously, we can see that $\frac{\partial^2 F_{\text{post}}}{\partial g^2}>\frac{\partial^2 P_{\text{post}}}{\partial g^2}$ for all $\kappa_i,g, \chi_c$ in the over-coupled regime $\kappa_c>\kappa_i$. Therefore, to optimize overall performance, we select the driving strength that maximizes fidelity.}. At the optimal driving strength, both fidelity and probability can be expressed as functions of the dimensionless ratios $\kappa_i/\kappa_c$ and $\chi_c/\kappa_c$, as shown in Fig.~\ref{Fig3}(g)(h). Requiring both $F_{\text{post}}$ and $P_{\text{post}}$ to be greater than those of the SPDC scheme at any given driving strength sets constraints on $\kappa_i/\kappa_c$ and $\chi_c/\kappa_c$. This is shown by the red curve in Fig.~\ref{Fig3}(g)(h). Above the red curve is the region where our scheme outperforms the SPDC scheme in both fidelity and probability. The curve shows that as long as the intrinsic loss rate is smaller than the coupling loss rate by $\kappa_i/\kappa_c<0.6$, it suffices to have nonlinearity $\chi_c/\kappa_c \gtrsim 6$. If $\kappa_c$ is on the order of a few megahertz, this requirement can be met by current superconducting platforms \cite{Blais2021,Chakram2022,Gao2019, Ye2021, PhysRevA.76.042319}.

\emph{Discussion.}\textemdash
We propose a nonlinear microwave-optical Bell pair generation scheme for entanglement-based transduction. Thanks to the blockade effect of strong Kerr nonlinearity, the system is capable of generating high-fidelity Bell pairs. We optimize the pulse length and driving strength, demonstrating that our scheme can outperform the SPDC scheme in the presence of intrinsic photon loss and limited Kerr nonlinearity. 

A future research direction involves embedding our scheme into currently available experimental platforms \cite{Han2020,Xu2021, Sahu2023, Meesala2024, Hana2025}. If the system contains any intermediary mode that mediates the interaction between optical and microwave photons, we may also introduce nonlinearity by hybridizing the intermediary and the microwave mode \cite{Zhong2020}. It is noteworthy that $\kappa_o=\kappa_e$ is not required for our system as long as $\kappa_o + \kappa_e \ll \chi_e \sim \chi_c$, where the detuning continues to suppress higher excitation by shifting the photon frequency outside the linewidth. Although it is currently challenging to have the coexistence of small optical linewidth and large microwave nonlinearity in a system, experiments are rapidly progressing towards that regime \cite{konradnature, PhysRevX.12.021062, Hana2025}. 

Other important experimental considerations include quantifying the influence of thermal noise on our protocol, designing entanglement verification procedures with mode filtering \cite{PhysRevApplied.22.064047, PhysRevX.14.031055, Meesala2024}, and optimizing the pulse shape using the DRAG scheme \cite{PhysRevLett.103.110501} to further suppress higher excitations.

A theoretical direction involves studying transduction protocols using information-theoretic tools. Transduction may be completed through quantum teleportation, consuming the generated Bell pairs. The performance of transduction is then measured by the quantum capacity of the teleportation-induced channel \cite{Wang2022,Zhong2022, PhysRevA.109.022619, Wu2021}. It will be instructive to compare the quantum capacity of our scheme and the SPDC scheme and to identify the high-capacity regime for both protocols.

\begin{acknowledgments}
We acknowledge support from the ARO(W911NF-23-1-0077), ARO MURI (W911NF-21-1-0325), AFOSR MURI (FA9550-19-1-0399, FA9550-21-1-0209, FA9550-23-1-0338), DARPA (HR0011-24-9-0359, HR0011-24-9-0361), NSF (OMA-1936118, ERC-1941583, OMA-2137642, OSI-2326767, CCF-2312755), Packard Foundation (2020-71479), and the Marshall and Arlene Bennett Family Research Program. This material is based upon work supported by the U.S. Department of Energy, Office of Science, National Quantum Information Science Research Centers and Advanced Scientific Computing Research (ASCR) program under contract number DE-AC02-06CH11357 as part of the InterQnet quantum networking project. This work was completed with resources provided by the University of Chicago’s Research Computing Center. Outside of the author list, F.L. personally thanks Qiyuan Hu for his patient answers to coding-related questions. 
\end{acknowledgments}

\bibliography{cite}

\begin{thebibliography}{42}%
\makeatletter
\providecommand \@ifxundefined [1]{%
 \@ifx{#1\undefined}
}%
\providecommand \@ifnum [1]{%
 \ifnum #1\expandafter \@firstoftwo
 \else \expandafter \@secondoftwo
 \fi
}%
\providecommand \@ifx [1]{%
 \ifx #1\expandafter \@firstoftwo
 \else \expandafter \@secondoftwo
 \fi
}%
\providecommand \natexlab [1]{#1}%
\providecommand \enquote  [1]{``#1''}%
\providecommand \bibnamefont  [1]{#1}%
\providecommand \bibfnamefont [1]{#1}%
\providecommand \citenamefont [1]{#1}%
\providecommand \href@noop [0]{\@secondoftwo}%
\providecommand \href [0]{\begingroup \@sanitize@url \@href}%
\providecommand \@href[1]{\@@startlink{#1}\@@href}%
\providecommand \@@href[1]{\endgroup#1\@@endlink}%
\providecommand \@sanitize@url [0]{\catcode `\\12\catcode `\$12\catcode
  `\&12\catcode `\#12\catcode `\^12\catcode `\_12\catcode `\%12\relax}%
\providecommand \@@startlink[1]{}%
\providecommand \@@endlink[0]{}%
\providecommand \url  [0]{\begingroup\@sanitize@url \@url }%
\providecommand \@url [1]{\endgroup\@href {#1}{\urlprefix }}%
\providecommand \urlprefix  [0]{URL }%
\providecommand \Eprint [0]{\href }%
\providecommand \doibase [0]{https://doi.org/}%
\providecommand \selectlanguage [0]{\@gobble}%
\providecommand \bibinfo  [0]{\@secondoftwo}%
\providecommand \bibfield  [0]{\@secondoftwo}%
\providecommand \translation [1]{[#1]}%
\providecommand \BibitemOpen [0]{}%
\providecommand \bibitemStop [0]{}%
\providecommand \bibitemNoStop [0]{.\EOS\space}%
\providecommand \EOS [0]{\spacefactor3000\relax}%
\providecommand \BibitemShut  [1]{\csname bibitem#1\endcsname}%
\let\auto@bib@innerbib\@empty
%</preamble>
\bibitem [{\citenamefont {Kimble}(2008)}]{Kimble2008}%
  \BibitemOpen
  \bibfield  {author} {\bibinfo {author} {\bibfnamefont {H.~J.}\ \bibnamefont
  {Kimble}},\ }\href {https://doi.org/10.1038/nature07127} {\bibinfo {title}
  {The quantum internet}} (\bibinfo {year} {2008})\BibitemShut {NoStop}%
\bibitem [{\citenamefont {Cirac}\ \emph {et~al.}(1997)\citenamefont {Cirac},
  \citenamefont {Zoller}, \citenamefont {Kimble},\ and\ \citenamefont
  {Mabuchi}}]{Cirac1997}%
  \BibitemOpen
  \bibfield  {author} {\bibinfo {author} {\bibfnamefont {J.~I.}\ \bibnamefont
  {Cirac}}, \bibinfo {author} {\bibfnamefont {P.}~\bibnamefont {Zoller}},
  \bibinfo {author} {\bibfnamefont {H.~J.}\ \bibnamefont {Kimble}},\ and\
  \bibinfo {author} {\bibfnamefont {H.}~\bibnamefont {Mabuchi}},\ }\bibfield
  {title} {\bibinfo {title} {Quantum state transfer and entanglement
  distribution among distant nodes in a quantum network},\ }\bibfield
  {journal} {\bibinfo  {journal} {Physical Review Letters}\ }\textbf {\bibinfo
  {volume} {78}},\ \href {https://doi.org/10.1103/PhysRevLett.78.3221}
  {10.1103/PhysRevLett.78.3221} (\bibinfo {year} {1997})\BibitemShut {NoStop}%
\bibitem [{\citenamefont {Lauk}\ \emph {et~al.}(2020)\citenamefont {Lauk},
  \citenamefont {Sinclair}, \citenamefont {Barzanjeh}, \citenamefont {Covey},
  \citenamefont {Saffman}, \citenamefont {Spiropulu},\ and\ \citenamefont
  {Simon}}]{Lauk2020}%
  \BibitemOpen
  \bibfield  {author} {\bibinfo {author} {\bibfnamefont {N.}~\bibnamefont
  {Lauk}}, \bibinfo {author} {\bibfnamefont {N.}~\bibnamefont {Sinclair}},
  \bibinfo {author} {\bibfnamefont {S.}~\bibnamefont {Barzanjeh}}, \bibinfo
  {author} {\bibfnamefont {J.~P.}\ \bibnamefont {Covey}}, \bibinfo {author}
  {\bibfnamefont {M.}~\bibnamefont {Saffman}}, \bibinfo {author} {\bibfnamefont
  {M.}~\bibnamefont {Spiropulu}},\ and\ \bibinfo {author} {\bibfnamefont
  {C.}~\bibnamefont {Simon}},\ }\href
  {https://doi.org/10.1088/2058-9565/ab788a} {\bibinfo {title} {Perspectives on
  quantum transduction}} (\bibinfo {year} {2020})\BibitemShut {NoStop}%
\bibitem [{\citenamefont {Lambert}\ \emph {et~al.}(2020)\citenamefont
  {Lambert}, \citenamefont {Rueda}, \citenamefont {Sedlmeir},\ and\
  \citenamefont {Schwefel}}]{Lambert2020}%
  \BibitemOpen
  \bibfield  {author} {\bibinfo {author} {\bibfnamefont {N.~J.}\ \bibnamefont
  {Lambert}}, \bibinfo {author} {\bibfnamefont {A.}~\bibnamefont {Rueda}},
  \bibinfo {author} {\bibfnamefont {F.}~\bibnamefont {Sedlmeir}},\ and\
  \bibinfo {author} {\bibfnamefont {H.~G.}\ \bibnamefont {Schwefel}},\
  }\bibfield  {title} {\bibinfo {title} {Coherent conversion between microwave
  and optical photons—an overview of physical implementations},\ }\bibfield
  {journal} {\bibinfo  {journal} {Advanced Quantum Technologies}\ }\textbf
  {\bibinfo {volume} {3}},\ \href {https://doi.org/10.1002/qute.201900077}
  {10.1002/qute.201900077} (\bibinfo {year} {2020})\BibitemShut {NoStop}%
\bibitem [{\citenamefont {Han}\ \emph {et~al.}(2021)\citenamefont {Han},
  \citenamefont {Fu}, \citenamefont {Zou}, \citenamefont {Jiang},\ and\
  \citenamefont {Tang}}]{Han2021}%
  \BibitemOpen
  \bibfield  {author} {\bibinfo {author} {\bibfnamefont {X.}~\bibnamefont
  {Han}}, \bibinfo {author} {\bibfnamefont {W.}~\bibnamefont {Fu}}, \bibinfo
  {author} {\bibfnamefont {C.-L.}\ \bibnamefont {Zou}}, \bibinfo {author}
  {\bibfnamefont {L.}~\bibnamefont {Jiang}},\ and\ \bibinfo {author}
  {\bibfnamefont {H.~X.}\ \bibnamefont {Tang}},\ }\bibfield  {title} {\bibinfo
  {title} {Microwave-optical quantum frequency conversion},\ }\bibfield
  {journal} {\bibinfo  {journal} {Optica}\ }\textbf {\bibinfo {volume} {8}},\
  \href {https://doi.org/10.1364/optica.425414} {10.1364/optica.425414}
  (\bibinfo {year} {2021})\BibitemShut {NoStop}%
\bibitem [{\citenamefont {Andrews}\ \emph {et~al.}(2014)\citenamefont
  {Andrews}, \citenamefont {Peterson}, \citenamefont {Purdy}, \citenamefont
  {Cicak}, \citenamefont {Simmonds}, \citenamefont {Regal},\ and\ \citenamefont
  {Lehnert}}]{Andrews2014}%
  \BibitemOpen
  \bibfield  {author} {\bibinfo {author} {\bibfnamefont {R.~W.}\ \bibnamefont
  {Andrews}}, \bibinfo {author} {\bibfnamefont {R.~W.}\ \bibnamefont
  {Peterson}}, \bibinfo {author} {\bibfnamefont {T.~P.}\ \bibnamefont {Purdy}},
  \bibinfo {author} {\bibfnamefont {K.}~\bibnamefont {Cicak}}, \bibinfo
  {author} {\bibfnamefont {R.~W.}\ \bibnamefont {Simmonds}}, \bibinfo {author}
  {\bibfnamefont {C.~A.}\ \bibnamefont {Regal}},\ and\ \bibinfo {author}
  {\bibfnamefont {K.~W.}\ \bibnamefont {Lehnert}},\ }\bibfield  {title}
  {\bibinfo {title} {Bidirectional and efficient conversion between microwave
  and optical light},\ }\bibfield  {journal} {\bibinfo  {journal} {Nature
  Physics}\ }\textbf {\bibinfo {volume} {10}},\ \href
  {https://doi.org/10.1038/nphys2911} {10.1038/nphys2911} (\bibinfo {year}
  {2014})\BibitemShut {NoStop}%
\bibitem [{\citenamefont {Fan}\ \emph {et~al.}(2018)\citenamefont {Fan},
  \citenamefont {Zou}, \citenamefont {Cheng}, \citenamefont {Guo},
  \citenamefont {Han}, \citenamefont {Gong}, \citenamefont {Wang},\ and\
  \citenamefont {Tang}}]{Fan2018}%
  \BibitemOpen
  \bibfield  {author} {\bibinfo {author} {\bibfnamefont {L.}~\bibnamefont
  {Fan}}, \bibinfo {author} {\bibfnamefont {C.~L.}\ \bibnamefont {Zou}},
  \bibinfo {author} {\bibfnamefont {R.}~\bibnamefont {Cheng}}, \bibinfo
  {author} {\bibfnamefont {X.}~\bibnamefont {Guo}}, \bibinfo {author}
  {\bibfnamefont {X.}~\bibnamefont {Han}}, \bibinfo {author} {\bibfnamefont
  {Z.}~\bibnamefont {Gong}}, \bibinfo {author} {\bibfnamefont {S.}~\bibnamefont
  {Wang}},\ and\ \bibinfo {author} {\bibfnamefont {H.~X.}\ \bibnamefont
  {Tang}},\ }\bibfield  {title} {\bibinfo {title} {Superconducting cavity
  electro-optics: A platform for coherent photon conversion between
  superconducting and photonic circuits},\ }\bibfield  {journal} {\bibinfo
  {journal} {Science Advances}\ }\textbf {\bibinfo {volume} {4}},\ \href
  {https://doi.org/10.1126/sciadv.aar4994} {10.1126/sciadv.aar4994} (\bibinfo
  {year} {2018})\BibitemShut {NoStop}%
\bibitem [{\citenamefont {Rueda}\ \emph {et~al.}(2019)\citenamefont {Rueda},
  \citenamefont {Hease}, \citenamefont {Barzanjeh},\ and\ \citenamefont
  {Fink}}]{Rueda2019}%
  \BibitemOpen
  \bibfield  {author} {\bibinfo {author} {\bibfnamefont {A.}~\bibnamefont
  {Rueda}}, \bibinfo {author} {\bibfnamefont {W.}~\bibnamefont {Hease}},
  \bibinfo {author} {\bibfnamefont {S.}~\bibnamefont {Barzanjeh}},\ and\
  \bibinfo {author} {\bibfnamefont {J.~M.}\ \bibnamefont {Fink}},\ }\bibfield
  {title} {\bibinfo {title} {Electro-optic entanglement source for microwave to
  telecom quantum state transfer},\ }\bibfield  {journal} {\bibinfo  {journal}
  {npj Quantum Information}\ }\textbf {\bibinfo {volume} {5}},\ \href
  {https://doi.org/10.1038/s41534-019-0220-5} {10.1038/s41534-019-0220-5}
  (\bibinfo {year} {2019})\BibitemShut {NoStop}%
\bibitem [{\citenamefont {Jiang}\ \emph {et~al.}(2020)\citenamefont {Jiang},
  \citenamefont {Sarabalis}, \citenamefont {Dahmani}, \citenamefont {Patel},
  \citenamefont {Mayor}, \citenamefont {McKenna}, \citenamefont {Laer},\ and\
  \citenamefont {Safavi-Naeini}}]{Jiang2020}%
  \BibitemOpen
  \bibfield  {author} {\bibinfo {author} {\bibfnamefont {W.}~\bibnamefont
  {Jiang}}, \bibinfo {author} {\bibfnamefont {C.~J.}\ \bibnamefont
  {Sarabalis}}, \bibinfo {author} {\bibfnamefont {Y.~D.}\ \bibnamefont
  {Dahmani}}, \bibinfo {author} {\bibfnamefont {R.~N.}\ \bibnamefont {Patel}},
  \bibinfo {author} {\bibfnamefont {F.~M.}\ \bibnamefont {Mayor}}, \bibinfo
  {author} {\bibfnamefont {T.~P.}\ \bibnamefont {McKenna}}, \bibinfo {author}
  {\bibfnamefont {R.~V.}\ \bibnamefont {Laer}},\ and\ \bibinfo {author}
  {\bibfnamefont {A.~H.}\ \bibnamefont {Safavi-Naeini}},\ }\bibfield  {title}
  {\bibinfo {title} {Efficient bidirectional piezo-optomechanical transduction
  between microwave and optical frequency},\ }\bibfield  {journal} {\bibinfo
  {journal} {Nature Communications}\ }\textbf {\bibinfo {volume} {11}},\ \href
  {https://doi.org/10.1038/s41467-020-14863-3} {10.1038/s41467-020-14863-3}
  (\bibinfo {year} {2020})\BibitemShut {NoStop}%
\bibitem [{\citenamefont {Mirhosseini}\ \emph {et~al.}(2020)\citenamefont
  {Mirhosseini}, \citenamefont {Sipahigil}, \citenamefont {Kalaee},\ and\
  \citenamefont {Painter}}]{Mirhosseini2020}%
  \BibitemOpen
  \bibfield  {author} {\bibinfo {author} {\bibfnamefont {M.}~\bibnamefont
  {Mirhosseini}}, \bibinfo {author} {\bibfnamefont {A.}~\bibnamefont
  {Sipahigil}}, \bibinfo {author} {\bibfnamefont {M.}~\bibnamefont {Kalaee}},\
  and\ \bibinfo {author} {\bibfnamefont {O.}~\bibnamefont {Painter}},\
  }\bibfield  {title} {\bibinfo {title} {Superconducting qubit to optical
  photon transduction},\ }\bibfield  {journal} {\bibinfo  {journal} {Nature}\
  }\textbf {\bibinfo {volume} {588}},\ \href
  {https://doi.org/10.1038/s41586-020-3038-6} {10.1038/s41586-020-3038-6}
  (\bibinfo {year} {2020})\BibitemShut {NoStop}%
\bibitem [{\citenamefont {Zhong}\ \emph
  {et~al.}(2020{\natexlab{a}})\citenamefont {Zhong}, \citenamefont {Wang},
  \citenamefont {Zou}, \citenamefont {Zhang}, \citenamefont {Han},
  \citenamefont {Fu}, \citenamefont {Xu}, \citenamefont {Shankar},
  \citenamefont {Devoret}, \citenamefont {Tang},\ and\ \citenamefont
  {Jiang}}]{Zhong2020PRL}%
  \BibitemOpen
  \bibfield  {author} {\bibinfo {author} {\bibfnamefont {C.}~\bibnamefont
  {Zhong}}, \bibinfo {author} {\bibfnamefont {Z.}~\bibnamefont {Wang}},
  \bibinfo {author} {\bibfnamefont {C.}~\bibnamefont {Zou}}, \bibinfo {author}
  {\bibfnamefont {M.}~\bibnamefont {Zhang}}, \bibinfo {author} {\bibfnamefont
  {X.}~\bibnamefont {Han}}, \bibinfo {author} {\bibfnamefont {W.}~\bibnamefont
  {Fu}}, \bibinfo {author} {\bibfnamefont {M.}~\bibnamefont {Xu}}, \bibinfo
  {author} {\bibfnamefont {S.}~\bibnamefont {Shankar}}, \bibinfo {author}
  {\bibfnamefont {M.~H.}\ \bibnamefont {Devoret}}, \bibinfo {author}
  {\bibfnamefont {H.~X.}\ \bibnamefont {Tang}},\ and\ \bibinfo {author}
  {\bibfnamefont {L.}~\bibnamefont {Jiang}},\ }\bibfield  {title} {\bibinfo
  {title} {Proposal for heralded generation and detection of entangled
  microwave-optical-photon pairs},\ }\bibfield  {journal} {\bibinfo  {journal}
  {Physical Review Letters}\ }\textbf {\bibinfo {volume} {124}},\ \href
  {https://doi.org/10.1103/PhysRevLett.124.010511}
  {10.1103/PhysRevLett.124.010511} (\bibinfo {year}
  {2020}{\natexlab{a}})\BibitemShut {NoStop}%
\bibitem [{\citenamefont {Krastanov}\ \emph {et~al.}(2021)\citenamefont
  {Krastanov}, \citenamefont {Raniwala}, \citenamefont {Holzgrafe},
  \citenamefont {Jacobs}, \citenamefont {Lončar}, \citenamefont {Reagor},\
  and\ \citenamefont {Englund}}]{Krastanov2021}%
  \BibitemOpen
  \bibfield  {author} {\bibinfo {author} {\bibfnamefont {S.}~\bibnamefont
  {Krastanov}}, \bibinfo {author} {\bibfnamefont {H.}~\bibnamefont {Raniwala}},
  \bibinfo {author} {\bibfnamefont {J.}~\bibnamefont {Holzgrafe}}, \bibinfo
  {author} {\bibfnamefont {K.}~\bibnamefont {Jacobs}}, \bibinfo {author}
  {\bibfnamefont {M.}~\bibnamefont {Lončar}}, \bibinfo {author} {\bibfnamefont
  {M.~J.}\ \bibnamefont {Reagor}},\ and\ \bibinfo {author} {\bibfnamefont
  {D.~R.}\ \bibnamefont {Englund}},\ }\bibfield  {title} {\bibinfo {title}
  {Optically heralded entanglement of superconducting systems in quantum
  networks},\ }\bibfield  {journal} {\bibinfo  {journal} {Physical Review
  Letters}\ }\textbf {\bibinfo {volume} {127}},\ \href
  {https://doi.org/10.1103/PhysRevLett.127.040503}
  {10.1103/PhysRevLett.127.040503} (\bibinfo {year} {2021})\BibitemShut
  {NoStop}%
\bibitem [{\citenamefont {Wu}\ \emph {et~al.}(2021)\citenamefont {Wu},
  \citenamefont {Cui}, \citenamefont {Fan},\ and\ \citenamefont
  {Zhuang}}]{Wu2021}%
  \BibitemOpen
  \bibfield  {author} {\bibinfo {author} {\bibfnamefont {J.}~\bibnamefont
  {Wu}}, \bibinfo {author} {\bibfnamefont {C.}~\bibnamefont {Cui}}, \bibinfo
  {author} {\bibfnamefont {L.}~\bibnamefont {Fan}},\ and\ \bibinfo {author}
  {\bibfnamefont {Q.}~\bibnamefont {Zhuang}},\ }\bibfield  {title} {\bibinfo
  {title} {Deterministic microwave-optical transduction based on quantum
  teleportation},\ }\bibfield  {journal} {\bibinfo  {journal} {Physical Review
  Applied}\ }\textbf {\bibinfo {volume} {16}},\ \href
  {https://doi.org/10.1103/PhysRevApplied.16.064044}
  {10.1103/PhysRevApplied.16.064044} (\bibinfo {year} {2021})\BibitemShut
  {NoStop}%
\bibitem [{\citenamefont {Zhong}\ \emph {et~al.}(2024)\citenamefont {Zhong},
  \citenamefont {Li}, \citenamefont {Meesala}, \citenamefont {Wood},
  \citenamefont {Lake}, \citenamefont {Painter},\ and\ \citenamefont
  {Jiang}}]{PhysRevApplied.22.064047}%
  \BibitemOpen
  \bibfield  {author} {\bibinfo {author} {\bibfnamefont {C.}~\bibnamefont
  {Zhong}}, \bibinfo {author} {\bibfnamefont {F.}~\bibnamefont {Li}}, \bibinfo
  {author} {\bibfnamefont {S.}~\bibnamefont {Meesala}}, \bibinfo {author}
  {\bibfnamefont {S.}~\bibnamefont {Wood}}, \bibinfo {author} {\bibfnamefont
  {D.}~\bibnamefont {Lake}}, \bibinfo {author} {\bibfnamefont {O.}~\bibnamefont
  {Painter}},\ and\ \bibinfo {author} {\bibfnamefont {L.}~\bibnamefont
  {Jiang}},\ }\bibfield  {title} {\bibinfo {title} {Microwave-optical
  entanglement from pulse-pumped electro-optomechanics},\ }\href
  {https://doi.org/10.1103/PhysRevApplied.22.064047} {\bibfield  {journal}
  {\bibinfo  {journal} {Phys. Rev. Appl.}\ }\textbf {\bibinfo {volume} {22}},\
  \bibinfo {pages} {064047} (\bibinfo {year} {2024})}\BibitemShut {NoStop}%
\bibitem [{\citenamefont {Zhong}\ \emph {et~al.}(2022)\citenamefont {Zhong},
  \citenamefont {Han},\ and\ \citenamefont {Jiang}}]{Zhong2022}%
  \BibitemOpen
  \bibfield  {author} {\bibinfo {author} {\bibfnamefont {C.}~\bibnamefont
  {Zhong}}, \bibinfo {author} {\bibfnamefont {X.}~\bibnamefont {Han}},\ and\
  \bibinfo {author} {\bibfnamefont {L.}~\bibnamefont {Jiang}},\ }\bibfield
  {title} {\bibinfo {title} {Microwave and optical entanglement for quantum
  transduction with electro-optomechanics},\ }\bibfield  {journal} {\bibinfo
  {journal} {Physical Review Applied}\ }\textbf {\bibinfo {volume} {18}},\
  \href {https://doi.org/10.1103/PhysRevApplied.18.054061}
  {10.1103/PhysRevApplied.18.054061} (\bibinfo {year} {2022})\BibitemShut
  {NoStop}%
\bibitem [{\citenamefont {Duan}\ \emph {et~al.}(2001)\citenamefont {Duan},
  \citenamefont {Lukin}, \citenamefont {Cirac},\ and\ \citenamefont
  {Zoller}}]{Duan2001}%
  \BibitemOpen
  \bibfield  {author} {\bibinfo {author} {\bibfnamefont {L.~M.}\ \bibnamefont
  {Duan}}, \bibinfo {author} {\bibfnamefont {M.~D.}\ \bibnamefont {Lukin}},
  \bibinfo {author} {\bibfnamefont {J.~I.}\ \bibnamefont {Cirac}},\ and\
  \bibinfo {author} {\bibfnamefont {P.}~\bibnamefont {Zoller}},\ }\bibfield
  {title} {\bibinfo {title} {Long-distance quantum communication with atomic
  ensembles and linear optics},\ }\bibfield  {journal} {\bibinfo  {journal}
  {Nature}\ }\textbf {\bibinfo {volume} {414}},\ \href
  {https://doi.org/10.1038/35106500} {10.1038/35106500} (\bibinfo {year}
  {2001})\BibitemShut {NoStop}%
\bibitem [{\citenamefont {Sahu}\ \emph {et~al.}(2023)\citenamefont {Sahu},
  \citenamefont {Qiu}, \citenamefont {Hease}, \citenamefont {Arnold},
  \citenamefont {Minoguchi}, \citenamefont {Rabl},\ and\ \citenamefont
  {Fink}}]{Sahu2023}%
  \BibitemOpen
  \bibfield  {author} {\bibinfo {author} {\bibfnamefont {R.}~\bibnamefont
  {Sahu}}, \bibinfo {author} {\bibfnamefont {L.}~\bibnamefont {Qiu}}, \bibinfo
  {author} {\bibfnamefont {W.}~\bibnamefont {Hease}}, \bibinfo {author}
  {\bibfnamefont {G.}~\bibnamefont {Arnold}}, \bibinfo {author} {\bibfnamefont
  {Y.}~\bibnamefont {Minoguchi}}, \bibinfo {author} {\bibfnamefont
  {P.}~\bibnamefont {Rabl}},\ and\ \bibinfo {author} {\bibfnamefont {J.~M.}\
  \bibnamefont {Fink}},\ }\bibfield  {title} {\bibinfo {title} {Entangling
  microwaves with light},\ }\bibfield  {journal} {\bibinfo  {journal}
  {Science}\ }\textbf {\bibinfo {volume} {380}},\ \href
  {https://doi.org/10.1126/SCIENCE.ADG3812} {10.1126/SCIENCE.ADG3812} (\bibinfo
  {year} {2023})\BibitemShut {NoStop}%
\bibitem [{\citenamefont {Meesala}\ \emph
  {et~al.}(2024{\natexlab{a}})\citenamefont {Meesala}, \citenamefont {Wood},
  \citenamefont {Lake}, \citenamefont {Chiappina}, \citenamefont {Zhong},
  \citenamefont {Beyer}, \citenamefont {Shaw}, \citenamefont {Jiang},\ and\
  \citenamefont {Painter}}]{Meesala2024}%
  \BibitemOpen
  \bibfield  {author} {\bibinfo {author} {\bibfnamefont {S.}~\bibnamefont
  {Meesala}}, \bibinfo {author} {\bibfnamefont {S.}~\bibnamefont {Wood}},
  \bibinfo {author} {\bibfnamefont {D.}~\bibnamefont {Lake}}, \bibinfo {author}
  {\bibfnamefont {P.}~\bibnamefont {Chiappina}}, \bibinfo {author}
  {\bibfnamefont {C.}~\bibnamefont {Zhong}}, \bibinfo {author} {\bibfnamefont
  {A.~D.}\ \bibnamefont {Beyer}}, \bibinfo {author} {\bibfnamefont {M.~D.}\
  \bibnamefont {Shaw}}, \bibinfo {author} {\bibfnamefont {L.}~\bibnamefont
  {Jiang}},\ and\ \bibinfo {author} {\bibfnamefont {O.}~\bibnamefont
  {Painter}},\ }\bibfield  {title} {\bibinfo {title} {Non-classical
  microwave–optical photon pair generation with a chip-scale transducer},\
  }\bibfield  {journal} {\bibinfo  {journal} {Nature Physics}\ }\href
  {https://doi.org/10.1038/s41567-024-02409-z} {10.1038/s41567-024-02409-z}
  (\bibinfo {year} {2024}{\natexlab{a}})\BibitemShut {NoStop}%
\bibitem [{\citenamefont {Meesala}\ \emph
  {et~al.}(2024{\natexlab{b}})\citenamefont {Meesala}, \citenamefont {Lake},
  \citenamefont {Wood}, \citenamefont {Chiappina}, \citenamefont {Zhong},
  \citenamefont {Beyer}, \citenamefont {Shaw}, \citenamefont {Jiang},\ and\
  \citenamefont {Painter}}]{PhysRevX.14.031055}%
  \BibitemOpen
  \bibfield  {author} {\bibinfo {author} {\bibfnamefont {S.}~\bibnamefont
  {Meesala}}, \bibinfo {author} {\bibfnamefont {D.}~\bibnamefont {Lake}},
  \bibinfo {author} {\bibfnamefont {S.}~\bibnamefont {Wood}}, \bibinfo {author}
  {\bibfnamefont {P.}~\bibnamefont {Chiappina}}, \bibinfo {author}
  {\bibfnamefont {C.}~\bibnamefont {Zhong}}, \bibinfo {author} {\bibfnamefont
  {A.~D.}\ \bibnamefont {Beyer}}, \bibinfo {author} {\bibfnamefont {M.~D.}\
  \bibnamefont {Shaw}}, \bibinfo {author} {\bibfnamefont {L.}~\bibnamefont
  {Jiang}},\ and\ \bibinfo {author} {\bibfnamefont {O.}~\bibnamefont
  {Painter}},\ }\bibfield  {title} {\bibinfo {title} {Quantum entanglement
  between optical and microwave photonic qubits},\ }\href
  {https://doi.org/10.1103/PhysRevX.14.031055} {\bibfield  {journal} {\bibinfo
  {journal} {Phys. Rev. X}\ }\textbf {\bibinfo {volume} {14}},\ \bibinfo
  {pages} {031055} (\bibinfo {year} {2024}{\natexlab{b}})}\BibitemShut
  {NoStop}%
\bibitem [{\citenamefont {Kiilerich}\ and\ \citenamefont
  {Mølmer}(2019)}]{Kiilerich2019}%
  \BibitemOpen
  \bibfield  {author} {\bibinfo {author} {\bibfnamefont {A.~H.}\ \bibnamefont
  {Kiilerich}}\ and\ \bibinfo {author} {\bibfnamefont {K.}~\bibnamefont
  {Mølmer}},\ }\bibfield  {title} {\bibinfo {title} {Input-output theory with
  quantum pulses},\ }\bibfield  {journal} {\bibinfo  {journal} {Physical Review
  Letters}\ }\textbf {\bibinfo {volume} {123}},\ \href
  {https://doi.org/10.1103/PhysRevLett.123.123604}
  {10.1103/PhysRevLett.123.123604} (\bibinfo {year} {2019})\BibitemShut
  {NoStop}%
\bibitem [{\citenamefont {Kiilerich}\ and\ \citenamefont
  {Mølmer}(2020)}]{Kiilerich2020}%
  \BibitemOpen
  \bibfield  {author} {\bibinfo {author} {\bibfnamefont {A.~H.}\ \bibnamefont
  {Kiilerich}}\ and\ \bibinfo {author} {\bibfnamefont {K.}~\bibnamefont
  {Mølmer}},\ }\bibfield  {title} {\bibinfo {title} {Quantum interactions with
  pulses of radiation},\ }\bibfield  {journal} {\bibinfo  {journal} {Physical
  Review A}\ }\textbf {\bibinfo {volume} {102}},\ \href
  {https://doi.org/10.1103/PhysRevA.102.023717} {10.1103/PhysRevA.102.023717}
  (\bibinfo {year} {2020})\BibitemShut {NoStop}%
\bibitem [{\citenamefont {Lambert}\ \emph {et~al.}(2024)\citenamefont
  {Lambert}, \citenamefont {Giguère}, \citenamefont {Menczel}, \citenamefont
  {Li}, \citenamefont {Hopf}, \citenamefont {Suárez}, \citenamefont {Gali},
  \citenamefont {Lishman}, \citenamefont {Gadhvi}, \citenamefont {Agarwal},
  \citenamefont {Galicia}, \citenamefont {Shammah}, \citenamefont {Nation},
  \citenamefont {Johansson}, \citenamefont {Ahmed}, \citenamefont {Cross},
  \citenamefont {Pitchford},\ and\ \citenamefont
  {Nori}}]{lambert2024qutip5quantumtoolbox}%
  \BibitemOpen
  \bibfield  {author} {\bibinfo {author} {\bibfnamefont {N.}~\bibnamefont
  {Lambert}}, \bibinfo {author} {\bibfnamefont {E.}~\bibnamefont {Giguère}},
  \bibinfo {author} {\bibfnamefont {P.}~\bibnamefont {Menczel}}, \bibinfo
  {author} {\bibfnamefont {B.}~\bibnamefont {Li}}, \bibinfo {author}
  {\bibfnamefont {P.}~\bibnamefont {Hopf}}, \bibinfo {author} {\bibfnamefont
  {G.}~\bibnamefont {Suárez}}, \bibinfo {author} {\bibfnamefont
  {M.}~\bibnamefont {Gali}}, \bibinfo {author} {\bibfnamefont {J.}~\bibnamefont
  {Lishman}}, \bibinfo {author} {\bibfnamefont {R.}~\bibnamefont {Gadhvi}},
  \bibinfo {author} {\bibfnamefont {R.}~\bibnamefont {Agarwal}}, \bibinfo
  {author} {\bibfnamefont {A.}~\bibnamefont {Galicia}}, \bibinfo {author}
  {\bibfnamefont {N.}~\bibnamefont {Shammah}}, \bibinfo {author} {\bibfnamefont
  {P.}~\bibnamefont {Nation}}, \bibinfo {author} {\bibfnamefont {J.~R.}\
  \bibnamefont {Johansson}}, \bibinfo {author} {\bibfnamefont {S.}~\bibnamefont
  {Ahmed}}, \bibinfo {author} {\bibfnamefont {S.}~\bibnamefont {Cross}},
  \bibinfo {author} {\bibfnamefont {A.}~\bibnamefont {Pitchford}},\ and\
  \bibinfo {author} {\bibfnamefont {F.}~\bibnamefont {Nori}},\ }\href
  {https://arxiv.org/abs/2412.04705} {\bibinfo {title} {Qutip 5: The quantum
  toolbox in python}} (\bibinfo {year} {2024}),\ \Eprint
  {https://arxiv.org/abs/2412.04705} {arXiv:2412.04705 [quant-ph]} \BibitemShut
  {NoStop}%
\bibitem [{\citenamefont {Gao}\ \emph {et~al.}(2019)\citenamefont {Gao},
  \citenamefont {Lester}, \citenamefont {Chou}, \citenamefont {Frunzio},
  \citenamefont {Devoret}, \citenamefont {Jiang}, \citenamefont {Girvin},\ and\
  \citenamefont {Schoelkopf}}]{Gao2019}%
  \BibitemOpen
  \bibfield  {author} {\bibinfo {author} {\bibfnamefont {Y.~Y.}\ \bibnamefont
  {Gao}}, \bibinfo {author} {\bibfnamefont {B.~J.}\ \bibnamefont {Lester}},
  \bibinfo {author} {\bibfnamefont {K.~S.}\ \bibnamefont {Chou}}, \bibinfo
  {author} {\bibfnamefont {L.}~\bibnamefont {Frunzio}}, \bibinfo {author}
  {\bibfnamefont {M.~H.}\ \bibnamefont {Devoret}}, \bibinfo {author}
  {\bibfnamefont {L.}~\bibnamefont {Jiang}}, \bibinfo {author} {\bibfnamefont
  {S.~M.}\ \bibnamefont {Girvin}},\ and\ \bibinfo {author} {\bibfnamefont
  {R.~J.}\ \bibnamefont {Schoelkopf}},\ }\bibfield  {title} {\bibinfo {title}
  {Entanglement of bosonic modes through an engineered exchange interaction},\
  }\bibfield  {journal} {\bibinfo  {journal} {Nature}\ }\textbf {\bibinfo
  {volume} {566}},\ \href {https://doi.org/10.1038/s41586-019-0970-4}
  {10.1038/s41586-019-0970-4} (\bibinfo {year} {2019})\BibitemShut {NoStop}%
\bibitem [{\citenamefont {Blais}\ \emph {et~al.}(2021)\citenamefont {Blais},
  \citenamefont {Grimsmo}, \citenamefont {Girvin},\ and\ \citenamefont
  {Wallraff}}]{Blais2021}%
  \BibitemOpen
  \bibfield  {author} {\bibinfo {author} {\bibfnamefont {A.}~\bibnamefont
  {Blais}}, \bibinfo {author} {\bibfnamefont {A.~L.}\ \bibnamefont {Grimsmo}},
  \bibinfo {author} {\bibfnamefont {S.~M.}\ \bibnamefont {Girvin}},\ and\
  \bibinfo {author} {\bibfnamefont {A.}~\bibnamefont {Wallraff}},\ }\bibfield
  {title} {\bibinfo {title} {Circuit quantum electrodynamics},\ }\bibfield
  {journal} {\bibinfo  {journal} {Reviews of Modern Physics}\ }\textbf
  {\bibinfo {volume} {93}},\ \href
  {https://doi.org/10.1103/RevModPhys.93.025005} {10.1103/RevModPhys.93.025005}
  (\bibinfo {year} {2021})\BibitemShut {NoStop}%
\bibitem [{\citenamefont {Ye}\ \emph {et~al.}(2021)\citenamefont {Ye},
  \citenamefont {Peng}, \citenamefont {Naghiloo}, \citenamefont {Cunningham},\
  and\ \citenamefont {O'Brien}}]{Ye2021}%
  \BibitemOpen
  \bibfield  {author} {\bibinfo {author} {\bibfnamefont {Y.}~\bibnamefont
  {Ye}}, \bibinfo {author} {\bibfnamefont {K.}~\bibnamefont {Peng}}, \bibinfo
  {author} {\bibfnamefont {M.}~\bibnamefont {Naghiloo}}, \bibinfo {author}
  {\bibfnamefont {G.}~\bibnamefont {Cunningham}},\ and\ \bibinfo {author}
  {\bibfnamefont {K.~P.}\ \bibnamefont {O'Brien}},\ }\bibfield  {title}
  {\bibinfo {title} {Engineering purely nonlinear coupling between
  superconducting qubits using a quarton},\ }\bibfield  {journal} {\bibinfo
  {journal} {Physical Review Letters}\ }\textbf {\bibinfo {volume} {127}},\
  \href {https://doi.org/10.1103/PhysRevLett.127.050502}
  {10.1103/PhysRevLett.127.050502} (\bibinfo {year} {2021})\BibitemShut
  {NoStop}%
\bibitem [{\citenamefont {Chakram}\ \emph {et~al.}(2022)\citenamefont
  {Chakram}, \citenamefont {He}, \citenamefont {Dixit}, \citenamefont {Oriani},
  \citenamefont {Naik}, \citenamefont {Leung}, \citenamefont {Kwon},
  \citenamefont {Ma}, \citenamefont {Jiang},\ and\ \citenamefont
  {Schuster}}]{Chakram2022}%
  \BibitemOpen
  \bibfield  {author} {\bibinfo {author} {\bibfnamefont {S.}~\bibnamefont
  {Chakram}}, \bibinfo {author} {\bibfnamefont {K.}~\bibnamefont {He}},
  \bibinfo {author} {\bibfnamefont {A.~V.}\ \bibnamefont {Dixit}}, \bibinfo
  {author} {\bibfnamefont {A.~E.}\ \bibnamefont {Oriani}}, \bibinfo {author}
  {\bibfnamefont {R.~K.}\ \bibnamefont {Naik}}, \bibinfo {author}
  {\bibfnamefont {N.}~\bibnamefont {Leung}}, \bibinfo {author} {\bibfnamefont
  {H.}~\bibnamefont {Kwon}}, \bibinfo {author} {\bibfnamefont {W.~L.}\
  \bibnamefont {Ma}}, \bibinfo {author} {\bibfnamefont {L.}~\bibnamefont
  {Jiang}},\ and\ \bibinfo {author} {\bibfnamefont {D.~I.}\ \bibnamefont
  {Schuster}},\ }\bibfield  {title} {\bibinfo {title} {Multimode photon
  blockade},\ }\bibfield  {journal} {\bibinfo  {journal} {Nature Physics}\
  }\textbf {\bibinfo {volume} {18}},\ \href
  {https://doi.org/10.1038/s41567-022-01630-y} {10.1038/s41567-022-01630-y}
  (\bibinfo {year} {2022})\BibitemShut {NoStop}%
\bibitem [{Note1()}]{Note1}%
  \BibitemOpen
  \bibinfo {note} {In principle, there also exists an optimal driving strength
  for maximizing the generation probability, given by \( g = \left ( 0.09 \chi
  _c^2 \kappa _i \right )^{1/3} \). However, Fig.~\ref {Fig3}(d)(f) indicates
  that fidelity is more sensitive to variations in \( g \) than the probability
  is. More rigorously, we can see that $\protect \frac {\partial ^2 F_{\protect
  \text {post}}}{\partial g^2}>\protect \frac {\partial ^2 P_{\protect \text
  {post}}}{\partial g^2}$ for all $\kappa _i,g, \chi _c$ in the over-coupled
  regime $\kappa _c>\kappa _i$. Therefore, to optimize overall performance, we
  select the driving strength that maximizes fidelity.}\BibitemShut {Stop}%
\bibitem [{\citenamefont {Koch}\ \emph {et~al.}(2007)\citenamefont {Koch},
  \citenamefont {Yu}, \citenamefont {Gambetta}, \citenamefont {Houck},
  \citenamefont {Schuster}, \citenamefont {Majer}, \citenamefont {Blais},
  \citenamefont {Devoret}, \citenamefont {Girvin},\ and\ \citenamefont
  {Schoelkopf}}]{PhysRevA.76.042319}%
  \BibitemOpen
  \bibfield  {author} {\bibinfo {author} {\bibfnamefont {J.}~\bibnamefont
  {Koch}}, \bibinfo {author} {\bibfnamefont {T.~M.}\ \bibnamefont {Yu}},
  \bibinfo {author} {\bibfnamefont {J.}~\bibnamefont {Gambetta}}, \bibinfo
  {author} {\bibfnamefont {A.~A.}\ \bibnamefont {Houck}}, \bibinfo {author}
  {\bibfnamefont {D.~I.}\ \bibnamefont {Schuster}}, \bibinfo {author}
  {\bibfnamefont {J.}~\bibnamefont {Majer}}, \bibinfo {author} {\bibfnamefont
  {A.}~\bibnamefont {Blais}}, \bibinfo {author} {\bibfnamefont {M.~H.}\
  \bibnamefont {Devoret}}, \bibinfo {author} {\bibfnamefont {S.~M.}\
  \bibnamefont {Girvin}},\ and\ \bibinfo {author} {\bibfnamefont {R.~J.}\
  \bibnamefont {Schoelkopf}},\ }\bibfield  {title} {\bibinfo {title}
  {Charge-insensitive qubit design derived from the cooper pair box},\ }\href
  {https://doi.org/10.1103/PhysRevA.76.042319} {\bibfield  {journal} {\bibinfo
  {journal} {Phys. Rev. A}\ }\textbf {\bibinfo {volume} {76}},\ \bibinfo
  {pages} {042319} (\bibinfo {year} {2007})}\BibitemShut {NoStop}%
\bibitem [{\citenamefont {Han}\ \emph {et~al.}(2020)\citenamefont {Han},
  \citenamefont {Fu}, \citenamefont {Zhong}, \citenamefont {Zou}, \citenamefont
  {Xu}, \citenamefont {Sayem}, \citenamefont {Xu}, \citenamefont {Wang},
  \citenamefont {Cheng}, \citenamefont {Jiang},\ and\ \citenamefont
  {Tang}}]{Han2020}%
  \BibitemOpen
  \bibfield  {author} {\bibinfo {author} {\bibfnamefont {X.}~\bibnamefont
  {Han}}, \bibinfo {author} {\bibfnamefont {W.}~\bibnamefont {Fu}}, \bibinfo
  {author} {\bibfnamefont {C.}~\bibnamefont {Zhong}}, \bibinfo {author}
  {\bibfnamefont {C.~L.}\ \bibnamefont {Zou}}, \bibinfo {author} {\bibfnamefont
  {Y.}~\bibnamefont {Xu}}, \bibinfo {author} {\bibfnamefont {A.~A.}\
  \bibnamefont {Sayem}}, \bibinfo {author} {\bibfnamefont {M.}~\bibnamefont
  {Xu}}, \bibinfo {author} {\bibfnamefont {S.}~\bibnamefont {Wang}}, \bibinfo
  {author} {\bibfnamefont {R.}~\bibnamefont {Cheng}}, \bibinfo {author}
  {\bibfnamefont {L.}~\bibnamefont {Jiang}},\ and\ \bibinfo {author}
  {\bibfnamefont {H.~X.}\ \bibnamefont {Tang}},\ }\bibfield  {title} {\bibinfo
  {title} {Cavity piezo-mechanics for superconducting-nanophotonic quantum
  interface},\ }\bibfield  {journal} {\bibinfo  {journal} {Nature
  Communications}\ }\textbf {\bibinfo {volume} {11}},\ \href
  {https://doi.org/10.1038/s41467-020-17053-3} {10.1038/s41467-020-17053-3}
  (\bibinfo {year} {2020})\BibitemShut {NoStop}%
\bibitem [{\citenamefont {Xu}\ \emph {et~al.}(2021)\citenamefont {Xu},
  \citenamefont {Sayem}, \citenamefont {Fan}, \citenamefont {Zou},
  \citenamefont {Wang}, \citenamefont {Cheng}, \citenamefont {Fu},
  \citenamefont {Yang}, \citenamefont {Xu},\ and\ \citenamefont
  {Tang}}]{Xu2021}%
  \BibitemOpen
  \bibfield  {author} {\bibinfo {author} {\bibfnamefont {Y.}~\bibnamefont
  {Xu}}, \bibinfo {author} {\bibfnamefont {A.~A.}\ \bibnamefont {Sayem}},
  \bibinfo {author} {\bibfnamefont {L.}~\bibnamefont {Fan}}, \bibinfo {author}
  {\bibfnamefont {C.~L.}\ \bibnamefont {Zou}}, \bibinfo {author} {\bibfnamefont
  {S.}~\bibnamefont {Wang}}, \bibinfo {author} {\bibfnamefont {R.}~\bibnamefont
  {Cheng}}, \bibinfo {author} {\bibfnamefont {W.}~\bibnamefont {Fu}}, \bibinfo
  {author} {\bibfnamefont {L.}~\bibnamefont {Yang}}, \bibinfo {author}
  {\bibfnamefont {M.}~\bibnamefont {Xu}},\ and\ \bibinfo {author}
  {\bibfnamefont {H.~X.}\ \bibnamefont {Tang}},\ }\bibfield  {title} {\bibinfo
  {title} {Bidirectional interconversion of microwave and light with thin-film
  lithium niobate},\ }\bibfield  {journal} {\bibinfo  {journal} {Nature
  Communications}\ }\textbf {\bibinfo {volume} {12}},\ \href
  {https://doi.org/10.1038/s41467-021-24809-y} {10.1038/s41467-021-24809-y}
  (\bibinfo {year} {2021})\BibitemShut {NoStop}%
\bibitem [{\citenamefont {Warner}\ \emph {et~al.}(2025)\citenamefont {Warner},
  \citenamefont {Holzgrafe}, \citenamefont {Yankelevich}, \citenamefont
  {Barton}, \citenamefont {Poletto}, \citenamefont {Xin}, \citenamefont
  {Sinclair}, \citenamefont {Zhu}, \citenamefont {Sete}, \citenamefont
  {Langley}, \citenamefont {Batson}, \citenamefont {Colangelo}, \citenamefont
  {Shams-Ansari}, \citenamefont {Joe}, \citenamefont {Berggren}, \citenamefont
  {Jiang}, \citenamefont {Reagor},\ and\ \citenamefont {Lon{\v
  c}ar}}]{Hana2025}%
  \BibitemOpen
  \bibfield  {author} {\bibinfo {author} {\bibfnamefont {H.~K.}\ \bibnamefont
  {Warner}}, \bibinfo {author} {\bibfnamefont {J.}~\bibnamefont {Holzgrafe}},
  \bibinfo {author} {\bibfnamefont {B.}~\bibnamefont {Yankelevich}}, \bibinfo
  {author} {\bibfnamefont {D.}~\bibnamefont {Barton}}, \bibinfo {author}
  {\bibfnamefont {S.}~\bibnamefont {Poletto}}, \bibinfo {author} {\bibfnamefont
  {C.~J.}\ \bibnamefont {Xin}}, \bibinfo {author} {\bibfnamefont
  {N.}~\bibnamefont {Sinclair}}, \bibinfo {author} {\bibfnamefont
  {D.}~\bibnamefont {Zhu}}, \bibinfo {author} {\bibfnamefont {E.}~\bibnamefont
  {Sete}}, \bibinfo {author} {\bibfnamefont {B.}~\bibnamefont {Langley}},
  \bibinfo {author} {\bibfnamefont {E.}~\bibnamefont {Batson}}, \bibinfo
  {author} {\bibfnamefont {M.}~\bibnamefont {Colangelo}}, \bibinfo {author}
  {\bibfnamefont {A.}~\bibnamefont {Shams-Ansari}}, \bibinfo {author}
  {\bibfnamefont {G.}~\bibnamefont {Joe}}, \bibinfo {author} {\bibfnamefont
  {K.~K.}\ \bibnamefont {Berggren}}, \bibinfo {author} {\bibfnamefont
  {L.}~\bibnamefont {Jiang}}, \bibinfo {author} {\bibfnamefont {M.~J.}\
  \bibnamefont {Reagor}},\ and\ \bibinfo {author} {\bibfnamefont
  {M.}~\bibnamefont {Lon{\v c}ar}},\ }\bibfield  {title} {\bibinfo {title}
  {Coherent control of a superconducting qubit using light},\ }\bibfield
  {journal} {\bibinfo  {journal} {Nature Physics}\ }\href
  {https://doi.org/10.1038/s41567-025-02812-0} {10.1038/s41567-025-02812-0}
  (\bibinfo {year} {2025})\BibitemShut {NoStop}%
\bibitem [{\citenamefont {Zhong}\ \emph
  {et~al.}(2020{\natexlab{b}})\citenamefont {Zhong}, \citenamefont {Han},
  \citenamefont {Tang},\ and\ \citenamefont {Jiang}}]{Zhong2020}%
  \BibitemOpen
  \bibfield  {author} {\bibinfo {author} {\bibfnamefont {C.}~\bibnamefont
  {Zhong}}, \bibinfo {author} {\bibfnamefont {X.}~\bibnamefont {Han}}, \bibinfo
  {author} {\bibfnamefont {H.~X.}\ \bibnamefont {Tang}},\ and\ \bibinfo
  {author} {\bibfnamefont {L.}~\bibnamefont {Jiang}},\ }\bibfield  {title}
  {\bibinfo {title} {Entanglement of microwave-optical modes in a strongly
  coupled electro-optomechanical system},\ }\bibfield  {journal} {\bibinfo
  {journal} {Physical Review A}\ }\textbf {\bibinfo {volume} {101}},\ \href
  {https://doi.org/10.1103/PhysRevA.101.032345} {10.1103/PhysRevA.101.032345}
  (\bibinfo {year} {2020}{\natexlab{b}})\BibitemShut {NoStop}%
\bibitem [{\citenamefont {Delaney}\ \emph {et~al.}(2022)\citenamefont
  {Delaney}, \citenamefont {Urmey}, \citenamefont {Mittal}, \citenamefont
  {Brubaker}, \citenamefont {Kindem}, \citenamefont {Burns}, \citenamefont
  {Regal},\ and\ \citenamefont {Lehnert}}]{konradnature}%
  \BibitemOpen
  \bibfield  {author} {\bibinfo {author} {\bibfnamefont {R.~D.}\ \bibnamefont
  {Delaney}}, \bibinfo {author} {\bibfnamefont {M.~D.}\ \bibnamefont {Urmey}},
  \bibinfo {author} {\bibfnamefont {S.}~\bibnamefont {Mittal}}, \bibinfo
  {author} {\bibfnamefont {B.~M.}\ \bibnamefont {Brubaker}}, \bibinfo {author}
  {\bibfnamefont {J.~M.}\ \bibnamefont {Kindem}}, \bibinfo {author}
  {\bibfnamefont {P.~S.}\ \bibnamefont {Burns}}, \bibinfo {author}
  {\bibfnamefont {C.~A.}\ \bibnamefont {Regal}},\ and\ \bibinfo {author}
  {\bibfnamefont {K.~W.}\ \bibnamefont {Lehnert}},\ }\bibfield  {title}
  {\bibinfo {title} {Superconducting-qubit readout via low-backaction
  electro-optic transduction},\ }\href
  {https://doi.org/10.1038/s41586-022-04720-2} {\bibfield  {journal} {\bibinfo
  {journal} {Nature}\ }\textbf {\bibinfo {volume} {606}},\ \bibinfo {pages}
  {489} (\bibinfo {year} {2022})}\BibitemShut {NoStop}%
\bibitem [{\citenamefont {Brubaker}\ \emph {et~al.}(2022)\citenamefont
  {Brubaker}, \citenamefont {Kindem}, \citenamefont {Urmey}, \citenamefont
  {Mittal}, \citenamefont {Delaney}, \citenamefont {Burns}, \citenamefont
  {Vissers}, \citenamefont {Lehnert},\ and\ \citenamefont
  {Regal}}]{PhysRevX.12.021062}%
  \BibitemOpen
  \bibfield  {author} {\bibinfo {author} {\bibfnamefont {B.~M.}\ \bibnamefont
  {Brubaker}}, \bibinfo {author} {\bibfnamefont {J.~M.}\ \bibnamefont
  {Kindem}}, \bibinfo {author} {\bibfnamefont {M.~D.}\ \bibnamefont {Urmey}},
  \bibinfo {author} {\bibfnamefont {S.}~\bibnamefont {Mittal}}, \bibinfo
  {author} {\bibfnamefont {R.~D.}\ \bibnamefont {Delaney}}, \bibinfo {author}
  {\bibfnamefont {P.~S.}\ \bibnamefont {Burns}}, \bibinfo {author}
  {\bibfnamefont {M.~R.}\ \bibnamefont {Vissers}}, \bibinfo {author}
  {\bibfnamefont {K.~W.}\ \bibnamefont {Lehnert}},\ and\ \bibinfo {author}
  {\bibfnamefont {C.~A.}\ \bibnamefont {Regal}},\ }\bibfield  {title} {\bibinfo
  {title} {Optomechanical ground-state cooling in a continuous and efficient
  electro-optic transducer},\ }\href
  {https://doi.org/10.1103/PhysRevX.12.021062} {\bibfield  {journal} {\bibinfo
  {journal} {Phys. Rev. X}\ }\textbf {\bibinfo {volume} {12}},\ \bibinfo
  {pages} {021062} (\bibinfo {year} {2022})}\BibitemShut {NoStop}%
\bibitem [{\citenamefont {Motzoi}\ \emph {et~al.}(2009)\citenamefont {Motzoi},
  \citenamefont {Gambetta}, \citenamefont {Rebentrost},\ and\ \citenamefont
  {Wilhelm}}]{PhysRevLett.103.110501}%
  \BibitemOpen
  \bibfield  {author} {\bibinfo {author} {\bibfnamefont {F.}~\bibnamefont
  {Motzoi}}, \bibinfo {author} {\bibfnamefont {J.~M.}\ \bibnamefont
  {Gambetta}}, \bibinfo {author} {\bibfnamefont {P.}~\bibnamefont
  {Rebentrost}},\ and\ \bibinfo {author} {\bibfnamefont {F.~K.}\ \bibnamefont
  {Wilhelm}},\ }\bibfield  {title} {\bibinfo {title} {Simple pulses for
  elimination of leakage in weakly nonlinear qubits},\ }\href
  {https://doi.org/10.1103/PhysRevLett.103.110501} {\bibfield  {journal}
  {\bibinfo  {journal} {Phys. Rev. Lett.}\ }\textbf {\bibinfo {volume} {103}},\
  \bibinfo {pages} {110501} (\bibinfo {year} {2009})}\BibitemShut {NoStop}%
\bibitem [{\citenamefont {Wang}\ \emph {et~al.}(2022)\citenamefont {Wang},
  \citenamefont {Li},\ and\ \citenamefont {Jiang}}]{Wang2022}%
  \BibitemOpen
  \bibfield  {author} {\bibinfo {author} {\bibfnamefont {C.~H.}\ \bibnamefont
  {Wang}}, \bibinfo {author} {\bibfnamefont {F.}~\bibnamefont {Li}},\ and\
  \bibinfo {author} {\bibfnamefont {L.}~\bibnamefont {Jiang}},\ }\bibfield
  {title} {\bibinfo {title} {Quantum capacities of transducers},\ }\bibfield
  {journal} {\bibinfo  {journal} {Nature Communications}\ }\textbf {\bibinfo
  {volume} {13}},\ \href {https://doi.org/10.1038/s41467-022-34373-8}
  {10.1038/s41467-022-34373-8} (\bibinfo {year} {2022})\BibitemShut {NoStop}%
\bibitem [{\citenamefont {Wu}\ \emph {et~al.}(2024)\citenamefont {Wu},
  \citenamefont {Fan},\ and\ \citenamefont {Zhuang}}]{PhysRevA.109.022619}%
  \BibitemOpen
  \bibfield  {author} {\bibinfo {author} {\bibfnamefont {J.}~\bibnamefont
  {Wu}}, \bibinfo {author} {\bibfnamefont {L.}~\bibnamefont {Fan}},\ and\
  \bibinfo {author} {\bibfnamefont {Q.}~\bibnamefont {Zhuang}},\ }\bibfield
  {title} {\bibinfo {title} {Teleportation-based microwave-to-optical quantum
  transduction: The limited role of single-mode squeezing},\ }\href
  {https://doi.org/10.1103/PhysRevA.109.022619} {\bibfield  {journal} {\bibinfo
   {journal} {Phys. Rev. A}\ }\textbf {\bibinfo {volume} {109}},\ \bibinfo
  {pages} {022619} (\bibinfo {year} {2024})}\BibitemShut {NoStop}%
\bibitem [{\citenamefont {Reiter}\ and\ \citenamefont
  {Sørensen}(2012)}]{reiter_effective_2012}%
  \BibitemOpen
  \bibfield  {author} {\bibinfo {author} {\bibfnamefont {F.}~\bibnamefont
  {Reiter}}\ and\ \bibinfo {author} {\bibfnamefont {A.~S.}\ \bibnamefont
  {Sørensen}},\ }\bibfield  {title} {\bibinfo {title} {Effective operator
  formalism for open quantum systems},\ }\href
  {https://doi.org/10.1103/PhysRevA.85.032111} {\bibfield  {journal} {\bibinfo
  {journal} {Physical Review A}\ }\textbf {\bibinfo {volume} {85}},\ \bibinfo
  {pages} {032111} (\bibinfo {year} {2012})}\BibitemShut {NoStop}%
\bibitem [{\citenamefont {Fischer}\ \emph {et~al.}(2018)\citenamefont
  {Fischer}, \citenamefont {Trivedi}, \citenamefont {Ramasesh}, \citenamefont
  {Siddiqi},\ and\ \citenamefont {Vuckovic}}]{Fischer2018}%
  \BibitemOpen
  \bibfield  {author} {\bibinfo {author} {\bibfnamefont {K.~A.}\ \bibnamefont
  {Fischer}}, \bibinfo {author} {\bibfnamefont {R.}~\bibnamefont {Trivedi}},
  \bibinfo {author} {\bibfnamefont {V.}~\bibnamefont {Ramasesh}}, \bibinfo
  {author} {\bibfnamefont {I.}~\bibnamefont {Siddiqi}},\ and\ \bibinfo {author}
  {\bibfnamefont {J.}~\bibnamefont {Vuckovic}},\ }\bibfield  {title} {\bibinfo
  {title} {Scattering into one-dimensional waveguides from a coherently-driven
  quantum-optical system},\ }\bibfield  {journal} {\bibinfo  {journal}
  {Quantum}\ }\textbf {\bibinfo {volume} {2}},\ \href
  {https://doi.org/10.22331/q-2018-05-28-69} {10.22331/q-2018-05-28-69}
  (\bibinfo {year} {2018})\BibitemShut {NoStop}%
\bibitem [{\citenamefont {Raymer}\ and\ \citenamefont
  {Walmsley}(2020)}]{Raymer2020}%
  \BibitemOpen
  \bibfield  {author} {\bibinfo {author} {\bibfnamefont {M.~G.}\ \bibnamefont
  {Raymer}}\ and\ \bibinfo {author} {\bibfnamefont {I.~A.}\ \bibnamefont
  {Walmsley}},\ }\bibfield  {title} {\bibinfo {title} {Temporal modes in
  quantum optics: Then and now},\ }\bibfield  {journal} {\bibinfo  {journal}
  {Physica Scripta}\ }\textbf {\bibinfo {volume} {95}},\ \href
  {https://doi.org/10.1088/1402-4896/ab6153} {10.1088/1402-4896/ab6153}
  (\bibinfo {year} {2020})\BibitemShut {NoStop}%
\bibitem [{\citenamefont {Gardiner}\ and\ \citenamefont
  {Zoller}(2004)}]{gardiner_quantum_2004}%
  \BibitemOpen
  \bibfield  {author} {\bibinfo {author} {\bibfnamefont {C.}~\bibnamefont
  {Gardiner}}\ and\ \bibinfo {author} {\bibfnamefont {P.}~\bibnamefont
  {Zoller}},\ }\href@noop {} {\emph {\bibinfo {title} {Quantum {Noise}: {A}
  {Handbook} of {Markovian} and {Non}-{Markovian} {Quantum} {Stochastic}
  {Methods} with {Applications} to {Quantum} {Optics}}}}\ (\bibinfo
  {publisher} {Springer Science \& Business Media},\ \bibinfo {year}
  {2004})\BibitemShut {NoStop}%
\bibitem [{\citenamefont {Gambetta}\ \emph {et~al.}(2011)\citenamefont
  {Gambetta}, \citenamefont {Motzoi}, \citenamefont {Merkel},\ and\
  \citenamefont {Wilhelm}}]{gambetta_analytic_2011}%
  \BibitemOpen
  \bibfield  {author} {\bibinfo {author} {\bibfnamefont {J.~M.}\ \bibnamefont
  {Gambetta}}, \bibinfo {author} {\bibfnamefont {F.}~\bibnamefont {Motzoi}},
  \bibinfo {author} {\bibfnamefont {S.~T.}\ \bibnamefont {Merkel}},\ and\
  \bibinfo {author} {\bibfnamefont {F.~K.}\ \bibnamefont {Wilhelm}},\
  }\bibfield  {title} {\bibinfo {title} {Analytic control methods for
  high-fidelity unitary operations in a weakly nonlinear oscillator},\ }\href
  {https://doi.org/10.1103/PhysRevA.83.012308} {\bibfield  {journal} {\bibinfo
  {journal} {Physical Review A}\ }\textbf {\bibinfo {volume} {83}},\ \bibinfo
  {pages} {012308} (\bibinfo {year} {2011})}\BibitemShut {NoStop}%
\end{thebibliography}%

\onecolumngrid

\appendix
  
\section{Photon Occupation in the Large Loss Regime}\label{AE}

In the following, we show that if the hierarchy $\kappa\ll g \ll \chi$ is violated by having a large optical linewidth $\kappa_a \gg \chi$, the strong Kerr blockade cannot be established. We consider a SPDC with Kerr nonlinearity with optical mode $\hat{a}$ and microwave mode $\hat{b}$. The Hamiltonian is $\hat{H}=\Delta \hat{a}^{\dagger} \hat{a}+g\left(\hat{a} \hat{b}+\hat{a}^{\dagger} \hat{b}^{\dagger}\right)+\chi \hat{b}^{\dagger} \hat{b}^{\dagger} \hat{b} \hat{b}$ and the dissipators are $\hat{L}_{a}=\sqrt{\kappa_{a}} \hat{a}$ and $\hat{L}_{b}=\sqrt{\kappa_{b}} \hat{b}$. We assume $\kappa_{a}$ to be large such that $\hat{a}$ is only weakly excited and can be treated as a two-level system $\{|g\rangle,|e\rangle\}$. Our goal is to adiabatically eliminate $\hat{a}$ and then solve for the steady state photon occupation for $\hat{b}$. In the case of $\kappa_a \gg \chi$, the higher excitation in the microwave mode remains unsuppressed. 

Our derivation below follows \cite{reiter_effective_2012}. The Hamiltonian projected to the ground and excited subspaces are $\hat{H}_{g}=\chi \hat{b}^{\dagger} \hat{b}^{\dagger} \hat{b} \hat{b}$ and $\hat{H}_{e}=\Delta+\chi \hat{b}^{\dagger} \hat{b}^{\dagger} \hat{b} \hat{b}$, and the coupling between the two subspaces are given by $\hat{V}_{+}=g \hat{b}^{\dagger}$ and $\hat{V}_{-}=g \hat{b}$. The non-Hermitian Hamiltonian is $H_{\mathrm{NH}}=\Delta+\chi \hat{b}^{\dagger} \hat{b}^{\dagger} \hat{b} \hat{b}-\frac{i}{2} \kappa_{a}$ and the initial-state dependent propagators are
\begin{equation}
\left(H_{\mathrm{NH}}^{(n)}\right)^{-1}=\left(H_{\mathrm{NH}}-E_{n}\right)^{-1}, 
\end{equation}
where $|n\rangle$ is the eigenstate of $\hat{H}_{g}$ with eigen-energy $E_{n}=\chi\left(n^{2}-n\right)$.\\
Notice that
\begin{equation}
\begin{aligned}
& \sum_n\left(H_{\mathrm{NH}}^{(n)}\right)^{-1} \hat{V}_{+}^n=\sum_n \frac{1}{\Delta+\chi \hat{b}^\dagger\hat{b}^\dagger\hat{b} \hat{b}-\frac{i}{2} \kappa_a-E_n} g \sqrt{n+1}|n+1\rangle\langle n| \\
= & \sum_n \frac{g \sqrt{n+1}|n+1\rangle\langle n|}{\Delta+E_{n+1}-\frac{i}{2} \kappa_a-E_n}=\sum_n \frac{g \sqrt{n+1}|n+1\rangle\langle n|}{\Delta+2 n \chi-\frac{i}{2} \kappa_a} \\
= & g \hat{b}^{\dagger} \frac{1}{\Delta+2 \chi \hat{n}-\frac{i}{2} \kappa_a},
\end{aligned}
\end{equation}
where $\hat{n}=\hat{b}^{\dagger} \hat{b}$. Therefore the effective Hamiltonian is
\begin{equation}
\hat{H}_{\mathrm{eff}}=\chi \hat{n}(\hat{n}-1)-\frac{g^{2}(\hat{n}+1)(\Delta+2 \chi \hat{n})}{(\Delta+2 \chi \hat{n})^{2}+\left(\kappa_{a} / 2\right)^{2}}, 
\end{equation}
and the effective dissipator is
\begin{equation}
\hat{L}_{a,\text{eff}}=g \hat{b}^{\dagger} \frac{\sqrt{\kappa_{a}}}{\Delta+2 \chi \hat{n}-\frac{i}{2} \kappa_{a}} . 
\end{equation}
The term $\sqrt{\kappa_{a}} /\left(\Delta+2 \chi \hat{n}-\frac{i}{2} \kappa_{a}\right)$ comes from the Lorentzian lineshape of the optical mode $\hat{a}$, since different microwave transitions have frequencies spaced by $2 \chi$.

The steady state $\hat{\rho}$ of mode $\hat{b}$ satisfies
\begin{equation}
i\left[\hat{H}_{\mathrm{eff}}, \hat{\rho}\right]=D\left[\hat{L}_{a, \mathrm{eff}}\right](\hat{\rho})+D\left[\hat{L}_{b}\right](\hat{\rho}) . 
\end{equation}
The photon occupations $p_{n}=\langle n| \hat{\rho}|n\rangle$ satisfy
\begin{equation}
\kappa_{b}\left((n+1) p_{n+1}-n p_{n}\right)+g^{2} \kappa_{a}\left(\frac{n p_{n-1}}{(\Delta+2 \chi(n-1))^{2}+\left(\kappa_{a} / 2\right)^{2}}-\frac{(n+1) p_{n}}{(\Delta+2 \chi n)^{2}+\left(\kappa_{a} / 2\right)^{2}}\right)=0 . 
\end{equation}
Notice that
\begin{equation}
\kappa_{b}(n+1) p_{n+1}-g^{2} \kappa_{a} \frac{(n+1) p_{n}}{(\Delta+2 \chi n)^{2}+\left(\kappa_{a} / 2\right)^{2}}=\kappa_{b} n p_{n}-g^{2} \kappa_{a} \frac{n p_{n-1}}{(\Delta+2 \chi(n-1))^{2}+\left(\kappa_{a} / 2\right)^{2}}=\text { const. } 
\end{equation}
Using the boundary condition that $p_{n} \rightarrow 0$ when $n$ is large, the constant must be 0. Therefore, we have
\begin{equation}
\frac{p_{n+1}}{p_{n}}=\frac{g^{2} \kappa_{a}}{\kappa_{b}\left((\Delta+2 \chi n)^{2}+\left(\kappa_{a} / 2\right)^{2}\right)} .
\end{equation}
To see the effect of the photon number blockade, we compare the following two special cases:
\begin{enumerate}
  \item When $\Delta=\chi=0$, we have $p_{n+1} / p_{n}=4 g^{2} / \kappa_{a} \kappa_{b} \equiv C$ is the cooperativity, and therefore $p_{n} \propto C^{-n}$. The derivation above only holds if $C<1$.
  \item When $\Delta=0$ and $\chi \neq 0$, we have
\end{enumerate}
\begin{equation}
\frac{p_{2}}{p_{1}}=\frac{4 g^{2} \kappa_{a}}{\kappa_{b}\left(16 \chi^{2}+\kappa_{a}^{2}\right)} \approx C\left[1-\left(\frac{4 \chi}{\kappa_{a}}\right)^{2}\right] 
\end{equation}
when $\kappa_{a} \gg \chi$. Therefore, if the linewidth is much larger than the nonlinearity, $\chi/\kappa_a\ll1$, the ratio $p_2/p_1=C$ is the same as in the $\chi=0$ case. Effective blockade cannot be established.

\section{Input-Output Theory}\label{iot}
Due to the large system nonlinearity, we do not have a closed set of Heisenberg-Langevin equations for mode operators. So, the traditional approach of combining the equation of motion and the input-output relation, as prevalently employed in the quantum transduction literature, does not work here. Instead, we use an input-output relation developed in \cite{Kiilerich2019,Kiilerich2020}. Similar methods are reviewed and developed in \cite{Fischer2018}. We outline the formalism we used for our system in the following, and all numerical solutions of quantum dynamics in this work rely on the QuTiP software package \cite{lambert2024qutip5quantumtoolbox}.

We start with our system Hamiltonian Eq.~\ref{entangler}. The driven-dissipative dynamics can be described by the Lindblad master equation: 
\begin{equation} \label{mast}
    \dot{\rho}=-\frac{i}{\hbar}[\hat{H}_{\text{ent}},\rho]+\sum_{i=1}^{4}\frac{1}{2}\left(2\hat{L}_i\rho\hat{L}_i^\dagger-\{\hat{L}_i^\dagger\hat{L}_i,\rho\}\right),
\end{equation}
where $\rho(t)$ is the intra-cavity system density operator including the optical and microwave modes. The jump operators are $\hat{L}_1=\sqrt{\kappa_o}\hat{a}$, $\hat{L}_2=\sqrt{\kappa_{e,c}}\hat{b}$, $\hat{L}_3=\sqrt{\kappa_{e,i}(1+n_{th})}\hat{b}$ and $\hat{L}_4=\sqrt{\kappa_{e,i}n_{th}}\hat{b}^\dagger$, where $\kappa_{o}=\kappa_{o,c}+\kappa_{o,i}$, $\kappa_e=\kappa_{e,c}+\kappa_{e,i}$ are the total dissipation rates for the optical and microwave mode, respectively. The sub-indexes ``c" and ``i" denote the corresponding coupling and intrinsic loss rates. $n_{th}$ is the average thermal photon of the bath, which couples to the microwave mode intrinsically. The thermal excitation of the optical mode is strongly suppressed by the Bose-Einstein distribution even at room temperature due to its high frequency, so only the microwave mode is assumed to couple to a thermal bath. In the main text, we take $n_{th}=0$ for convenience. But the input-output theory applies for general $n_{th}\neq0$ case as well. 

Due to the strong nonlinearity, the output from the system occupies multiple modes in nature. To understand the coherence properties of the optical or microwave output, we want to calculate the first-order correlation function for each of the mode operators, $g^{(1)}_{\text{opt}}(t,t')=\langle \hat{a}^{\dagger}_{\text{out}}(t)\hat{a}_{\text{out}}(t')\rangle$, and $g^{(1)}_{\text{mw}}(t,t')=\langle \hat{b}^{\dagger}_{\text{out}}(t)\hat{b}_{\text{out}}(t')\rangle$ which could be understood as the temporal interference pattern of the output. The temporal mode information in the emission is encoded in the interference. To calculate that, we use the input-output relation $\hat{a}_{out}=\hat{a}_{in}+\sqrt{\kappa_{o,c}}\hat{a}$ and $\hat{b}_{out}=\hat{b}_{in}+\sqrt{\kappa_{e,c}}\hat{b}$ and note that the input operators for both the modes are in the vacuum as the system is driven solely by a parametric pump included in the Hamiltonian. 

The time-dependence that appeared in the first-order correlation function comes from the time evolution of the operators in the Heisenberg picture. Solving the time evolution from the Lindblad master equation Eq.~\ref{mast}, the correlation functions can be calculated using the Quantum Regression Theorem. To understand the emitted modes of the system, it is best to decompose the temporal interference in a set of orthogonal mode bases by noticing the Hermitian property $g^{(1)}(t,t') = (g^{(1)}(t',t))^*$:
\begin{equation}\label{eq:corr-decomp}
\begin{split}
        &g^{(1)}_{\text{opt}}(t,t')=\sum_i n^a_i v_i^*(t) v_i(t') \\ 
        &g^{(1)}_{\text{mw}}(t,t')=\sum_i n^b_i u_i^*(t) u_i(t'),
\end{split}
\end{equation}
where $\{v_i(t)\}$ and $\{u_i(t)\}$ are the output temporal mode bases for optical and microwave respectively. The reality of the eigenvalues $n_i^a$ and $n_i^b$ is guaranteed due to the Hermitian property. Furthermore, they can be shown to take on the meaning of occupation of the corresponding $i$-th temporal modes, which makes such a basis decomposition meaningful \cite{Raymer2020}. When we numerically apply such a mode decomposition to our system, the temporal mode with the largest $n_i^a (n_i^b)$ already accounts for more than $90\%$ of the output photon population, so we approximate the output as solely consisting of the pair of most occupied modes $v_1(t)$ and $u_1(t)$. This is also justified by considering the zero-th order perturbation theory in the Appendix.~\ref{PT}. 

The seminal argument in \cite{Kiilerich2019} is that we can choose an emitted mode $v_i(t)$(or $ u_i(t)$ for microwave), and absorb all the photons in that mode by cascading the system into a downstream cavity with time-dependent couplings. All other modes will be treated as losses. For each temporal mode we want to absorb, we need an additional downstream cavity with cavity mode $\hat{a}_{v_i}$(or $\hat{a}_{u_i}$ for microwave). In our case, we want to absorb the two dominant temporal modes $v_1(t)$ and $u_1(t)$ into cavity mode $\hat{a}_{v}$ and $\hat{a}_{u}$. The cascade system is described by a Lindblad master equation on the extended Hilbert space including the two auxiliary cavities: $\mathcal{H}_{\text{tot}} = \mathcal{H}_{\text{sys}}\otimes \mathcal{H}_{\text{cav},v}\otimes \mathcal{H}_{\text{cav},u}$. In addition to the original system Hamiltonian and Lindbladians, the cascade dynamics introduce a new time-dependent cascade part of the Hamiltonian:   
\begin{equation}
\hat{H}_{\text{casc}}(t)=  \frac{i}{2}\left[\sqrt{\kappa_{o,c}} g_v^*(t) \hat{a}^{\dagger} \hat{a}_v + \sqrt{\kappa_{e,c}} g_u^*(t) \hat{b}^{\dagger} \hat{b}_u -\text { h.c. }\right],
\end{equation}
\begin{figure*}[t]
\centering
\includegraphics[width=\textwidth]{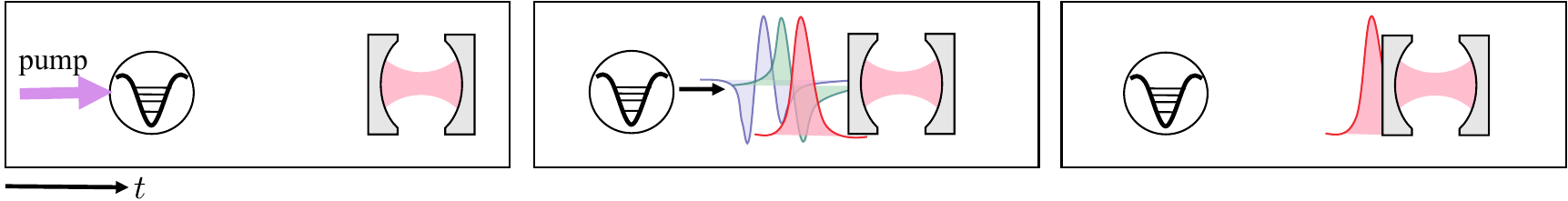}
\caption{Cartoon illustration of the input-output theory. Time goes from left to right. In the non-linear scattering setting, to know the state in one of the output modes, we need a downstream cavity (drawn on the right side of each frame above) coupled uni-directionally to the system (drawn as an anharmonic potential enclosed by a circle). In frame 1, the system is pumped by laser. In frame 2, the system emits in multiple orthogonal modes. In the last frame, the downstream cavity selectively absorbs one of the output modes whiling treating the others as loss. \label{Fig4}}
\end{figure*}
and a pair of new time-dependent Lindbladians $\hat{L}_5(t)=\sqrt{\kappa_{o,c}} \hat{a}+ g_v^*(t) \hat{a}_v$, $\hat{L}_6(t)=\sqrt{\kappa_{e,c}} \hat{b}+ g_u^*(t) \hat{b}_u$, replacing the original jump operators for coupling loss. The time-dependent couplings are calculated from the chosen temporal mode as
\begin{equation}\label{eq:gv_def}
g_v(t)=-\frac{v^*(t)}{\sqrt{\int_0^t d t^{\prime}\left|v\left(t^{\prime}\right)\right|^2}}
\end{equation}
for optical, and the same for microwave, so as to transfer all the photons in the mode $v(t)$ and $u(t)$ into the downstream cavity. This cascade system Lindblad master equation can be solved numerically over a time interval $[0,T]$, and we monitor the expected photon number in the downstream cavities $\overline{n}_v=\langle \hat{a}_v^\dagger(T)\hat{a}_v(T)\rangle$ and $\overline{n}_u=\langle\hat{b}_u^\dagger(T)\hat{b}_u(T)\rangle$. These approach a constant as $T\rightarrow \infty$ and the system approaches a steady state $\rho(T)$ defined on the total Hilbert space. The output state then is understood as the state in the downstream cavities after tracing out the system: $\rho_{\text{out}}(T) = tr_{\text{sys}}(\rho_{\text{tot}}(T))$. The system cavity modes are expected to be in the vacuum. The essence of the idea is captured by the cartoon in Fig.~\ref{Fig4}. In the large separation of scales $\kappa \ll g \ll \chi_e$, we also use the formalism to obtain the output state analytically via perturbation theory. 

\section{Numerical Simulation}\label{NS}
In the numerical treatment of the Lindblad master equation, we need to specify a truncation on the Hilbert space dimension, so that we get a finite set of differential equations. In the single-rail case, both Fig.~\ref{Fig2} and Eq.~\ref{empirical} are calculated with a dimension truncation $D=3$ for a single mode, which means that each mode has at most two photons. The Kerr nonlinearity is seen by the first-to-second excited state transition. The total dimension of the Hilbert space including the auxiliary downstream cavity is $D=3^4=81$. We have also tries higher dimension truncation, which gives us similar results due to blockade effect of the large nonlinearity. 

In the dual-rail case, Fig.~\ref{Fig3} is plotted using the numerical simulation that constrains the total photon number to be $n=6$. Since there are in total 8 modes in the system, each with $D=3$, the total Hilbert dimension is $D=1711$. We stay in the over-coupled regime $\kappa_c > \kappa_i$ to reduce the simulation time, which could still take $4$ hours to compute one point in the parameter space. For parameter scanning to get Eq.~\ref{dualP} and Eq.~\ref{dualF}, we simplify by taking infinite self-Kerr. The only coherent error comes from the cross-Kerr. In that way, the size of the Hilbert space is $D=2^8=256$. 

To get the fitting formula in Eq.~\ref{empirical}, we first find an empirical formula for the optimal pulse duration. We assume that the optimal pulse duration deviates from the ideal $\pi/2$-pulse in the closed system by a small amount governed by a combination of dimensionless small parameters in the system. This leads to a least-square fitting that yields: 
\begin{equation}\label{dhalf}
    \frac{\tau_*-\tau_{\pi/2}}{2\pi} = (0.014\pm 0.005)\frac{\kappa_c}{g^2} - (0.223\pm0.05)\frac{\kappa_i}{g^2}-(0.006\pm0.002)\frac{\kappa_i}{(\kappa_c+\kappa_i)^2}, 
\end{equation}
where, for system parameters in the range of megahertz, the quotient has the dimension of microseconds. We have a pulse length deviation on the order of $1$ns. This is shown in Fig.~\ref{Fig2}(c). 

Similarly in the dual-rail case, the optimal pulse length for the generation probability deviates from a $\pi$-pulse by
\begin{equation}\label{dpi}
    \frac{\tau_*-\tau_{\pi}}{2\pi} = (0.036\pm 0.001)\frac{\kappa_c}{g^2} - (0.060\pm0.004)\frac{\kappa_i}{g^2},
\end{equation}
where the deviation is again on the order of nano-second as illustrated in Fig.~\ref{Fig3}(c). 

In the single-rail case, we scan across parameter space $\{g, \kappa_c ,\kappa_i, \chi_e\}$ at the optimal pulse length given in Eq.~\ref{dhalf} with the values from the following range: $\kappa_c /2\pi \in \{ 1, 1.5, 2, 2.5, 3, 3.5, 4,5 \}\text{MHz}$; $\kappa_i/2\pi \in \{ 0.5, 0.6, 0.7, 0.8, 0.9, 1   \}\text{MHz}$; $g/2\pi \in \{ 8, 10,12,14, 16  \}\text{MHz}$; $\chi_e/2\pi \in \{ 50, 60, 70, 80, 90, 100   \}\text{MHz}$. The least-square fitting gives Eq.~\ref{empirical} with one standard deviation: 
\begin{equation}\label{empirical2}
\begin{aligned}
1-F \approx  (0.02 \pm 0.01) \frac{\kappa_c}{g}+(0.36 \pm 0.07) \frac{\kappa_i}{g} 
 +(0.68 \pm 0.02) \frac{\kappa_i}{\kappa_c+\kappa_i}+(0.18 \pm 0.08) \frac{g^2}{\chi_e^2}.
\end{aligned}
\end{equation}

For the dual-rail parameter space scanning $\{g, \kappa_c, \kappa_i, \chi_c\}$, we utilize the optimal pulse length in Eq.~\ref{dpi}. The parameter range is chosen as follows: $\kappa_c /2\pi \in \{ 2,3,4 \}\text{MHz}$; $\kappa_i/2\pi \in \{ 0.5, 1.25, 2.0    \}\text{MHz}$; $g/2\pi \in \{ 10, 20, 30  \}\text{MHz}$; $\chi_c/2\pi \in \{ 100, 150, 200   \}\text{MHz}$. The fitting result gives Eq.~\ref{dualP} and Eq.~\ref{dualF} with one standard deviation: 
\begin{equation}
    1-P_{\text{post}} \approx (0.23 \pm 0.10)\frac{\kappa_i}{g}+(1.51\pm0.03)\frac{\kappa_i}{\kappa_i+\kappa_c} + (0.85\pm 0.16)\frac{g^2}{\chi_c^2},
\end{equation}
\begin{equation}
    1-F_{\text{post}} \approx (0.07\pm 0.01)\frac{\kappa_c}{g} + (0.17\pm 0.03)\frac{\kappa_i}{g} + (1.29\pm0.04)\frac{g^2}{\chi_c^2}.
\end{equation}

Overall, there are three major contributing factors to errors in numerical simulation: the dimension truncation, the time resolution, and the total simulation time. The simulation becomes incorrect when photons start to populate truncated states. A large nonlinearity will ensure that the mean photon number never exceeds the number of photons allowed by the truncated Hamiltonian, and that the photon number distribution has a fast decaying tail due to the blockade. The time resolution matters when we try to numerically find the temporal modes by diagonalizing Eq.~\ref{eq:corr-decomp}. A coarser time step may miss some features in the functional shape of the temporal modes. Last but not least, a long simulation time ensures that the output state is completely absorbed by the downstream cavity. A complete state transfer is when the system cavity is in the vacuum, and all the photons are in the downstream cavity. In principle, it only becomes true in the asymptotic limit. We adopt a simulation time such that the error will is below $0.1\%$ fidelity.

\section{Microwave Hybridization}\label{hybrid}
In practice, the cross-Kerr can be accomplished by hybridizing the two microwave cavity modes with a common nonlinear Josephson device. The Hamiltonian of the two cavity modes $\hat{b}_1$ and $\hat{b}_2$ coupled to a qubit $\hat{q} $ with self-Kerr nonlinearity $\chi_q$ is given by
\begin{equation}
    H = \omega_1 \hat{b}_1^\dagger \hat{b}_1 +  \omega_2 \hat{b}_2^\dagger \hat{b}_2 + \omega_q \hat{q}^\dagger \hat{q} + \chi_q \hat{q}^\dagger \hat{q}^\dagger \hat{q}\hat{q} + g_1(\hat{b}_1^\dagger\hat{q} + \text{h.c.} ) + g_2(\hat{b}_2^\dagger\hat{q} + \text{h.c.} ) . 
\end{equation}
In the frame rotating at the frequency $\omega_q$, the Hamiltonian is rewritten as
\begin{equation}
    H = \Delta_1 \hat{b}_1^\dagger \hat{b}_1 +  \Delta_2 \hat{b}_2^\dagger \hat{b}_2 +  \chi_q \hat{q}^\dagger \hat{q}^\dagger \hat{q}\hat{q} + g_1(\hat{b}_1^\dagger\hat{q} + \text{h.c.} ) + g_2(\hat{b}_2^\dagger\hat{q} + \text{h.c.} ),
\end{equation}
where $\Delta_1 = \omega_1 - \omega_q$ and $\Delta_2 = \omega_2 - \omega_q$. We consider the dispersive regime $|\Delta_1|, |\Delta_2| \gg g_1, g_2$, and diagonalize the linear part of the Hamiltonian. To first order, the bare modes obtain a small admixture of the qubit mode, and the hybridized modes are 
\begin{equation}
    \tilde{b}_1 = \hat{b}_1 + \frac{g_1}{\Delta_1} \hat{q}; \quad \tilde{b}_2 = \hat{b}_2 + \frac{g_2}{\Delta_2} \hat{q}; \quad \tilde{q} = \hat{q} - \frac{g_1}{\Delta_1} \hat{b}_1 - \frac{g_2}{\Delta_2}\hat{b}_2. 
\end{equation}
Due to the mixing of modes, the self-Kerr nonlinearity in $\hat{q}$ now induces a cross-Kerr nonlinearity between the modes $\hat{b}_1 $ and $\hat{b}_2$ on the order of $\chi_q \frac{g_1^2}{\Delta_1^2} \frac{g_2^2}{\Delta_2^2}$. 

\section{Perturbation Theory and Fidelity Scaling}\label{PT}

In this appendix, we will apply the perturbative method to evaluate the infidelity of the produced optical-microwave entangled Bell-pair utilizing 2-mode squeezed drive and strong Kerr nonlinearity in microwave mode, as explained in Sec.~\ref{sec:EO-model}. The perturbative hierarchy we adopt here is the following: we assume the driving amplitude $g$ is much smaller than the microwave self-Kerr $\chi_e$ to reduce the production of more photons. We also assume that the total loss rate $\kappa$ for both optical and microwave modes is much smaller than $g$, so that after the 2-mode squeezed pulse the quantum state left will be closer to the pure Bell pair.

The structure of the derivation is as follows: we first calculate the correlation function $g^{(1)}_{\text{opt/mw}}(t,t')$ at the 0-th order, i.e., the dynamics we assume during the 2-mode squeezing process ($t\in [0,\tau]$) is a unitary evolution perfectly blockaded in the 2-level subspace spanned by $\{\ket{0_o, 0_e}, \ket{1_o, 1_e}\}$, and after that ($t>\tau$) the state converts into the downstream cavities with internal loss. After extracting $v(t)$ and $u(t)$ from the correlation function we get, we start to compute the dynamics with finite blockade $\chi_e$ and nonzero loss $\kappa$. The derivation is again separated into $t<\tau$ and $t>\tau$ parts as the dynamics is quite different. The target is to get the expression for the infidelity of the final Bell pair in the downstream cavities up to $O(\kappa/g)$ and $O(g^2/\chi^2_e)$ order. For simplicity, we fix $\tau = \pi/4g$ in the derivation, though in principle we can further optimize the choice of $\tau$ for slightly better performance.

\subsection{Derivation of correlation functions at 0-th order}

To derive the correlation function, we need to specify the dynamics of the system we are considering. First, here we can only focus on the central optical-microwave modes without taking the downstream part into account. Regarding the dynamics, we consider it separately within different time regimes $t<\tau$ and $t>\tau$. During the time $t<\tau$, we implement the 2-mode squeezed drive. At the 0-th order, we assume the blockade is perfect as $\chi_e \to \infty$, so that the population of the quantum state will not leak into the subspace with microwave excitations larger than $1$. Also, there will be no loss from the central system ($\kappa = 0$). After the drive ($t>\tau$), we consider the excitation loss from the central system with the rate characterized by $\kappa$. Therefore, we can write down the master equation of the system as
\begin{equation}
    \dot \rho^{(0)}(t) = \mathcal{L}(t)\rho^{(0)}(t) = \begin{cases}
 -i[\hat H_\text{bk}, \rho^{(0)}(t)]&  (t<\tau), \\
 \kappa \mathcal{D}[\hat a]\rho^{(0)}(t) + \kappa \mathcal{D}[\ketbra{0_e}{1_e}]\rho^{(0)}(t) & (t>\tau), \end{cases}
\end{equation}
where $\mathcal{D}[\hat J] \bullet := \hat J \bullet \hat J^\da - \frac{1}{2}\{\hat J^\da \hat J, \bullet\}$ stands for the Lindblad dissipator, and the blockade Hamiltonian $\hat H_\bk$ can be written as
\begin{equation}
    \hat H_\bk = ig\sum_n (\sqrt{n+1}\ketbra{(n+1)_o, 1_e}{n_o, 0_e} - \hc).
\end{equation}
We further denote the propagator $\Lambda(t,t')$ as
\begin{equation}
    \Lambda(t,t') = \mathcal{T}e^{\int_{t'}^t \mathcal{L}(T) \dif T},
\end{equation}
which maps the density matrix from $\rho^{(0)}(t')$ to $\rho^{(0)}(t)$ when evolving under the dissipator $\mathcal{L}(t)$. $\mathcal{T}$ in the equation above is the time-ordering operator. Then, we can write down the correlation function $g^{(1)}_{\text{opt}}(t,t')$ as follows~\cite{gardiner_quantum_2004}:
\begin{equation}
    g^{(1)}_{\text{opt}}(t,t') = \kappa_c \cdot \Tr{\hat a^\dagger \Lambda(t,t')[\hat a \rho^{(0)}(t')]}.
\end{equation}

For $t'<\tau$, we have
\begin{equation}\label{eq:dm-zth-tls}
    \rho^{(0)}(t') = \ketbra{\psi^{(0)}(t')}{\psi^{(0)}(t')}, \quad \text{where} \  \ket{\psi^{(0)}(t')} = \cos(gt')\ket{0_o, 0_e} + \sin(gt')\ket{1_o, 1_e}.
\end{equation}
As a result, $\hat a\rho^{(0)}(t') = \sin(gt')\ketbra{0_o, 1_e}{\psi^{(0)}(t')}$. In the strong blockade limit, $\ket{0_o, 1_e}$ does not couple to any other state via $\hat H_\bk$, and we denote $\ket{\phi_v} := \ket{0_o, 1_e}$ for later use. Therefore, when $t'<t<\tau$, we can get the correlation function $g^{(1)}_{\text{opt}}(t,t')$ as
\begin{equation}
    g^{(1)}_{\text{opt}}(t,t') = \kappa_c \sin(gt')\cdot \tr{\hat a^\dagger\ketbra{0_o, 1_e}{\psi^{(0)}(t)}} = \kappa_c \sin(gt')\sin(gt).
\end{equation}

For $t'<\tau<t$, we can separate the evolution operator into two parts: $\Lambda(t,t') = \Lambda(t,\tau)\Lambda(\tau,t')$. Notice that, after time $\tau$, the optical and microwave modes evolve separately without any coupling between them, so after time $\tau$ we may trace out the microwave mode to simplify the calculations. Denote $\Lambda_o(t,\tau) = e^{\kappa(t-\tau) \mathcal{D}[\hat a]}$, we can write 
\begin{equation}
\begin{split}
    g^{(1)}_{\text{opt}}(t,t') &= \kappa_c \sin(gt')\cdot \tr_o\left\{\hat a^\dagger \Lambda_o(t,\tau)\left[\tr_e\left\{\ketbra{0_o, 1_e}{\psi^{(0)}(\tau)}\right\}\right]\right\} = \frac{\kappa_c \sin(gt')}{\sqrt{2}} \tr_o\left\{\hat a^\dagger \Lambda_o(t,\tau) [\ketbra{0_o}{1_o}]\right\} \\
    &=\kappa_c \sin(gt') \frac{e^{-\kappa(t-\tau)/2}}{\sqrt{2}}.
\end{split}
\end{equation}
For $\tau < t' < t$, we can use similar tricks. Notice that $\tr_e\left\{\ketbra{\psi^{(0)}(\tau)}{\psi^{(0)}(\tau)}\right\} = \frac{\ketbra{0_o}{0_o} + \ketbra{1_o}{1_o}}{2}$, we can get
\begin{equation}
\begin{split}
    g^{(1)}_{\text{opt}}(t,t') &= \kappa_c \cdot \tr_o\left\{\hat a^\dagger \Lambda_o(t,t')\left\{\hat a \Lambda_o(t',\tau)\left[\frac{\ketbra{0_o}{0_o} + \ketbra{1_o}{1_o}}{2}\right]\right\}\right\} = \frac{\kappa_c e^{-\kappa(t'-\tau)}}{2} \tr_o\left\{\hat a^\dagger \Lambda_o(t,t') [\ketbra{0_o}{1_o}]\right\} \\
    &=\kappa_c \frac{e^{-\kappa(t'+t-2\tau)/2}}{2}.
\end{split}
\end{equation}
From the expression of $g^{(1)}_{\text{opt}}(t,t')$ at 0-th order, we can find only one output temporal mode $v(t)$ that contributes to the eigenmode decomposition as shown in Eq.~\eqref{eq:corr-decomp}, where
\begin{equation}\label{eq:vt_explicit}
    v(t) =  \begin{cases}
        \mu_v\sin(gt) & (t<\tau),\\
        \frac{\mu_v}{\sqrt{2}}e^{-\kappa(t-\tau)/2} & (t> \tau).
    \end{cases}
\end{equation}
Here $\mu_v = \sqrt{\frac{2\kappa}{1 + 2C_v\kappa/g}} $ is introduced as the normalization factor, and $C_v = \frac{\pi-2}{8}$ is a constant. For future convenience, we also denote
\begin{equation}
\epsilon_v := \int_0^\tau |v(t')|^2 \dif t' = \frac{C_v \mu_v^2}{g} = \frac{2C_v\kappa / g}{1 + 2C_v \kappa / g}.
\end{equation}
We can see that $\epsilon_v$ is a perturbative number with the order of $O(\kappa/g)$.

Then we switch to the microwave mode and calculate the correlation function there. Similar to the optical case, $ g^{(1)}_{\text{mw}}(t,t')$ satisfies
\begin{equation}
    g^{(1)}_{\text{mw}}(t,t') = \kappa_c \cdot \Tr{\hat b^\dagger \Lambda(t,t')[\hat b \rho^{(0)}(t')]}.
\end{equation}
For $t'<\tau$, we have $\hat b\rho^{(0)}(t') = \sin(gt')\ketbra{1_o, 0_e}{\psi^{(0)}(t')}$. Different from the optical situation, here $\ket{1_o, 0_e}$ couples with $\ket{2_o, 1_e}$ via the 2-mode squeezed drive $\hat H_\bk$. We can also denote
\begin{equation}
\ket{\phi_u(T)} := e^{-i\hat H_\bk T} \ket{1_o, 0_e} = \cos(\sqrt{2} gT)\ket{1_o, 0_e} + \sin(\sqrt{2} gT)\ket{2_o, 1_e}.
\end{equation}
This difference leads to the modification of the form of the solution for $g^{(1)}_{\text{mw}}(t,t')$ in comparison with its optical counterpart, and now we can write it down explicitly as
\begin{equation}
    g^{(1)}_{\text{mw}}(t,t') = \begin{cases}
        \kappa_c \sin(gt')\sin(gt) \cos[\sqrt{2}g(t-t')] & (t'<t<\tau),\\
        \frac{\kappa_c}{\sqrt{2}} \sin(gt') \cos[\sqrt{2}g(\tau-t')] e^{-\kappa(t-\tau)/2} & (t' < \tau < t),\\
        \frac{\kappa_c}{2} e^{-\kappa(t'+t-2\tau)/2} & (\tau<t'<t).
    \end{cases}
\end{equation}
We can easily verify that this $g^{(1)}_{\text{mw}}(t,t')$ can be decomposed as $g^{(1)}_{\text{mw}}(t,t') = n^b_1 u^*_1(t) u_1(t') + n^b_2 u^{\prime *}_2(t) u'_2(t')$, with
\begin{equation}\label{eq:ut_explicit}
    u_1(t) =  \begin{cases}
        \mu_u\sin(gt)\cos[\sqrt{2}g(\tau-t)] & (t<\tau),\\
        \frac{\mu_u}{\sqrt{2}}e^{-\kappa(t-\tau)/2} & (t> \tau);
    \end{cases}\qquad
    u'_2(t)= \begin{cases}
        \mu_{u,2}\sin(gt)\sin[\sqrt{2}g(\tau-t)] & (t<\tau),\\
        0 & (t> \tau).
    \end{cases}
\end{equation}
Here $\mu_u = \sqrt{\frac{2\kappa}{1 + 2C_{u,1}\kappa/g}}$ and $\mu_{u,2} = \sqrt{g/C_{u,2}}$ are both normalization factors again with constants $C_{u,1} = \frac{\pi - \sqrt{2}\sin(\pi/\sqrt{2})}{16}$ and $C_{u,2} = C_v - C_{u,1}$. We notice that $u_1(t)$ and $u'_2(t)$ are not orthogonal, however, since the occupation number $n^b_2 = \frac{\kappa_c}{\mu_{u,2}^2}\ll n^b_1 = \frac{\kappa_c}{\mu_u^2}$ in the limit of $\kappa \ll g$, we will only keep $u_1(t)$ as the dominant temporal mode and denote $u(t) := u_1(t)$ for future use. We also denote 
\begin{equation}
\epsilon_u := \int_0^\tau |u(t')|^2 \dif t' = \frac{C_{u,1} \mu_u^2}{g} = \frac{2C_{u,1}\kappa / g}{1 + 2C_{u,1} \kappa / g},
\end{equation}
which is again of the order $O(\kappa/g)$ in the perturbative regime.

% \begin{equation}
%     \rho(t') = (\cos(gt)\ket{0_o0_e} + \sin(gt)\ket{1_o1_e}) (\cos(gt)\bra{0_o0_e} + \sin(gt) \bra{1_o1_e}),
% \end{equation}

\subsection{Dynamics during $t<\tau$}\label{subsec:tls}

Now we start to evaluate the effects from the imperfections during the 2-mode squeezed drive ($t<\tau$), including the finite blockade ($\chi_e < \infty$) and non-zero loss from the central system ($\kappa > 0$). As we only target for the lowest-order corrections, in the following discussion we will treat the two noise mechanisms separately. First, we consider the case with no loss but the blockade is not perfect, and we use the Schrieffer-Wolff transformation to calculate the amount of the population leakage into the subspace with higher excitations at time $t=\tau$. Later, we focus on another situation where the blockade is perfect but the loss is in existence. Here, both the loss into the downstream cavities and the environment are considered, where the coupling strengths $g_{v/u}(t)$ between central and downstream modes are achieved using the 0-th order correlation functions $g^{(1)}_{\text{opt/mw}}(t,t')$ that were derived in the previous part.

\subsubsection{Corrections from finite self-Kerr $\chi_e$}\label{subsubsec:tls_chi}

In this part, we consider the effect from finite self-Kerr $\chi_e$ where higher excitations could be produced due to the imperfect blockade. Since the 2-mode squeezed drive preserves the photon number difference between optical and microwave modes, and the effect from the loss is ignored (i.e., $\kappa=0$), here we are allowed to only focus on the subspace spanned by $\{ \ket{\bar{n}} := \ket{n_o, n_e} \}$.

A standard treatment to calculate the dynamics is to use the Schrieffer-Wolff transformation, which is widely used to investigate the leakage effect during gate execution in superconducting qubits~\cite{gambetta_analytic_2011}. Specifically, here we want to find a transformation $e^{-i\hat S}$ such that in the new frame the dynamics within the lowest 2 levels are well isolated with the levels above, with the coupling strength in between as small as powers of perturbation parameter $g/\chi_e$.

The Hamiltonian we are considering here is 
\begin{equation}
    \hat H_\text{truc} = \chi_e \sum_{n=2}^N n(n-1) \ketbra{\bar n}{\bar n} + ig\sum_{n=1}^N [n\ketbra{\bar n}{\overline{n-1}} - \hc].
\end{equation}
In principle we have $N\to +\infty$, but to simplify the calculations, we make the truncation up to $N = 3$. This truncation will not affect the correction of the final density matrix at time $\tau$ up to $O(g^2/\chi_e^2)$ order. Then, following the procedure outlined in~\cite{gambetta_analytic_2011}, we find
\begin{equation}
    \hat S = \frac{g}{\chi_e} \hat S^{(1)} + \left(\frac{g}{\chi_e}\right)^2 \hat S^{(2)} + \left(\frac{g}{\chi_e}\right)^3 \hat S^{(3)},
\end{equation}
where
\begin{equation}
    \hat S^{(1)} = \ketbra{\bar 2}{\bar 1} + \hc, \quad \hat S^{(2)} = i\left[(\ketbra{\bar 2}{\bar 0} - \ketbra{\bar 3}{\bar 1}) - \hc\right]/2, \quad \hat S^{(3)} = \left[(\ketbra{\bar 3}{\bar 0} - \ketbra{\bar 2}{\bar 1}) + \hc\right]/3.
\end{equation} 
Using $\hat S$ discovered above, we can write down the Hamiltonian in the transformed frame $\hat H_\eff := e^{i\hat S}\hat H_\text{truc} e^{-i\hat S}$. After introducing a projection operator $\hat \Pi = \ketbra{\bar 0}{\bar 0} + \ketbra{\bar 1}{\bar 1}$, we have
\begin{equation}
    \hat \Pi \hat H_\eff \hat \Pi \simeq g \begin{pmatrix}
        0 & -i (1 - \frac{g^2}{2\chi_e^2})\\
        i (1 - \frac{g^2}{2\chi_e^2}) & -2\frac{g}{\chi_e}
    \end{pmatrix},\quad (\hat I - \hat \Pi) \hat H_\eff \hat \Pi \sim O\left(\frac{g^4}{\chi_e^3}\right).
\end{equation}
Similarly, the initial state $\ket{\psi(0)} = \ket{\bar 0}$ will be transformed into $\ket{\psi_\eff(0)} = e^{i\hat S}\ket{\bar 0} \simeq \ket{\bar 0} - \frac{g^2}{2\chi_e^2}\ket{\bar 2}$. After evolving time $\tau = \frac{\pi}{4g}$ and throwing away those fast-oscillating terms whose amplitude is at most $O(g^2/\chi_e^2)$, we have $\ket{\psi_\eff(\tau)} \simeq [(1-\frac{ig}{\chi_e})\ket{\bar{0}} + (1-\frac{g^2}{2\chi_e^2})\ket{\bar 1}] / \sqrt{2}$. Finally, when going back to the original frame, we will achieve
\begin{equation}
    \ket{\psi(\tau)} \simeq \ket{\psi^{(0)}(\tau)} + \frac{g}{\chi_e}\ket{\psi_\chi^{(1)}} + \left(\frac{g}{\chi_e}\right)^2\ket{\psi_\chi^{(2)}},
\end{equation}
where
\begin{equation}
    \ket{\psi^{(0)}(\tau)} = \frac{\ket{\bar 0} + \ket{\bar 1}}{\sqrt{2}}, \quad \ket{\psi_\chi^{(1)}} = (-i)\frac{\ket{\bar 0} + \ket{\bar 2}}{\sqrt{2}}, \quad \ket{\psi_\chi^{(2)}} = \frac{-2\ket{\bar 1} + \ket{\bar 2} - \ket{\bar 3}}{2\sqrt{2}}.
\end{equation}
Therefore, the correction of the achieved density matrix at time $\tau$ due to the finite Kerr $\chi_e$ effect is
\begin{equation}\label{eq:rho_chi_sigma}
    \rho_\chi(\tau) \simeq \rho_\chi^{(1)}(\tau) + \rho_\chi^{(2)}(\tau), \quad \text{where} \quad \left\{ \begin{aligned}&\rho_\chi^{(1)}(\tau) = \frac{g}{\chi_e}\ketbra{\psi_\chi^{(1)}}{\psi^{(0)}(\tau)} + \hc, \\&\rho_\chi^{(2)}(\tau) = \frac{g^2}{\chi^2_e}\left(\ketbra{\psi_\chi^{(2)}}{\psi^{(0)}(\tau)} + \hc \right) + \frac{g^2}{\chi^2_e}\ketbra{\psi_\chi^{(1)}}{\psi_\chi^{(1)}}.\end{aligned}\right.
\end{equation}

We would like to point out that, in practice, we can also use the DRAG correction scheme to reduce the leakage into higher excited levels at the end of the drive, but for simplicity we just keep the imperfection here where the population of the leakage scales as $O(g^2/\chi_e^2)$.

\subsubsection{Corrections from nonzero loss $\kappa$}

In this part, we consider the effects of the excitation loss, i.e., a photon may go to either the downstream cavities associated with the optical/microwave modes (via $\kappa_c$ terms) or the environment (via $\kappa_i$ terms) directly during the 2-mode squeezed drive. In contrast to the previous part, we choose $\chi_e = +\infty$ in the derivation below so that the driving Hamiltonian will be $\hat H_\bk$ again. The perturbative parameter here is $\kappa/g \ll 1$, and we will only keep the terms up to the first order $O(\kappa/g)$.

Since we have assumed the perfect blockade condition, the master equation we will use is
\begin{equation}
    \frac{\dif \rho}{\dif t} = -i[\hat H_\bk, \rho] -\sqrt{\kappa_c}\sum_\jal (g_\alpha(t) \hat c_{j} \hat c^\da_{j,\alpha}\rho + \hc) + \sqrt{\kappa_c} \sum_\jal (g_\alpha(t)\hat c_j\rho \hat c_{j,\alpha}^\da + \hc) + \kappa \sum_j \mathcal{D}[\hat c_j]\rho + \sum_\jal |g_\alpha(t)|^2 \mathcal{D}[\hat c_{j,\alpha}] \rho,
\end{equation}
where $\hat c_j \in \{\hat a, \hat b\}$, $\hat c_{j,\alpha} \in \{\hat a_v, \hat b_u\}$ are annihilation operators for optical/microwave modes and their associated downstream cavity modes respectively. The $\alpha$ indexes in coupling functions $g_\alpha(t)$ run over $\alpha \in \{v,u\}$.

To extract the first-order correction, we need to investigate how that perturbation term affects the 0-th order results. Similar to the treatment in Ref.~\cite{reiter_effective_2012}, we introduce the transformation operator $\hat O(t)$:
\begin{equation}
    \hat O(t) := \mathcal{T}e^{-i\int_0^t \hat H_\bknh(t') \dif t'}, \quad \text{where} \quad \hat H_\bknh(t) := \hat H_\bk - \frac{i}{2} \sum_\jal |g_\alpha(t)|^2 \hat c^\da_{j,\alpha} \hat c_{j,\alpha}. 
\end{equation}
In our problem, $\hat O(t)$ can be simplified as 
\begin{equation}
    \hat O(t) := \hat O_S(t) \cdot \hat O_\text{NH}(t), \quad \text{where} \quad \hat O_S(t) := e^{-i\hat H_\bk t}, \quad \hat O_\text{NH}(t) := e^{-\frac{1}{2}\sum_\jal \int_0^t |g_\alpha(t')|^2 \dif t'\ \hat c^\da_{j,\alpha} \hat c_{j,\alpha}}.
\end{equation}
This observation enables us to calculate the transformation of the operators for the central optical/microwave modes and those downstream modes separately. we also note that $\hat O_S(t)$ is a unitary operator while $\hat O_\text{NH}(t)$ is not.

The transformation operator $\hat O(t)$ is used to capture the 0-th order dynamics ($\kappa = 0$). Therefore, we further define the following transformations:
\begin{equation}
\left\{\begin{aligned}
    & \tilde \rho(t) := \hat O^{-1}(t) \rho(t) [\hat O^\da(t)]^{-1},\\
    & \tilde c_j(t) := \hat O_S^{-1} (t) \hat c_j \hat O_S(t),\\
    & \tilde c_{j,\alpha}(t) := \hat O_\nh^{-1}(t) \hat c_{j,\alpha}\hat O_\nh(t) = e^{-\frac{1}{2}\int_0^t |g_\alpha(t')|^2\dif t'} \hat c_{j,\alpha},\\
    & \tilde c^\da_{j,\alpha}(t) := \hat O_\nh^{-1}(t) \hat c^\da_{j,\alpha}\hat O_\nh(t) = e^{\frac{1}{2}\int_0^t |g_\alpha(t')|^2\dif t'} \hat c^\da_{j,\alpha}.
\end{aligned}\right.
\end{equation}
To avoid confusion, we want to point out that in this convention $\tilde c^\da_{j,\alpha}(t) \neq [\tilde c_{j,\alpha}(t)]^\da$ since $\hat O_\nh(t)$ is not a unitary operator. After this transformation, we will get a new master equation:
\begin{equation}
    \frac{\dif \tilde \rho(t)}{\dif t} = [\tcaL^{(0)}(t) + \sqrt{\kappa_c}\tcaL^{(1/2)}(t) + \kappa\tcaL^{(1)}(t)]\tilde\rho(t),
\end{equation}
where
\begin{equation}
    \left\{\begin{aligned}
        & \tcaL^{(0)}(t)\tilde\rho := \sum_\jal |g_\alpha(t)|^2 \tilde c_{j,\alpha}(t) \tilde\rho [\tilde c_{j,\alpha}(t)]^\dag = \sum_\jal |g_\alpha(t)|^2 e^{-\int_0^t |g_\alpha(t')|^2\dif t'} \hat c_{j,\alpha} \tilde\rho \hat c^\da_{j,\alpha},\\
        & \tcaL^{(1/2)}(t)\tilde\rho := -\sum_\jal (g_\alpha(t) \tilde c_{j}(t) \tilde c^\da_{j,\alpha}(t)\tilde\rho + \hc) + \sum_\jal (g_\alpha(t)\tilde c_j(t)\tilde\rho [\tilde c_{j,\alpha}(t)]^\da + \hc),\\
        & \tcaL^{(1)}(t)\tilde\rho := \sum_j \mathcal{D}[\tilde c_j(t)]\tilde \rho.
    \end{aligned}\right.
\end{equation}
We can also write down the perturbative expansion for $\tilde \rho(t)$ as the following:
\begin{equation}
    \tilde \rho(t) = \tilde\rho^{(0)}(t) + \tilde \rho_\kappa^{(1/2)}(t) + \tilde \rho_\kappa^{(1)}(t) + o(\kappa/g),
\end{equation}
where the subscript $\kappa$ indicates those terms are corrections caused by nonzero $\kappa$, and $o(\kappa/g)$ indicates that higher-order terms are ignored. We finally get the following three equations:
\begin{subequations}\label{eq:lind-kp-expan}
\begin{empheq}[left=\empheqlbrace]{align}
    & \dot{\tilde\rho}^{(0)}(t) =  \tcaL^{(0)}(t)\tilde\rho^{(0)}(t), \label{eq:lind-kp-0}\\
    & \dot{\tilde\rho}^{(1/2)}_\kappa (t) = \sqrt{\kappa_c}\tcaL^{(1/2)}(t)\tilde\rho^{(0)}(t) + \tcaL^{(0)}(t) \tilde\rho^{(1/2)}_\kappa (t), \label{eq:lind-kp-hf}\\
    & \dot{\tilde\rho}^{(1)}_\kappa (t) = \kappa \tcaL^{(1)}(t)\tilde\rho^{(0)}(t) + \sqrt{\kappa_c}\tcaL^{(1/2)}(t)\tilde\rho^{(1/2)}_\kappa(t) + \tcaL^{(0)}(t) \tilde\rho^{(1)}_\kappa (t), \label{eq:lind-kp-1}
\end{empheq}
\end{subequations}
with the initial condition
% \begin{equation}
%     \tilde\rho^{(0)}(t=0) = \ketbra{0_o0_e}{0_o0_e} \otimes \ketbra{0_v0_u}{0_v0_u}, \quad \tilde\rho^{(k>0)}_\kappa (t=0) = 0.
% \end{equation}
\begin{equation}
    \tilde\rho^{(0)}(t=0) = \ketbra{0_o, 0_e, 0_v, 0_u}{0_o, 0_e, 0_v, 0_u}, \quad \tilde\rho^{(k>0)}_\kappa (t=0) = 0.
\end{equation}

Finally, we discuss the way to solve the equations in Eq.~\eqref{eq:lind-kp-expan}. The first two of them are relatively easy. For Eq.~\eqref{eq:lind-kp-0}, we know that the state is initially a vacuum state while the $\tcaL^{(0)}$ process does not create any excitation. So, the solution will be trivially
\begin{equation}
    \tilde\rho^{(0)}(t) = \tilde\rho^{(0)}(0) = \ketbra{0_o, 0_e, 0_v, 0_u}{0_o, 0_e, 0_v, 0_u}.
\end{equation}
When going back to the original frame, we can see that indeed this gives us the 0-th order solution we arrived before:
\begin{equation}
    \rho^{(0)}(t) = \hat O(t)\tilde\rho^{(0)} \hat O^\da(t) = \hat O_S(t)\ketbra{0_o, 0_e}{0_o, 0_e}\hat O_S^\da(t) \otimes \ketbra{0_v, 0_u}{0_v, 0_u} = \ketbra{\psi^{(0)}(t)}{\psi^{(0)}(t)} \otimes \ketbra{0_v, 0_u}{0_v, 0_u},
\end{equation}
where $\ket{\psi^{(0)}(t)}$ is first introduced in Eq.~\eqref{eq:dm-zth-tls}. Besides, since $\tilde\rho^{(0)}(t)$ is constant in time, we simply omit its time dependence in the equations below.

The Eq.~\eqref{eq:lind-kp-hf} can be solved in a self-consistent manner. First, we claim that the solution for $\tilde\rho^{(1/2)}_\kappa (t)$ is
\begin{equation}
    \tilde\rho^{(1/2)}_\kappa (t) = \int_0^t \sqrt{\kappa_c}\tcaL^{(1/2)}(t')\tilde\rho^{(0)} \dif t'.
\end{equation}
And then, we need to verify that again $\tcaL^{(0)}(t)$ act trivially on the solution for $\tilde\rho^{(1/2)}_\kappa (t)$. This can be simply argued from the observation that $\tcaL^{(1/2)}(t')$ can create an excitation in the downstream modes from either the left or the right side of the density operator, but not both. However, $\tcaL^{(0)}(t)$ will destroy a downstream excitation on both sides of the density operator, which results in a zero outcome. Specifically, one can write down the expression for $\tilde\rho^{(1/2)}_\kappa (t)$ straightforwardly as 
\begin{equation}
\begin{split}
    \tilde\rho^{(1/2)}_\kappa (t) ={}& -\sqrt{\kappa_c}\int_0^t g_v(t') e^{\frac{1}{2}\int_0^{t'} |g_v|^2 \dif t^\pp}\hat O_S^{-1}(t')\hat a \ketbra{\psi^{(0)}(t')}{0_o, 0_e}\otimes \ketbra{1_v, 0_u}{0_v, 0_u} \dif t' \\
& -\sqrt{\kappa_c}\int_0^t g_u(t') e^{\frac{1}{2}\int_0^{t'} |g_u|^2 \dif t^\pp}\hat O_S^{-1}(t')\hat b \ketbra{\psi^{(0)}(t')}{0_o, 0_e}\otimes \ketbra{0_v, 1_u}{0_v, 0_u} \dif t' + \hc\\
={}& e^{\frac{1}{2}\int_0^{+\infty} |g_v|^2 \dif t'} \sqrt{\kappa_c}\int_0^t v^*(t')\sin(gt') \ketbra{\phi_v}{0_o, 0_e}\otimes \ketbra{1_v, 0_u}{0_v, 0_u} \dif t'\\
& + e^{\frac{1}{2}\int_0^{+\infty} |g_u|^2 \dif t'} \sqrt{\kappa_c}\int_0^t u^*(t')\sin(gt') \ketbra{\phi_u(-t')}{0_o, 0_e}\otimes \ketbra{0_v, 1_u}{0_v, 0_u} \dif t' + \hc.
\end{split}
\end{equation}
In the derivation above we have used the identity
\begin{equation}
v(t) = -g^*_v(t) e^{-\frac{1}{2}\int_t^{+\infty} |g_v|^2 \dif t'}, \quad u(t) = -g^*_u(t) e^{-\frac{1}{2}\int_t^{+\infty} |g_u|^2 \dif t'},
\end{equation}
which is equivalent to Eq.~\eqref{eq:gv_def}. It is worth to mention that, although the terms $e^{\frac{1}{2}\int_0^{+\infty} |g_\alpha|^2 \dif t}$ in the expression of $\tilde\rho^{(1/2)}_\kappa (t)$ seem to cause singularity, however, this singularity will be gone when going back to the untransformed frame via $\rho^{(1/2)}_\kappa (t) = \hat O(t) \tilde\rho^{(1/2)}_\kappa (t) \hat O^\da(t)$. So, at time $t=\tau$, we will have
\begin{equation}\label{eq:rho_kappa_half_sigma}
\begin{split}
&\rho^{(1/2)}_\kappa (\tau) = e^{\frac{1}{2}\int_\tau^{+\infty} |g_v|^2 \dif t'} \sqrt{\kappa_c}\int_0^\tau v^*(t')\sin(gt') \ketbra{\phi_v}{\psi^{(0)}(\tau)}\otimes \ketbra{1_v, 0_u}{0_v, 0_u} \dif t'\\
& + e^{\frac{1}{2}\int_\tau^{+\infty} |g_u|^2 \dif t'} \sqrt{\kappa_c}\int_0^\tau u^*(t')\sin(gt') \ketbra{\phi_u(\tau-t')}{\psi^{(0)}(\tau)}\otimes \ketbra{0_v, 1_u}{0_v, 0_u} \dif t' + \hc \\
={}& \sqrt{\frac{\kappa_c C_v}{g}} \ketbra{\phi_v}{\psi^{(0)}(\tau)}\otimes \ketbra{1_v, 0_u}{0_v, 0_u} + \sqrt{\frac{\kappa_c C_{u,1}}{g}}  \left(\ket{1_o, 0_e} + \frac{C_{u,3}}{C_{u,1}}\ket{2_o, 1_e}\right)\bra{\psi^{(0)}(\tau)}\otimes \ketbra{0_v, 1_u}{0_v, 0_u} \dif t' + \hc,
\end{split}
\end{equation}
where $C_{u,3} = \frac{1+\cos(\pi/\sqrt{2})}{8\sqrt{2}}$.

Similarly, $\tilde\rho^{(1)}_\kappa (t)$ can be written as
\begin{equation}
\tilde\rho^{(1)}_\kappa (t) = \int_0^t \kappa \tcaL^{(1)}(t')\tilde\rho^{(0)} \dif t' + \int_0^t \sqrt{\kappa_c}\tcaL^{(1/2)}(t')\tilde\rho^{(1/2)}_\kappa(t') \dif t' + \int_0^t \tcaL^{(0)}(t')\tilde\rho^{(1)}_\kappa(t') \dif t'.
\end{equation}
The first two terms can be calculated directly, but the last term does not give us a zero result again. This is because that the second term $\tcaL^{(1/2)}(t)\tilde\rho^{(1/2)}_\kappa(t)$ can provide one (and at most one) downstream excitation on both sides of the density operator together. We can put the the full expression of $\tilde\rho^{(1)}_\kappa (t)$ into the last term above, so that
\begin{equation}
\int_0^t \tcaL^{(0)}(t')\tilde\rho^{(1)}_\kappa(t') \dif t' = \int_0^t \tcaL^{(0)}(t')\left[\int_0^{t'} \sqrt{\kappa_c}\tcaL^{(1/2)}(t^{\prime\prime})\tilde\rho^{(1/2)}_\kappa(t^{\prime\prime}) \dif t^{\prime\prime} \right] \dif t' + \int_0^t \tcaL^{(0)}(t')\left[\int_0^{t'} \tcaL^{(0)}(t^{\prime\prime})\tilde\rho^{(1)}_\kappa(t^{\prime\prime}) \dif t^{\prime\prime}\right].
\end{equation}
However, this time the last term will be zero again. This claim can also be checked self-consistently, as the solution for $\tilde\rho^{(1)}_\kappa(t)$ will not contain the term that both left and right sides of the density operator contain at least 2 downstream excitations. Finally, we will have
\begin{equation}\label{eq:tilde_rho_1_final}
\tilde\rho^{(1)}_\kappa (t) = \int_0^t \kappa \tcaL^{(1)}(t')\tilde\rho^{(0)} \dif t' + \int_0^t \sqrt{\kappa_c}\tcaL^{(1/2)}(t')\tilde\rho^{(1/2)}_\kappa(t') \dif t' + \int_0^t \tcaL^{(0)}(t')\left[\int_0^{t'} \sqrt{\kappa_c}\tcaL^{(1/2)}(t^{\prime\prime})\tilde\rho^{(1/2)}_\kappa(t^{\prime\prime}) \dif t^{\prime\prime} \right] \dif t',
\end{equation}
and then we can transform back to $\rho^{(1)}_\kappa (\tau)$ via $\rho^{(1)}_\kappa (\tau) = \hat O(\tau) \tilde\rho^{(1)}_\kappa (\tau) \hat O^\da(\tau)$. For the first term in Eq.~\eqref{eq:tilde_rho_1_final}, we have
\begin{equation}
\begin{split}
& \hat O(\tau)\left[\int_0^\tau \kappa \tcaL^{(1)}(t)\tilde\rho^{(0)} \dif t \right]\hat O^\da(\tau) = \frac{\kappa}{g} \Big\{ C_{u,1} \ketbra{1_o, 0_e}{1_o, 0_e} + C_{u,2} \ketbra{2_o, 1_e}{2_o, 1_e} + C_{u,3} (\ketbra{1_o, 0_e}{2_o, 1_e} + \hc)  \\
& + C_v \ketbra{0_o, 1_e}{0_o, 1_e} - [(C_v-\frac{1}{4}) \ketbra{0_o, 0_e}{0_o, 0_e} + C_v (\ketbra{0_o, 0_e}{1_o, 1_e} + \hc) + (C_v+\frac{1}{4}) \ketbra{1_o, 1_e}{1_o, 1_e}]\Big\} \otimes \ketbra{0_v, 0_u}{0_v, 0_u},
\end{split}
\end{equation}
For the second term,
\begin{equation}
\begin{split}
& \hat O(\tau)\left[\int_0^\tau \sqrt{\kappa_c}\tcaL^{(1/2)}(t)\tilde\rho^{(1/2)}_\kappa(t) \dif t \right]\hat O^\da(\tau) = \frac{\kappa_c C_v}{g} \ketbra{0_o, 1_e}{0_o, 1_e} \otimes (\ketbra{1_v}{1_v} - 2\ketbra{0_v}{0_v}) \otimes \ketbra{0_u}{0_u}\\ 
& +  \frac{\kappa_c C_{u,1}}{g} \ketbra{1_o, 0_e}{1_o, 0_e} \otimes \ketbra{0_v}{0_v} \otimes (\ketbra{1_u}{1_u} - 2\ketbra{0_u}{0_u})+ \frac{\kappa_c C_{u,3}}{g} (\ketbra{1_o, 0_e}{2_o, 1_e} + \hc) \otimes \ketbra{0_v}{0_v} \otimes (\ketbra{1_u}{1_u} - \ketbra{0_u}{0_u})\\
& - \frac{\kappa_c}{\mu_u} (\ketbra{1_o, 0_e}{2_o, 1_e} + \hc) \otimes \int_0^\tau \dif t\  \frac{u^2(t) \int_0^t u(t') \sin(gt') \sin[\sqrt{2}g(\tau-t')] \dif t'}{\int_0^t u^2 (t') \dif t'}\ketbra{0_v, 0_u}{0_v, 0_u}\\
 & + \kappa_c \ketbra{2_o, 1_e}{2_o, 1_e} \otimes \ketbra{0_v}{0_v} \otimes \Big\{\frac{C^2_{u,3}}{g C_{u,1}} \ketbra{1_u}{1_u} - \int_0^\tau \frac{\dif \ \{\int_0^t u(t') \sin(gt') \sin[\sqrt{2}g(\tau-t')] \dif t'\}^2}{\int_0^t u^2 (t') \dif t'} \ketbra{0_u}{0_u}\Big\} + O(\kappa_c/g) \rho_\text{else}.
\end{split}
\end{equation}
Here, $\rho_\text{else}$ is a superposition of the Fock basis operators with the downstream parts among $\ketbra{1_v, 1_u}{0_v, 0_u}$, $\ketbra{0_v, 2_u}{0_v, 0_u}$, $\ketbra{1_v, 0_u}{0_v, 1_u}$ and their Hermitian conjugates, which as we will show later lead to higher order terms when evolving till $t\to +\infty$. This is the reason why we ignore the explicit expression for $\rho_\text{else}$ in the equation above. As for the last term in Eq.~\eqref{eq:tilde_rho_1_final}, we have
\begin{equation}
\begin{split}
& \hat O(\tau)\left\{ \int_0^\tau \tcaL^{(0)}(t)\left[\int_0^{t} \sqrt{\kappa_c}\tcaL^{(1/2)}(t^{\prime})\tilde\rho^{(1/2)}_\kappa(t^{\prime}) \dif t^{\prime} \right] \dif t \right\}\hat O^\da(\tau) = \Bigg\{\frac{\kappa_c C_v}{g} \ketbra{0_o, 1_e}{0_o, 1_e} + \frac{\kappa_c C_{u,1}}{g} \ketbra{1_o, 0_e}{1_o, 0_e}\\ 
& + \frac{\kappa_c}{\mu_u} \int_0^\tau \dif t\  \frac{u^2(t) \int_0^t u(t') \sin(gt') \sin[\sqrt{2}g(\tau-t')] \dif t'}{\int_0^t u^2 (t') \dif t'} (\ketbra{1_o, 0_e}{2_o, 1_e} + \hc)\\
& -\kappa_c \Big[\frac{C^2_{u,3}}{g C_{u,1}} - \int_0^\tau \frac{\dif \ \{\int_0^t u(t') \sin(gt') \sin[\sqrt{2}g(\tau-t')] \dif t'\}^2}{\int_0^t u^2 (t') \dif t'} \Big] \ketbra{2_o, 1_e}{2_o, 1_e}\Bigg\} \otimes \ketbra{0_v, 0_u}{0_v, 0_u}.
\end{split}
\end{equation}
This concludes our derivation of $\rho^{(1)}_\kappa(\tau)$, which can be written as
\begin{equation}\label{eq:rho_kappa_1_sigma}
\begin{split}
& \rho^{(1)}_\kappa(\tau) = \frac{\kappa}{g} \Big\{ C_{u,1} \ketbra{1_o, 0_e}{1_o, 0_e} + C_{u,2} \ketbra{2_o, 1_e}{2_o, 1_e} + C_{u,3} (\ketbra{1_o, 0_e}{2_o, 1_e} + \hc)  + C_v \ketbra{0_o, 1_e}{0_o, 1_e}\\
& - [(C_v-\frac{1}{4}) \ketbra{0_o, 0_e}{0_o, 0_e} + C_v (\ketbra{0_o, 0_e}{1_o, 1_e} + \hc) + (C_v+\frac{1}{4}) \ketbra{1_o, 1_e}{1_o, 1_e}]\Big\} \otimes \ketbra{0_v, 0_u}{0_v, 0_u} \\
&+\frac{\kappa_c}{g}\left[C_{u,1} \ketbra{1_o, 0_e}{1_o, 0_e} + C_{u,3} (\ketbra{1_o, 0_e}{2_o, 1_e} + \hc) + \frac{C^2_{u,3}}{C_{u,1}}\ketbra{2_o, 1_e}{2_o, 1_e}\right]\otimes \ketbra{0_v}{0_v} \otimes (\ketbra{1_u}{1_u} - \ketbra{0_u}{0_u})\\
&+\frac{\kappa_c C_v}{g} \ketbra{0_o, 1_e}{0_o, 1_e} \otimes (\ketbra{1_v}{1_v} - \ketbra{0_v}{0_v}) \otimes \ketbra{0_u}{0_u} + O(\kappa_c/g) \rho_\text{else}.
\end{split}
\end{equation}

\subsection{Dynamics during $t>\tau$}

Till now, we have achieved the density matrix of our system at time $t=\tau$ with the perturbative correction from finite $\chi_e$ and nonzero $\kappa$. Here we start to discuss the dynamics during $t>\tau$, where the two-mode squeezed drive is off. Notice that, in this time regime there is no coupling between optical and microwave modes, so we are allowed to calculate their dynamics separately. We first focus on the optical and its own downstream mode, which is simply a linear system as a whole. Then we will go to the microwave part, where there is a strong Kerr nonlinearity characterized by $\chi_e$. As discussed before, the occupation of more than one excitation in the microwave mode at time $\tau$ is already a perturbative effect, and after $t=\tau$ there is no mechanism to create more microwave excitations. So, in the later treatment we can simply use $\chi_e = +\infty$ to avoid tedious calculations on terms that will only cause higher-order effects.

\subsubsection{Dynamics on the optical side}

In this part, we focus on the dynamics within the optical mode $\hat a$ and the associated downstream cavity $\hat a_v$. Specifically, we focus on the final states at $t\to\infty$ given different inputs at $t=\tau$. Due to the linearity of the master equation, the ``input'' we mention here is not limited to a physical state. It can be any density matrix element instead.

The master equation we will solve in this part is:
\begin{equation}
\frac{\dif \rho_{ov}(t)}{\dif t} = -i[\hat H_{ov}(t), \rho_{ov}(t)] + \mathcal{D}[\hat L_{ov}(t)] \rho_{ov}(t) + \mathcal{D}[\hat L_{o,i}] \rho_{ov}(t),
\end{equation}
where
\begin{equation}
\hat H_{ov}(t) = \frac{i\sqrt{\kappa_c}}{2}(g^*_v(t) \hat a^\da \hat a_v - \hc), \quad \hat L_{ov}(t) = g^*_v(t) \hat a_v + \sqrt{\kappa_c} \hat a, \quad \hat L_{o,i} = \sqrt{\kappa_i} \hat a.
\end{equation}

The dynamics is solvable as there is no Kerr nonlinear term in the Hamiltonian. To solve the master equation, one can again introduce a frame transformation
\begin{equation}
\hat O_{ov}(t) = \mathcal{T} e^{-i\int_\tau^t \hat H_\eff^{ov}(t') \dif t'},
\end{equation}
where
\begin{equation}
\hat H_\eff^{ov}(t) = \hat H_{ov}(t) - \frac{i}{2}\hat L_{ov}^\da(t)\hat L_{ov}(t) - \frac{i}{2} \hat L_{o,i}^\da \hat L_{o,i} = -i \sqrt{\kappa_c} g_v(t)\hat a \hat a_v^\da - \frac{i}{2} |g_v(t)|^2 \hat a_v^\da \hat a_v - \frac{i\kappa}{2} \hat a^\da \hat a.
\end{equation}
Then we consider the transformation of the density matrix $\rho_{ov}$ and other operators $\hat r \in \{\hat a,\hat a_v, \hat a^\da, \hat a_v^\da\}$ as follows:
\begin{equation}
\tilde \rho_{ov}(t) := \hat O_{ov}^{-1}(t) \rho_{ov}(t) [\hat O_{ov}^\da(t)]^{-1}, \quad \tilde{r}(t) := \hat O_{ov}^{-1}(t) \hat r\hat O_{ov}(t).
\end{equation}
Notice that, $\tilde{r}(t)$ should satisfy the following equation due to its definition:
\begin{equation}
\frac{\dif \tilde r(t)}{\dif t} = \hat O_{ov}^{-1}(t) \cdot i[\hat H_\eff^{ov}, \hat r] \cdot \hat O_{ov}(t).
\end{equation}
Therefore, one will get the following set of equations:
\begin{equation}
\left\{\begin{aligned}
& \dot{\tilde a}(t) = -\kappa \tilde a(t)/2, \\
& \dot{\tilde a}_v(t)= -|g_v(t)|^2 \tilde a_v(t) / 2 - \sqrt{\kappa_c} g_v(t) \tilde a(t), \\
& \dot{\tilde a}^\da (t)= \sqrt{\kappa_c} g_v(t) \tilde a_v^\da(t) + \kappa \tilde a^\da(t)/2, \\
& \dot{\tilde a}_v^\da(t) = |g_v(t)|^2 \tilde a^\da_v(t) / 2. \\
\end{aligned}\right.
\end{equation}
These equations are all linear for $\tilde r(t)$, which makes them solvable. We will finally get
\begin{equation}\label{eq:tilde_ops_av_sol}
\left\{\begin{aligned}
& \tilde a(t) = e^{-\frac{1}{2}\kappa (t-\tau)} \hat a,\\
& \tilde a_v(t) = e^{-\frac{1}{2} \int_\tau^t |g_v|^2\dif t'} [\hat a_v-\sqrt{\kappa_c} \int_\tau^t  \dif t' \ g_v(t') e^{\frac{1}{2} \int_\tau^{t'} (|g_v|^2-\kappa)\dif t^\pp}\hat a] \\
& \tilde a^\da(t) = e^{\frac{1}{2}\kappa(t-\tau)}[\hat a^\da + \sqrt{\kappa_c} \int_\tau^t  \dif t' \ g_v(t') e^{\frac{1}{2} \int_\tau^{t'} (|g_v|^2-\kappa)\dif t^\pp} \hat a_v^\da],\\
& \tilde a_v^\da(t) = e^{\frac{1}{2} \int_\tau^{t} |g_v|^2\dif t'} \hat a_v^\da.
\end{aligned}\right.
\end{equation}

After achieving the explicit formula of the transformed operators, we need to look into the dynamics of the transformed state. The master equation after transformation can now be written as:
\begin{equation}\label{eq:dot_tilde_rho_ov}
\frac{\dif\tilde \rho_{ov}(t)}{\dif t} = \tilde L_{ov}(t) \tilde \rho_{ov}(t) [\tilde L_{ov}(t)]^\da + \tilde L_{o,i}(t) \tilde \rho_{ov}(t) [\tilde L_{o,i}(t)]^\da,
\end{equation}
where
\begin{equation}
\tilde L_{ov}(t) = g_v^*(t) \tilde a_v(t) + \sqrt{\kappa_c} \tilde a(t), \quad \tilde L_{o,i}(t) = \sqrt{\kappa_i} \tilde a(t).
\end{equation}
Further, by using the explicit formula of $v(t)$ during $t>\tau$ [See Eq.~\eqref{eq:vt_explicit}], we will achieve
\begin{equation}
\tilde L_{ov}(t) = j_{o,1}(t) \sqrt{\epsilon_v} [\sqrt{\epsilon_v \kappa_c} \hat a - \sqrt{(1-\epsilon_v)\kappa} \hat a_v], \quad \tilde L_{o,i}(t) = j_{o,2}(t) \sqrt{\kappa_i} \hat a,
\end{equation}
where
\begin{equation}
j_{o,1}(t) = \frac{e^{-\kappa(t-\tau)/2}}{1-(1-\epsilon_v)e^{-\kappa(t-\tau)}}, \quad j_{o,2}(t) = e^{-\kappa(t-\tau)/2}.
\end{equation}

To solve Eq.~\eqref{eq:dot_tilde_rho_ov}, we can first write it in the integral form:
\begin{equation}
\tilde \rho_{ov}(t) = \tilde \rho_{ov}(\tau) + \sum_\eta \int_\tau^t \tilde L_\eta(t')  \tilde \rho_{ov}(t') [\tilde L_\eta(t')]^\da \dif t',
\end{equation}
where the summation runs over $\tilde L_\eta \in \{\tilde L_{ov}, \tilde L_{o,i}\}$. We can again put the expression for $\tilde \rho_{ov}(t)$ into the right-hand-side of the equation above iteratively. Notice that the term $\tilde L_\eta(t)  \tilde \rho_{ov}(t) [\tilde L_\eta(t)]^\da$ will destroy an excitation on both sides of the density operator, so for a generic input $\tilde \rho_{ov}(\tau) =  \rho_{ov}(\tau) =\ketbra{(n_L)_o, (m_L)_v}{(n_R)_o, (m_R)_v}$, we have
\begin{equation}\label{eq:tilde_rho_ov_sol}
\begin{split}
\tilde \rho_{ov}(t) = \tilde \rho_{ov}(\tau) + \sum_{\eta_1, \dots, \eta_N} \int_\tau^t \dif t_N \int_\tau^{t_N} \dif t_{N-1} \dots &\int_\tau^{t_2} \dif t_1 \\
\tilde L_{\eta_N}(t_N) \tilde L_{\eta_{N-1}}(t_{N-1}) \dots \tilde L_{\eta_1}(t_1) &\tilde \rho_{ov}(\tau) [\tilde L_{\eta_1}(t_1)]^\da \dots [\tilde L_{\eta_{N-1}}(t_{N-1})]^\da [\tilde L_{\eta_{N}}(t_{N})]^\da,
\end{split}
\end{equation}
where $N = \min\{(n_L + m_L), (n_R + m_R)\}$ is the minimum excitations on either the left or the right side of the input $\tilde \rho_{ov}(\tau)= \ketbra{(n_L)_o, (m_L)_v}{(n_R)_o, (m_R)_v}$. We can use this expression to achieve $\tilde \rho_{ov}(+\infty)$ given different input $\tilde \rho_{ov}(\tau)$.

Finally, we need to convert $\tilde \rho_{ov}(+\infty)$ back to the untransformed $\rho_{ov}(+\infty) = \hat O_{ov}(+\infty) \tilde \rho_{ov}(+\infty) \hat O_{ov}^\da(+\infty)$. To solve this, we can again check how this transformation $\hat O_{ov}(+\infty)$ affects the operator $\hat r \in \{\hat a,\hat a_v, \hat a^\da, \hat a_v^\da\}$ first. Notice from Eq.~\eqref{eq:tilde_ops_av_sol} that
\begin{equation}
\left\{\begin{aligned}
& e^{-\frac{1}{2}\kappa(t-\tau)} \hat a^\da = \hat O_{ov}(t)\hat a^\da O_{ov}^{-1}(t)+ \sqrt{\kappa_c} \int_\tau^t  \dif t' \ g_v(t') e^{\frac{1}{2} \int_\tau^{t'} (|g_v|^2-\kappa)\dif t^\pp} \hat O_{ov}(t)\hat a_v^\da O_{ov}^{-1}(t),\\
& e^{-\frac{1}{2} \int_\tau^{t} |g_v|^2\dif t'} \hat a_v^\da =  \hat O_{ov}(t)\hat a_v^\da \hat O_{ov}^{-1}(t).
\end{aligned}\right.
\end{equation}
By setting $t\to \infty$, we will achieve
\begin{equation}
\left\{\begin{aligned}
& \hat O_{ov}(+\infty)\hat a_v^\da \hat O_{ov}^{-1}(+\infty) = \sqrt{\int_0^\tau |v(t)|^2 \dif t}\  \hat a_v^\da= \sqrt{\epsilon_v}\ \hat a_v^\da\\
& \hat O_{ov}(+\infty)\hat a^\da \hat O_{ov}^{-1}(+\infty) = \int_\tau^{+\infty} \sqrt{\kappa_c} v^*(t) e^{-\kappa(t-\tau)/2} \dif t \ \hat a_v^\da = \sqrt{\xi_v} \ \hat a_v^\da.
\end{aligned}\right.
\end{equation}
Here we denote $\xi_v := \frac{\kappa_c}{\kappa}(1-\epsilon_v)$. With these formulae, we can compute $\rho_{ov}(+\infty)$ from $\tilde \rho_{ov}(+\infty)$. For a given basis operator written as
\begin{equation}
\ketbra{(n_L)_o, (m_L)_v}{(n_R)_o, (m_R)_v} = \frac{(\hat a^{\da})^{n_L}}{\sqrt{n_L!}}\frac{(\hat a_v^{\da})^{m_L}}{\sqrt{m_L!}}\ketbra{0_o, 0_v}{0_o, 0_v} \frac{\hat a^{n_R}}{\sqrt{n_R!}}\frac{\hat a_v^{m_R}}{\sqrt{m_R!}},
\end{equation}
we will achieve
\begin{equation}
\begin{split}
 & \hat O_{ov}(+\infty)\ketbra{(n_L)_o, (m_L)_v}{(n_R)_o, (m_R)_v}  \hat O_{ov}^{-1}(+\infty)\\
  ={}& \hat O_{ov}(+\infty)  \frac{(\hat a^{\da})^{n_L}}{\sqrt{n_L!}}\frac{(\hat a_v^{\da})^{m_L}}{\sqrt{m_L!}} \hat O_{ov}^{-1}(+\infty) \hat O_{ov}(+\infty)\ketbra{0_o, 0_v}{0_o, 0_v} \hat O_{ov}^\da(+\infty)[\hat O_{ov}^{-1}(+\infty)]^\da \frac{\hat a^{n_R}}{\sqrt{n_R!}}\frac{\hat a_v^{m_R}}{\sqrt{m_R!}}\hat O_{ov}^\da(+\infty)\\
 ={}&\epsilon_v^{\frac{m_L+m_R}{2}} \xi_v^{\frac{n_L+n_R}{2}} \frac{(\hat a_v^{\da})^{n_L+m_L}}{\sqrt{(n_L)! (m_L)!}}\ketbra{0_o, 0_v}{0_o, 0_v}\frac{\hat a_v^{n_R+m_R}}{\sqrt{(n_R)! (m_R)!}} \\
 ={}& \sqrt{\frac{\epsilon_v^{m_L+m_R} \xi_v^{n_L+n_R}  (n_L+m_L)! (n_R+m_R)!}{(n_L)! (m_L)!(n_R)! (m_R)!}} \ketbra{0_o, (n_L+m_L)_v}{0_o, (n_R+m_R)_v}.
\end{split}
\end{equation}
Since the achieved $\tilde \rho_{ov}(+\infty)$ from Eq.~\eqref{eq:tilde_rho_ov_sol} is always a linear superposition of those basis operators, we can use the expression above to get $\rho_{ov}(+\infty)$. There will be no excitation left in the optical mode then.

In summary, we list the result of $\rho_{v}(+\infty)$ for a given Fock basis input $\rho_{ov}(\tau)$ in Table.~\ref{table:optical}.

\begin{table*}
\caption{The final optical downstream states $\rho_v(+\infty)$ given some Fock-basis input operators at time $t=\tau$, i.e. $\rho_{ov}(\tau) = \ketbra{(n_L)_o, (m_L)_v}{(n_R)_o, (m_R)_v}$.}
\begin{ruledtabular}\label{table:optical}
\begin{tabular}{|c|c|c||c|c|c|}
$\rho_{ov}(\tau)$ & $\tilde\rho_{ov}(+\infty)$ & $\rho_{v}(+\infty)$ & $\rho_{ov}(\tau)$ & $\tilde\rho_{ov}(+\infty)$ & $\rho_{v}(+\infty)$ \\ \hline
$\ketbra{0_o, 0_v}{0_o, 0_v}$ & $\ketbra{0_o, 0_v}{0_o, 0_v}$ & $\ketbra{0_v}{0_v}$ & $\ketbra{1_o, 0_v}{0_o, 0_v}$ & $\ketbra{1_o, 0_v}{0_o, 0_v}$ & $\sqrt{\xi_v}\ketbra{1_v}{0_v}$ \\ \hline
$\ketbra{1_o, 0_v}{1_o, 0_v}$  & $ \begin{aligned} &\ketbra{1_o, 0_v}{1_o, 0_v} \\ +{}& (1-\xi_v)\ketbra{0_o, 0_v}{0_o, 0_v} \end{aligned} $ & $\begin{aligned} &\xi_v\ketbra{1_v}{1_v} \\ +{}&(1-\xi_v)\ketbra{0_v}{0_v} \end{aligned}$ & $\ketbra{2_o, 0_v}{0_o, 0_v}$ & $\ketbra{2_o, 0_v}{0_o, 0_v}$ & $\xi_v\ketbra{2_v}{0_v}$ \\ \hline
$\ketbra{2_o, 0_v}{1_o, 0_v}$ & $ \begin{aligned} &\ketbra{2_o, 0_v}{1_o, 0_v} \\ +{}& \sqrt{2}(1 - \xi_v)\ketbra{1_o, 0_v}{0_o, 0_v} \end{aligned} $ & $\begin{aligned} &\sqrt{\xi_v}[\xi_v\ketbra{2_v}{1_v}+ \\ &\sqrt{2}(1 - \xi_v)\ketbra{1_v}{0_v}] \end{aligned}$ & $\ketbra{2_o, 0_v}{2_o, 0_v}$ & $ \begin{aligned} &\ketbra{2_o, 0_v}{2_o, 0_v} \\ +{}& \sqrt{2}(1 - \xi_v)\ketbra{1_o, 0_v}{1_o, 0_v} \\ +{}& (1 - \xi_v)^2 \ketbra{0_o, 0_v}{0_o, 0_v} \end{aligned} $ & $ \begin{aligned} &\xi_v^2\ketbra{2_v}{2_v} \\ +{}& 2\xi_v(1 - \xi_v)\ketbra{1_v}{1_v} \\ +{}& (1 - \xi_v)^2 \ketbra{0_v}{0_v} \end{aligned} $ \\ \hline
$\ketbra{3_o, 0_v}{0_o, 0_v}$ & $\ketbra{3_o, 0_v}{0_o, 0_v}$ & $\xi_v^{3/2} \ketbra{3_v}{0_v}$ & $\ketbra{3_o, 0_v}{1_o, 0_v}$ & $ \begin{aligned} &\ketbra{3_o, 0_v}{1_o, 0_v} \\ +{}& \sqrt{3}(1 - \xi_v)\ketbra{2_o, 0_v}{0_o, 0_v} \end{aligned} $ & $\begin{aligned} &\xi_v [\xi_v \ketbra{3_v}{1_v} +\\ &\sqrt{3}(1 - \xi_v)\ketbra{2_v}{0_v}] \end{aligned}$ \\ \hline
$\ketbra{0_o, 1_v}{0_o, 0_v}$ & $\ketbra{0_o, 1_v}{0_o, 0_v}$ & $\sqrt{\epsilon_v} \ketbra{1_v}{0_v}$ & $\ketbra{0_o, 1_v}{1_o, 0_v}$ &  $ \begin{aligned} &\ketbra{0_o, 1_v}{1_o, 0_v} \\ -{}& \sqrt{\xi_v \epsilon_v}\ketbra{0_o, 0_v}{0_o, 0_v} \end{aligned} $ & $ \begin{aligned} \sqrt{\xi_v \epsilon_v} (&\ketbra{1_v}{1_v} \\ &- \ketbra{0_v}{0_v}) \end{aligned}$ \\ \hline
$\ketbra{1_o, 1_v}{0_o, 0_v}$ & $\ketbra{1_o, 1_v}{0_o, 0_v}$ & $\sqrt{2 \xi_v \epsilon_v} \ketbra{2_v}{0_v}$ & $\ketbra{1_o, 1_v}{1_o, 0_v}$ &  $ \begin{aligned} &\ketbra{1_o, 1_v}{1_o, 0_v} \\ +{}& (1-\xi_v) \ketbra{0_o, 1_v}{0_o, 0_v}\\ -{}& \sqrt{\xi_v \epsilon_v}\ketbra{1_o, 0_v}{0_o, 0_v} \end{aligned} $ & $\begin{aligned} &\sqrt{\epsilon_v}[\sqrt{2}\xi_v \ketbra{2_v}{1_v}+ \\ &(1 - 2\xi_v)\ketbra{1_v}{0_v}] \end{aligned}$  \\ \hline
$\ketbra{0_o, 1_v}{2_o, 0_v}$ & $ \begin{aligned} &\ketbra{0_o, 1_v}{2_o, 0_v} \\ -{}& \sqrt{2\xi_v \epsilon_v}\ketbra{0_o, 0_v}{1_o, 0_v} \end{aligned} $ & $\begin{aligned} \xi_v \sqrt{\epsilon_v}(&\ketbra{1_v}{2_v}- \\ &\sqrt{2}\ketbra{0_v}{1_v}) \end{aligned}$ & $\ketbra{0_o, 1_v}{0_o, 1_v}$ & $\begin{aligned} &\ketbra{0_o, 1_v}{0_o, 1_v} \\ +{}& (1-\epsilon_v)\ketbra{0_o, 0_v}{0_o, 0_v} \end{aligned}$ & $\begin{aligned} &\epsilon_v\ketbra{1_v}{1_v} \\ +{}&(1-\epsilon_v)\ketbra{0_v}{0_v} \end{aligned}$ \\ \hline
$\ketbra{1_o, 1_v}{0_o, 1_v}$ & $ \begin{aligned} &\ketbra{1_o, 1_v}{0_o, 1_v} \\ +{}& (1-\epsilon_v) \ketbra{1_o, 0_v}{0_o, 0_v}\\ -{}& \sqrt{\xi_v \epsilon_v}\ketbra{0_o, 1_v}{0_o, 0_v} \end{aligned} $ & $\begin{aligned} &\sqrt{\xi_v}[\sqrt{2}\epsilon_v \ketbra{2_v}{1_v}+ \\ &(1 - 2\epsilon_v)\ketbra{1_v}{0_v}] \end{aligned}$  & $\ketbra{1_o, 1_v}{1_o, 1_v}$ & $ \begin{aligned} &\ketbra{1_o, 1_v}{1_o, 1_v} \\ +{}& (1-\xi_v) \ketbra{0_o, 1_v}{0_o, 1_v}\\ +{}& (1-\epsilon_v) \ketbra{1_o, 0_v}{1_o, 0_v}\\ -{}& \sqrt{\xi_v \epsilon_v}(\ketbra{0_o, 1_v}{1_o, 0_v}+ \\ &\ketbra{1_o, 0_v}{0_o, 1_v})\\ +{}&[\xi_v \epsilon_v + (1-\xi_v)(1-\epsilon_v)] \cdot \\ &\ketbra{0_o, 0_v}{0_o, 0_v}  \end{aligned} $ & $\begin{aligned}  &2\xi_v \epsilon_v\ketbra{2_v}{2_v} +\\ & (\xi_v + \epsilon_v - 4\xi_v \epsilon_v)\cdot \\ &\ketbra{1_v}{1_v} + \\ & (1 - \xi_v - \epsilon_v + 2\xi_v \epsilon_v)\cdot \\ & \ketbra{0_v}{0_v}  \end{aligned}$ \\ \hline
$\ketbra{0_o, 2_v}{0_o, 0_v}$ & $\ketbra{0_o, 2_v}{0_o, 0_v}$ & $\epsilon_v \ketbra{2_v}{0_v}$ & $\ketbra{0_o, 2_v}{1_o, 0_v}$ & $ \begin{aligned} &\ketbra{0_o, 2_v}{1_o, 0_v} \\ -{}& \sqrt{2\xi_v \epsilon_v}\ketbra{0_o, 1_v}{0_o, 0_v} \end{aligned} $ & $\begin{aligned} \epsilon_v \sqrt{\xi_v}(&\ketbra{2_v}{1_v}- \\ &\sqrt{2}\ketbra{1_v}{0_v}) \end{aligned}$ \\ \hline
$\ketbra{1_o, 2_v}{0_o, 0_v}$ & $\ketbra{1_o, 2_v}{0_o, 0_v}$ & $\epsilon_v\sqrt{3\xi_v/2} \ketbra{3_v}{0_v}$ & $\ketbra{1_o, 2_v}{1_o, 0_v}$ & $ \begin{aligned} &\ketbra{1_o, 2_v}{1_o, 0_v} \\ +{}& (1-\xi_v) \ketbra{0_o, 2_v}{0_o, 0_v}\\ -{}& \sqrt{2\xi_v \epsilon_v}\ketbra{1_o, 1_v}{0_o, 0_v} \end{aligned} $ & $\begin{aligned} &\xi_v \epsilon_v \sqrt{3/2}\ketbra{3_v}{1_v} \\ +{}&\epsilon_v(1 - 3\xi_v)\ketbra{2_v}{0_v} \end{aligned}$ 
\end{tabular}
\end{ruledtabular}
\end{table*}

\subsubsection{Dynamics on the microwave side}

Then we move forward and discuss the dynamics on the microwave side during time $t > \tau$. The difference between the microwave and optical sides is that in the former situation we have a strong Kerr term $\chi_e \hat b^{\da 2} \hat b^2$, which introduces extra difficulties to analytically solve the dynamics. On the other hand, due to the cascaded feature, there will be no excitation produced in the central microwave mode. The effect of Kerr starts to participate when $n_e \geq 2$, but as what we have shown in Appendix.~\ref{subsec:tls}, the occupation of the subspace with $n_e \geq 2$ for the achieved quantum state at $t=\tau$ has already been a perturbative effect. So, to avoid tedious calculations on higher order corrections, in this part we will simply set $\chi_e \to \infty$. 
%Besides, we only focus on the dynamics within the subspace $n_e \leq 3$, which also fits the result of the state at $t=\tau$, up to the $O(g^2/\chi_e^2)$ order. \cor{(may not need this sentence)}

We start with the master equation for the microwave mode and its downstream cavity first. After rotating wave approximation (which will be exact in the $\chi_e \to \infty$ limit), the quantum state will evolve as
\begin{equation}
\frac{\dif \rho_{eu}(t)}{\dif t} = -i[\hat H_{eu}(t), \rho_{eu}(t)] + \mathcal{D}[\hat L_{eu}(t)] \rho_{eu}(t) + \mathcal{D}[\sqrt{\kappa_i}\hat b_{0,1}] \rho_{eu}(t) + \sum_{n=2}^{+\infty} \mathcal{D}[\sqrt{\kappa n}\hat b_{n-1,n}] \rho_{eu}(t),
\end{equation}
where
\begin{equation}
\hat H_{eu}(t) = \frac{i\sqrt{\kappa_c}}{2}(g_u^*(t)\hat b_{0,1}^\da \hat b_u - \hc), \quad \hat L_{eu}(t) = g^*_u(t) \hat b_u + \sqrt{\kappa_c} \hat b_{0,1}, \quad \hat b_{m,n} = \ketbra{m_e}{n_e}.
\end{equation}
Due to the strong Kerr effect, the energy difference between nearest microwave levels $\omega_e^{(n,n+1)}$ now will be state-dependent, which results in the loss of coherence when the state decays from different levels. Therefore, each decay term $\hat b_{n-1,n}$ should be treated individually, as shown in the master equation above.

Similar to the treatment we did in the optical mode, here we also define the transformation operator
\begin{equation}
\hat O_{eu}(t) = \mathcal{T}e^{-i\int_\tau^t \hat H_\eff^{eu}(t') \dif t'},
\end{equation}
where
\begin{equation}
\begin{split}
\hat H_\eff^{eu}(t) &= \hat H_{eu}(t) - \frac{i}{2}\hat L_{eu}^\da(t)\hat L_{eu}(t) - \frac{i\kappa_i}{2} \hat b_{0,1}^\da \hat b_{0,1} - \frac{i\kappa }{2}\sum_{n=2}^{+\infty}  n \hat b_{n-1,n}^\da \hat b_{n-1,n} \\
&= -i \sqrt{\kappa_c} g_u(t)\hat b_{0,1} \hat b_u^\da - \frac{i}{2} |g_u(t)|^2 \hat b_u^\da \hat b_u - \frac{i\kappa}{2} \hat b^\da \hat b.
\end{split}
\end{equation}
The transformation of the density matrix $\rho_{eu}$ and other operators $\hat r \in \{\hat b_u, \hat b_u^\da, \hat b_{m,n}\}$ will again becomes:
\begin{equation}
\tilde \rho_{eu}(t) := \hat O_{eu}^{-1}(t) \rho_{eu}(t) [\hat O_{eu}^\da(t)]^{-1}, \quad \tilde{r}(t) := \hat O_{eu}^{-1}(t) \hat r\hat O_{eu}(t).
\end{equation}
To solve $\tilde r(t)$, first we notice that their time-derivatives satisfy
\begin{equation}\label{eq:dot_tilde_ops_bu}
\left\{\begin{aligned}
& \dot{\tilde b}_u(t) = -|g_u(t)|^2\tilde b_u(t)/2 - \sqrt{\kappa_c}g_u(t) \tilde b_{0,1}(t),\\
& \dot{\tilde b}^\da_u(t) = |g_u(t)|^2\tilde b^\da_u(t)/2,\\
& \dot{\tilde b}_{m,n}(t) = \kappa(m-n)\tilde b_{m,n}(t) / 2 + \sqrt{\kappa_c} g_u(t) [\delta_{m,1} \tilde b_{0,n}(t) - \tilde b_{m,1}(t) \delta_{0,n}] \tilde b_u^\da (t).
\end{aligned}\right.
\end{equation}
To solve the equations above, we can treat them sequentially. First we have the two simplest solutions
\begin{equation}\label{eq:tilde_b_sol_1}
\left\{\begin{aligned}
& \tilde b_u^\da(t) = e^{\frac{1}{2} \int_\tau^{t} |g_u|^2\dif t'} \hat b_u^\da, \\
& \tilde b_{m,n}(t) = e^{\frac{1}{2}(m-n)\kappa(t-\tau)} \hat b_{m,n} \quad (m\neq 1\ \text{and}\ n \neq 0).
\end{aligned}\right.
\end{equation}
By using these two equations, we can further achieve 
\begin{equation}\label{eq:tilde_b_sol_2}
\left\{\begin{aligned}
& \tilde b_u(t) = e^{-\frac{1}{2} \int_\tau^t |g_u|^2\dif t'} [\hat b_u - \sqrt{\kappa_c} \int_\tau^t  \dif t' \ f_u(t') \ \hat b_{0,1}],& & \\
& \tilde b_{1,n}(t) = e^{-\frac{1}{2}\kappa (n-1)(t-\tau)} [\hat b_{1,n} + \sqrt{\kappa_c} \int_\tau^t \dif t' \ f_u(t') \ \hat b_{0,n} \hat b_u^\da]& \quad &(n \neq 0),\\
& \tilde b_{m,0}(t) = e^{\frac{1}{2}\kappa m(t-\tau)} [\hat b_{m,0} - \sqrt{\kappa_c}  \int_\tau^t \dif t' \ f_u(t') \ \hat b_{m,1} \hat b_u^\da]& \quad &(m \neq 1),
\end{aligned}\right.
\end{equation}
where for simplicity we have denoted
\begin{equation}
f_u(t) := g_u(t) e^{\frac{1}{2} \int_\tau^{t} (|g_u|^2-\kappa)\dif t^\prime}.
\end{equation}
Finally, we can get the expression of $\tilde b_{1,0}(t)$ as
\begin{equation}\label{eq:tilde_b_sol_3}
\tilde b_{1,0}(t) = e^{\frac{1}{2}\kappa(t-\tau)} \left\{\hat b_{1,0} + \sqrt{\kappa_c} \int_\tau^t \dif t' \ f_u(t') (\hat b_{0,0} - \hat b_{1,1}) \hat b_u^\da  - \kappa_c \left[\int_\tau^t \dif t' \ f_u(t')\right]^2 \hat b_{0,1} \hat b_u^{\da 2}\right\},
\end{equation}
which concludes the solution for Eq.~\eqref{eq:dot_tilde_ops_bu}.

Now we can write down the master equation for the transformed density operator $\tilde \rho_{eu}(t)$. Similar to the optical case, here we have
\begin{equation}
\frac{\dif\tilde \rho_{eu}(t)}{\dif t} = \tilde L_{eu}(t) \tilde \rho_{eu}(t) [\tilde L_{eu}(t)]^\da + \kappa_i \tilde b_{0,1}(t) \tilde \rho_{eu}(t) [\tilde b_{0,1}(t)]^\da + \sum_{n=2}^{+\infty} \kappa n \tilde b_{n-1,n}(t) \tilde \rho_{eu}(t) [\tilde b_{n-1,n}(t)]^\da,
\end{equation}
where
\begin{equation}
\left\{\begin{aligned}
& \tilde L_{eu}(t) = \frac{e^{-\kappa(t-\tau)/2}}{1-(1-\epsilon_u)e^{-\kappa(t-\tau)}} \sqrt{\epsilon_u} [\sqrt{\epsilon_u \kappa_c} \hat b_{0,1} - \sqrt{(1-\epsilon_u)\kappa} \hat b_u],\\
& \tilde b_{1,2}(t) = e^{-\frac{1}{2}\kappa(t-\tau)} \left[\hat b_{1,2} - \sqrt{\xi_u/\epsilon_u} (1-e^{-\kappa(t-\tau)})\hat b_{0,2} \hat b_u^\da \right],\\
&\tilde b_{n-1,n}(t) = e^{-\frac{1}{2}\kappa(t-\tau)} \hat b_{n-1,n} \quad (n\neq 2).
\end{aligned}\right.
\end{equation}
Here we also denote $\xi_u := \frac{\kappa_c}{\kappa}(1-\epsilon_u)$. Notice that either $\tilde L_{eu}(t)$ or $\tilde b_{n-1,n}(t)$ will destroy an excitation in the microwave-downstream composite system, so given the initial Fock basis input $\tilde \rho_{eu}(\tau) =  \rho_{eu}(\tau) =\ketbra{(n_L)_e, (m_L)_u}{(n_R)_e, (m_R)_u}$, we can get $\tilde \rho_{eu}(t)$ as
\begin{equation}\label{eq:tilde_rho_eu_sol}
\begin{split}
\tilde \rho_{eu}(t) = \tilde \rho_{eu}(\tau) + \sum_{\eta_1, \dots, \eta_N} \int_\tau^t \dif t_N \int_\tau^{t_N} \dif t_{N-1} \dots &\int_\tau^{t_2} \dif t_1 \\
\tilde L_{\eta_N}(t_N) \tilde L_{\eta_{N-1}}(t_{N-1}) \dots \tilde L_{\eta_1}(t_1) &\tilde \rho_{eu}(\tau) [\tilde L_{\eta_1}(t_1)]^\da \dots [\tilde L_{\eta_{N-1}}(t_{N-1})]^\da [\tilde L_{\eta_{N}}(t_{N})]^\da,
\end{split}
\end{equation}
where $N = \min\{(n_L + m_L), (n_R + m_R)\}$ and $\tilde L_{\eta_i} \in \{ \tilde L_{eu}, \sqrt{\kappa_i} \tilde b_{0,1}, \sqrt{\kappa n} \tilde b_{n-1,n}\ (n>1)\}$, in accordance with Eq.~\eqref{eq:tilde_rho_ov_sol}. Notice that, if both $n_L, n_R < 2$, then we will get the solution with the same form as the optical case due to the fact that the effect of Kerr will not contribute to the dynamics in this situation. On the other hand, if $\max \{n_L, n_R\} \geq 2$ and $n_L \neq n_R$, then all the $\tilde L_{\eta_i}$ in Eq.~\eqref{eq:tilde_rho_eu_sol} can only be chosen as $\tilde L_{eu}$. This is because that we will get a trivial result when acting $\tilde b_{n-1, n}$ on both sides of $\ketbra{(n_1)_e}{(n_2)_e}$ if $n_1 \neq n_2$, and repetitively applying $\tilde L_{eu}(t_i)$ on both sides of $\tilde \rho_{eu}(\tau)$ cannot lead to a term with equal microwave excitations on both sides under the assumptions that $\max \{n_L, n_R\} \geq 2$ and $n_L \neq n_R$. These observations help to simplify the calculation for $\tilde \rho_{eu}(+\infty)$ given specific Fock basis input $\tilde \rho_{eu}(\tau)$.

In the end, if we get the expression for $\tilde \rho_{eu}(+\infty)$, we need to transform it back to $\rho_{eu}(+\infty)$. It will be helpful if we first get the expressions of the transformation $\hat O_{eu}(+\infty)\hat b_{n,0}\hat O_{eu}^{-1}(+\infty)$ and $\hat O_{eu}(+\infty)\hat b_u^\da \hat O_{eu}^{-1}(+\infty)$ first. From Eq.~\eqref{eq:tilde_b_sol_1}, \eqref{eq:tilde_b_sol_2} and \eqref{eq:tilde_b_sol_3}, it is easy to get
\begin{equation}
\left\{\begin{aligned}
& \hat O_{eu}(t)\hat b_u^\da \hat O_{eu}^{-1}(t) = e^{-\frac{1}{2} \int_\tau^{t} |g_u|^2\dif t'} \hat b_u^\da.\\
& \hat O_{eu}(t)\hat b_{m,n} \hat O_{eu}^{-1}(t) = e^{-\frac{1}{2}(m-n)\kappa(t-\tau)} \hat b_{m,n} \quad (m\neq 1\ \text{and}\ n \neq 0),\\
& \hat O_{eu}(t)\hat b_{m,0} \hat O_{eu}^{-1}(t) = e^{-\frac{1}{2}\kappa m(t-\tau)} \hat b_{m,0} + \sqrt{\kappa_c}  \int_\tau^t \dif t' \ f_u(t') \ \hat O_{eu}(t)\hat b_{m,1} \hat b_u^\da \hat O_{eu}^{-1}(t) \quad (m \neq 1),\\
& \hat O_{eu}(t) (\hat b_{0,0} - \hat b_{1,1}) \hat O_{eu}^{-1}(t) = (\hat b_{0,0} - \hat b_{1,1}) + 2\sqrt{\kappa_c} \int_\tau^t \dif t' \ f_u(t') \ \hat O_{eu}(t)\hat b_{0,1} \hat b_u^\da \hat O_{eu}^{-1}(t),\\
& \hat O_{eu}(t)\hat b_{1,0} \hat O_{eu}^{-1}(t) = e^{-\frac{1}{2}\kappa(t-\tau)} \hat b_{1,0} - \sqrt{\kappa_c} \int_\tau^t \dif t' \ f_u(t') \hat O_{eu}(t) (\hat b_{0,0} - \hat b_{1,1}) \hat b_u^\da \hat O_{eu}^{-1}(t)\\
&\qquad\qquad\qquad\qquad + \kappa_c \left[\int_\tau^t \dif t' \ f_u(t')\right]^2 \hat O_{eu}(t) \hat b_{0,1} \hat b_u^{\da 2} \hat O_{eu}^{-1}(t).
\end{aligned}\right.
\end{equation}
By taking $t \to +\infty$, we are able to get 
\begin{equation}
\left\{\begin{aligned}
& \hat O_{eu}(+\infty)\hat b_u^\da \hat O_{eu}^{-1}(+\infty) = \sqrt{\epsilon_u} \hat b_u^\da,\\
& \hat O_{eu}(+\infty)\hat b_{1,0}^\da \hat O_{eu}^{-1}(+\infty) \ket{0_e}= \ket{0_e} \otimes \sqrt{\xi_u} \hat b_u^\da,\\
& \hat O_{eu}(+\infty)\hat b_{m,0}^\da \hat O_{eu}^{-1}(+\infty) = 0 \quad (m \geq 2),
\end{aligned}\right.
\end{equation}
where we have used the explicit formula of $g_u(t)$ (and therefore $u(t)$, c.f. Eq.~\eqref{eq:ut_explicit}). Given this, we can achieve the transformation of the Fock basis operator
\begin{equation}
\begin{split}
 & \hat O_{eu}(+\infty)\ketbra{(n_L)_o, (m_L)_v}{(n_R)_o, (m_R)_v}  \hat O_{eu}^{-1}(+\infty)\\
  ={}& \hat O_{eu}(+\infty)   \hat b_{n_L,0}\frac{(\hat a_v^{\da})^{m_L}}{\sqrt{m_L!}} \hat O_{eu}^{-1}(+\infty) \hat O_{eu}(+\infty)\ketbra{0_e, 0_u}{0_e, 0_u} \hat O_{eu}^\da(+\infty)[\hat O_{eu}^{-1}(+\infty)]^\da \hat b_{0, n_R} \frac{\hat a_v^{m_R}}{\sqrt{m_R!}}\hat O_{eu}^\da(+\infty)\\
 ={}&\epsilon_u^{\frac{m_L+m_R}{2}} \xi_u^{\frac{\delta_{n_L,1} + \delta_{n_R,1}}{2}} \frac{(\hat b_u^{\da})^{m_L+\delta_{n_L,1}}}{\sqrt{m_L!}}\ketbra{0_e, 0_u}{0_e, 0_u}\frac{\hat b_u^{m_R+\delta_{n_R,1}}}{\sqrt{m_R!}} \\
 ={}& \sqrt{\frac{\epsilon_u^{m_L+m_R} \xi_u^{\delta_{n_L,1} + \delta_{n_R,1}}  (m_L+\delta_{n_L,1})! (m_R+\delta_{n_R,1})!}{(m_L)!(m_R)!}} \ketbra{0_e, (m_L+\delta_{n_L,1})_u}{0_e, (m_R+\delta_{n_R,1})_u}.
\end{split}
\end{equation}
This enables us to calculate the transformation from $\tilde \rho_{eu}(+\infty)$ to $\rho_{eu}(+\infty)$. Similar to the optical case, there will be no excitation in microwave mode at the end of the evolution. Here we also list the result of $\rho_{u}(+\infty)$ for a given Fock basis input $\rho_{eu}(\tau)$ in Table.~\ref{table:microwave}.

\begin{table*}
\caption{The final microwave downstream states $\rho_u(+\infty)$ given some Fock-basis input operators at time $t=\tau$, i.e. $\rho_{eu}(\tau) = \ketbra{(n_L)_e, (m_L)_u}{(n_R)_e, (m_R)_u}$. When both $n_L \leq 1$ and $n_R \leq 1$, it will lead to the result with the same form as the optical case (shown in Table.~\ref{table:optical}). The difference only comes from the substitution of all the subscripts from $v$ to $u$.}
\begin{ruledtabular}\label{table:microwave}
\begin{tabular}{|c|c|c|}
$\rho_{eu}(\tau)$ & $\tilde\rho_{eu}(+\infty)$ & $\rho_{u}(+\infty)$\\ \hline
\makecell{$\ketbra{(n_L)_e, 0_u}{(n_R)_e, 0_u}$, \\ $n_L \in \{2,3\}, n_R \in \{0,1\}$} & $\ketbra{(n_L)_e, 0_u}{(n_R)_e, 0_u}$ & 0\\ \hline
$\ketbra{2_e, 0_u}{2_e, 0_u}$ & $\begin{aligned} &\ketbra{2_e, 0_u}{2_e, 0_u} + 2\ketbra{1_e, 0_u}{1_e, 0_u} - \sqrt{\xi_u/\epsilon_u} (\ketbra{0_e, 1_u}{1_e, 0_u} + \hc) \\ +{}& (2\xi_u)/(3\epsilon_u) \ketbra{0_e, 1_u}{0_e, 1_u} +  (1 - 2\xi_u/3) \ketbra{0_e, 0_u}{0_e, 0_u}
\end{aligned}$
 & $2\xi_u/3 \ketbra{1_u}{1_u} + (1 - 2\xi_u/3) \ketbra{0_u}{0_u}$
\end{tabular}
\end{ruledtabular}
\end{table*}

\subsection{Infidelity on the final output state}

Ultimately, we can summarize the previous derivations and get the output Bell state fidelity. Due to the perturbative corrections from finite blockade and nonzero loss, the density matrix at time $t=\tau$ can be written as
\begin{equation}
    \rho(\tau) = [\rho^{(0)}(\tau) + \rho^{(1)}_\chi(\tau) + \rho^{(2)}_\chi(\tau)] \otimes \ketbra{0_v, 0_u}{0_v, 0_u} + \rho^{(1/2)}_\kappa(\tau) + \rho^{(1)}_\kappa(\tau),
\end{equation}
where the explicit formulae for each term above can be found in Eq.~\eqref{eq:dm-zth-tls}, \eqref{eq:rho_chi_sigma}, \eqref{eq:rho_kappa_half_sigma} and \eqref{eq:rho_kappa_1_sigma}. Then we use the results shown in Table.~\ref{table:optical} and \ref{table:microwave} to see how the quantum states at time $t=\tau$ evolve into the downstream modes as $t\to \infty$. We denote this process as $\mathcal{E}$, and list the relevant outcomes in the following:
\begin{equation}
\left\{\begin{aligned}
& \begin{aligned}
\bra{0_v,0_u}\mathcal{E}[\rho(\tau)]\ket{0_v,0_u} ={}& \frac{1+(1-\xi_v)(1-\xi_u)}{2} + \frac{g^2}{2\chi_e^2}[1-2(1-\xi_v)(1-\xi_u) + (1-\xi_v)^2(1-2\xi_u/3)]\\
&-\sqrt{\frac{2\kappa_c}{g}}[\sqrt{C_v\epsilon_v \xi_v}(1-\xi_u) + \sqrt{C_{u,1} \epsilon_u \xi_u}(1-\xi_v)]+\frac{\kappa}{g}\Big[C_v(1-\xi_u) + C_{u,1}(1-\xi_v)\\
& + C_{u,2}(1-\xi_v)^2 (1-\xi_u)-(C_v-1/4) - (C_v+1/4)(1-\xi_v)(1-\xi_u)\Big],
\end{aligned}\\
& \begin{aligned}
\bra{1_v,1_u}\mathcal{E}[\rho(\tau)]\ket{1_v,1_u} ={}& \frac{\xi_v \xi_u}{2} -\frac{g^2}{\chi_e^2}\xi_v\xi_u[1 - 2(1-\xi_v)/3] + \sqrt{\frac{2\kappa_c \xi_v \xi_u}{g}}(\sqrt{C_v\epsilon_v \xi_u} + \sqrt{C_{u,1} \epsilon_u \xi_v})\\
&+\frac{\kappa}{g}\xi_v \xi_u[2C_{u,2}(1-\xi_v) - (C_v+1/4)],
\end{aligned}\\
& \begin{aligned}
\text{Re}\{\bra{0_v,0_u}\mathcal{E}[\rho(\tau)]\ket{1_v,1_u}\} ={}& \frac{1}{2} \left(1-\frac{g^2}{\chi_e^2}\right)\sqrt{\xi_v\xi_u} + \sqrt{\frac{\kappa_c}{2g}}\Big[\sqrt{C_v\epsilon_v \xi_u} + \sqrt{C_{u,1} \epsilon_u \xi_v} \\
&+ \sqrt{2 C^2_{u,3} \epsilon_u \xi_v/C_{u,1}}(1-\xi_v)(1-2\xi_u)\Big] +\frac{\kappa}{g}\sqrt{\xi_v \xi_u}[\sqrt{2}C_{u,3}(1-\xi_v) - C_v],
\end{aligned}\\
&\text{Im}\{\bra{0_v,0_u}\mathcal{E}[\rho(\tau)]\ket{1_v,1_u}\} = -\frac{g\sqrt{\xi_v\xi_u}}{2\chi_e}.
\end{aligned}\right.
\end{equation}
Here we only keep the perturbative corrections with order up to $O(\kappa/g)$. Notice that, since $\epsilon_{v(u)} \sim O(\kappa/g)$, we can clearly see from the results above that the correction from $\rho_\kappa^{(1/2)}$ also scales as $O(\kappa/g)$ rather than $O(\sqrt{\kappa/g})$.

The fidelity $F$ of the output Bell state can now be expressed as 
\begin{equation}
    F = \frac{1}{2}\left(\bra{0_v,0_u}\mathcal{E}[\rho(\tau)]\ket{0_v,0_u} + \bra{1_v,1_u}\mathcal{E}[\rho(\tau)]\ket{1_v,1_u}\right) + \left|\bra{0_v,0_u}\mathcal{E}[\rho(\tau)]\ket{1_v,1_u}\right|,
\end{equation} 
where the fidelity is compared with the state $\frac{\ket{0_v,0_u} + e^{i\phi}\ket{1_v,1_u}}{\sqrt{2}}$ with optimally chosen relative phase $\phi$. Further, since $\text{Im}\{\bra{0_v,0_u}\mathcal{E}[\rho(\tau)]\ket{1_v,1_u}\} \sim O(g/\chi_e)$ and $\text{Re}\{\bra{0_v,0_u}\mathcal{E}[\rho(\tau)]\ket{1_v,1_u}\}\simeq 1/2$, the expression of $F$ can be simplified as
\begin{equation}
\begin{split}
    F \simeq{}& \frac{1}{2}\left(\bra{0_v,0_u}\mathcal{E}[\rho(\tau)]\ket{0_v,0_u} + \bra{1_v,1_u}\mathcal{E}[\rho(\tau)]\ket{1_v,1_u}\right)\\
    &+ \text{Re}\{\bra{0_v,0_u}\mathcal{E}[\rho(\tau)]\ket{1_v,1_u}\} + \left(\text{Im}\{\bra{0_v,0_u}\mathcal{E}[\rho(\tau)]\ket{1_v,1_u}\}\right)^2.    
\end{split}
\end{equation} 
We can also verify that $\rho_\chi^{(1)}$ leads to the correction of the final output fidelity with the perturbative order $O(g^2/\chi_e^2)$, rather than $O(g/\chi_e)$.

\begin{table}[t!]
\caption{Bell Fidelity at the $(\kappa_c,\chi_e)/2\pi=(2,100)$MHz point calculated from three different approaches.}
%\begin{center}
\begin{tabular}{| c| c|c | c |}
\hline
\hline
$\kappa_i$ (MHz) & Eq.~(\ref{empirical}) & Numerical Value &  First-order Perturbation     \\
\hline
0 &$ 0.993 \pm 0.002 $ &  $  0.995$   & $ 0.988$  \\
\hline
0.1 &$ 0.958 \pm 0.002 $ &  $  0.950$   & $ 0.940$  \\
\hline
0.2 &$ 0.926 \pm 0.003 $ &  $  0.911$   & $ 0.900$  \\
\hline
\hline
\end{tabular}
%\end{center}
\label{tab1}
\end{table}

To see Eq.~\ref{empirical} explicitly, we note that the expression for fidelity can be arranged as a linear combination of $\kappa/g$ terms and the $g^2/\chi_e^2$ term. To zeroth-order, the parameters can be simplified as $\epsilon_v  = \frac{2C_v\kappa / g}{1 + 2C_v \kappa / g} \sim 2C_v\kappa / g$  and $\epsilon_{u}  = \frac{2C_{u,1}\kappa / g}{1 + 2C_{u,1} \kappa / g} \sim 2C_{u,1}\kappa / g$. The $\xi_v := \frac{\kappa_c}{\kappa}(1-\epsilon_v)\sim \frac{\kappa_c}{\kappa}\sim 1$ and $\xi_u := \frac{\kappa_c}{\kappa}(1-\epsilon_u)\sim \frac{\kappa_c}{\kappa}\sim 1$ give the $\kappa_c/\kappa$ contribution, which is of order 1. The $\kappa_i/\kappa$ term comes from
\begin{equation}
    \frac{1+(1-\xi_v)(1-\xi_u)}{4} + \frac{\xi_v \xi_u}{4} + \frac{1}{2} \sqrt{\xi_v\xi_u} \sim -0.5\frac{\kappa_i}{\kappa}.
\end{equation}
The $g^2/\chi_e^2$ contribution comes from 
\begin{equation}
    \frac{g^2}{4\chi_e^2}[1-2(1-\xi_v)(1-\xi_u) + (1-\xi_v)^2(1-2\xi_u/3)] - \frac{g^2}{2\chi_e^2}\xi_v\xi_u[1 - 2(1-\xi_v)/3] - \frac{1}{2} \frac{g^2}{\chi_e^2}\sqrt{\xi_v\xi_u} + \frac{g^2\xi_v\xi_u}{4\chi^2_e}\sim -0.5\frac{g^2}{\chi_e^2}. 
\end{equation}
The $\kappa_c/g$ and $\kappa/g$ terms will combine to make $\kappa_i/g$ and $\kappa_c/g$, accounting for two different loss mechanisms. Firstly, we will have 
\begin{equation}
\begin{split}
    &\frac{1}{2}\left(-\sqrt{\frac{2\kappa_c}{g}}[\sqrt{C_v\epsilon_v \xi_v}(1-\xi_u) + \sqrt{C_{u,1} \epsilon_u \xi_u}(1-\xi_v)] + \sqrt{\frac{2\kappa_c \xi_v \xi_u}{g}}(\sqrt{C_v\epsilon_v \xi_u} + \sqrt{C_{u,1} \epsilon_u \xi_v})\right)  \\ & +\sqrt{\frac{\kappa_c}{2g}}\Big[\sqrt{C_v\epsilon_v \xi_u} + \sqrt{C_{u,1} \epsilon_u \xi_v} 
+ \sqrt{2 C^2_{u,3} \epsilon_u \xi_v/C_{u,1}}(1-\xi_v)(1-2\xi_u)\Big] \sim (C_v + C_{u,1})\frac{\kappa_c}{g}.
\end{split}
\end{equation}
Then, 
\begin{equation}
\begin{split}
    &\frac{\kappa}{2g}\Big[C_v(1-\xi_u) + C_{u,1}(1-\xi_v)
 + C_{u,2}(1-\xi_v)^2 (1-\xi_u)-(C_v-1/4) - (C_v+1/4)(1-\xi_v)(1-\xi_u)\Big] \\&+ \frac{\kappa}{2g}\xi_v \xi_u[2C_{u,2}(1-\xi_v) - (C_v+1/4)] + \frac{\kappa}{g}\sqrt{\xi_v \xi_u}[\sqrt{2}C_{u,3}(1-\xi_v) - C_v]\sim -2C_v\frac{\kappa}{g}.
\end{split}
\end{equation}
Plugging in the value for the constants, we have the last two equations give $-0.02\kappa_c/g-0.28\kappa_i/g$. The perturbation theory fixes the coefficient of the linear combination: 
\begin{equation}
    1-F \approx  0.02\frac{\kappa_c}{g} + 0.28
 \frac{\kappa_i}{g}  + 0.5 \frac{\kappa_i}{\kappa} + 0.5\frac{g^2}{\chi_e^2}. 
\end{equation}
The difference with Eq.~\ref{empirical} (and Eq.~\ref{empirical2}) can be explained by our choice of fitting parameter regime. In fact, we have $\kappa/g>0.1$ and $\kappa_i/\kappa>0.1$. Another reason is that the numerical fit automatically optimized the pulse duration, whereas the perturbation calculation has fixed $\tau=\pi/4g$, which could cause corrections from addition terms involving $g$. Taking $\kappa_c/\kappa \sim 0.75$, which is closer to the numerical range we used, the coefficient of $g^2/\chi_e^2$ is modified to $0.24$, which is closer to Eq.~\ref{empirical}. We can view Eq.~\ref{empirical} as capturing only the linear dependence of the fidelity on the small parameters. 
% \begin{equation}
% \begin{split}
%     &\mathcal{E}[\rho^{(0)}(\tau)\otimes\ketbra{0_v, 0_u}{0_v, 0_u}] = \frac{1}{2}\ketbra{0_v, 0_u}{0_v, 0_u} + \frac{\sqrt{\xi_v \xi_u}}{2} (\ketbra{1_v, 1_u}{0_v, 0_u} + \hc)\\
%     & + [\xi_v\ketbra{1_v}{1_v} + (1-\xi_v)\ketbra{0_v}{0_v}]\otimes [\xi_u\ketbra{1_u}{1_u} + (1-\xi_u)\ketbra{0_u}{0_u}],
% \end{split}
% \end{equation}
% \begin{equation}
%     \mathcal{E}[\rho^{(1)}_\chi(\tau)\otimes\ketbra{0_v, 0_u}{0_v, 0_u}] = \frac{(-i) g\sqrt{\xi_v \xi_u}}{2\chi_e} (\ketbra{0_v, 0_u}{1_v, 1_u} - \hc),
% \end{equation}
% \begin{equation}
% \begin{split}
%     &\mathcal{E}[\rho_\chi^{(2)}(\tau)\otimes\ketbra{0_v, 0_u}{0_v, 0_u}] = \frac{g^2}{2\chi_e^2}\Big\{\ketbra{0_v, 0_u}{0_v, 0_u} - \sqrt{\xi_v \xi_u} (\ketbra{1_v, 1_u}{0_v, 0_u} + \hc)\\
%     & -2[\xi_v\ketbra{1_v}{1_v} + (1-\xi_v)\ketbra{0_v}{0_v}]\otimes [\xi_u\ketbra{1_u}{1_u} + (1-\xi_u)\ketbra{0_u}{0_u}]\\
%     & +[\xi_v^2\ketbra{2_v}{2_v} + 2\xi_v(1-\xi_v)\ketbra{1_v}{1_v} + (1-\xi_v)^2\ketbra{0_v}{0_v}]\otimes [2\xi_u/3\ketbra{1_u}{1_u} + (1-2\xi_u/3)\ketbra{0_u}{0_u}]\Big\},
% \end{split}
% \end{equation}
% \begin{equation}
% \begin{split}
%     &\mathcal{E}[\rho_\kappa^{(1/2)}(\tau)] = \sqrt{\frac{\kappa_c \epsilon_v C_v}{2g}} \{ \sqrt{\xi_u}(\ketbra{1_v,1_u}{0_v,0_u}+\hc) + 2\sqrt{\xi_v} (\ketbra{1_v}{1_v} - \ketbra{0_v}{0_v})\otimes [\xi_u \ketbra{1_u}{1_u} + (1-\xi_u) \ketbra{0_u}{0_u}]
% \end{split}
% \end{equation}

\end{document}